\documentclass{aastex631}

\graphicspath{{./}{}}
\usepackage{xspace} 
\newcommand{\TESS}{TESS\xspace}
\newcommand{\Kepler}{\textit{Kepler}\xspace}

\received{Published in the Astronomical Journal, 167, \#202, 2024}

\begin{document}

\title{DIAmante \TESS AutoRegressive Planet Search (DTARPS): I. Analysis of 0.9 Million Light Curves}

\author{Elizabeth J. Melton}
\affiliation{Department of Astronomy \& Astrophysics, Pennsylvania State University, University Park, PA 16802, USA}
\affiliation{Center for Exoplanets and Habitable Worlds, 525 Davey Laboratory, The Pennsylvania State University, University Park, PA, 16802, USA.}

\author{Eric D. Feigelson}
\affiliation{Department of Astronomy \& Astrophysics, Pennsylvania State University, University Park, PA 16802, USA}
\affiliation{Center for Exoplanets and Habitable Worlds, 525 Davey Laboratory, The Pennsylvania State University, University Park, PA, 16802, USA.}
\affiliation{Center for Astrostatistics, 525 Davey Laboratory, The Pennsylvania State University, University Park, PA, 16802, USA.}

\author{Marco Montalto}
\affiliation{INAF - Osservatorio Astrofisico di Catania, Via S. Sofia 78, I-95123 Catania, Italy}

\author{Gabriel A. Caceres}
\affiliation{EY-Parthenon, 1540 Broadway, New York, NY 10036, USA}

\author{Andrew W. Rosenswie}
\affiliation{Department of Astronomy \& Astrophysics, Pennsylvania State University, University Park, PA 16802, USA}
\affiliation{Institut f\"{u}r Physik und Astronomie, Universit\"{a}t Potsdam, D-14476 Golm (Potsdam), Germany}
\affiliation{Leibniz-Institut f\"ur Astrophysik Potsdam (AIP), An der Sternwarte 16, D-14482 Potsdam, Germany}

\author{Cullen S. Abelson}
\affiliation{Department of Astronomy \& Astrophysics, Pennsylvania State University, University Park, PA 16802, USA}
\affiliation{Department of Physics and Astronomy, University of Pittsburgh, 100 Allen Hall, 
3941 O'Hara St.,
Pittsburgh, PA 15260, USA}

\begin{abstract}
Nearly one million light curves from the \TESS Year 1 southern hemisphere extracted from Full Field Images with the DIAmante pipeline are processed through the AutoRegressive Planet Search statistical procedure. ARIMA models remove lingering autocorrelated noise, the Transit Comb Filter identifies the strongest periodic signal in the light curve, and a Random Forest machine learning classifier is trained and applied to identify the best potential candidates. Classifier training sets are based on injections of planetary transit signals, eclipsing binaries and other variable stars. The optimized classifier has a True Positive Rate of 92.5\% and a False Positive Rate of 0.43\% from the labeled training set. The result of this DIAmante \TESS autoregressive planet search of the southern ecliptic hemisphere (DTARPS-S) analysis is a list of 7,377 potential exoplanet candidates. The classifier had a 64\% recall rate for previously confirmed exoplanets and a 78\% negative recall rate for known False Positives. The completeness map of the injected planetary signals shows high recall rates for planets with $8 - 30 R_\earth$ radii and periods $0.6-13$ days and poor completeness for planets with radii $< 2 R_\earth$ or periods $< 1$ day. The list has many False Alarms and False Positives that need to be culled with multifaceted vetting operations (Paper II). 
\end{abstract}


\section{Introduction}

\subsection{Challenges in \TESS Planet Discovery} \label{sec:challenges}

With the 2018 launching of the Transiting Exoplanet Survey Satellite (\TESS), scientists acquired a tool for in-depth analysis of rare phenomena such as transiting exoplanets, stellar superflares, and tidal disruption events \citep{Ricker15}. \TESS surveys the entire celestial sphere in month-long observations with four wide-field cameras with 21 \arcsec\/ pixels. During the prime mission (\TESS Years 1 and 2), over 200,000 bright stars were pre-chosen to have 2 minute observing cadence as prime transit targets, but millions of relatively bright stars are accessible from Full Frame Images (FFIs) with 30 minute cadence. 

The principal goal of the TESS mission is the identification of sub-Neptune ($R < 4$~R$_\oplus$) transiting planets around stars sufficiently bright for follow-up characterization of the planets' physical characteristics, including atmospheric composition. Quantitative calculations prior to the mission by \citet{Barclay18} predicted that $\sim  3100$ transiting planets would be found from light curves of $\sim 6$ million FFI stars acquired during prime mission; of these, $\sim 1100$ would be sub-Neptunes. They predicted that $\sim 12,000$ larger planets ($R > 4$~R$_\oplus$) would be discovered in the FFI database.  In a revised calculation, \citet{Kunimoto22} predict that $\sim 4000$ planets would be detected in the prime mission using FFI images.  

The predictions of \citet{Barclay18} and \citet{Kunimoto22} have been overly optimistic.  The \TESS Objects of Interest from the prime mission include 2241 Planet Candidates based on automated detection of a transit-like signal followed by review by the TOI Vetting Team \citep{Guerrero21}.  These include 1035 unique FFI stars obtained with MIT's Quick Look Processing (QLP) pipeline.  Nearly half of these have been subject to some follow-up observations (the community-based ExoFOP-TESS enterprise) of which most (88\% in the published 2021 catalog) have been redesignated False Positives.  Thus, only a few hundred FFI stars $-$ rather than thousands $-$ have emerged to date as reliable hosts of transiting planets from the \TESS prime mission. 

We can point to a variety of plausible contributions to this large discrepancy between predicted and actual performance in \TESS planet discovery:
\begin{enumerate}
    \item Barclay et al. assume a Gaussian noise model for the light curves, although with a generous 7.3$\sigma$ detection threshold and consideration of flux dilution by nearby stars blended in the large pixels. But the \Kepler mission showed that many stars are variable with amplitudes greater than planetary transit depths \citep{Gilliland11} and, even after detrending, exhibit autocorrelated structure \citep{Caceres19b}. Stellar variability has multiple sources: 'red noise' \citep{Pont06}, rotational modulation of starspots \citep{Boisse11}, magnetic reconnection flares \citep{Davenport16}, pulsations, supergranulation, stellar multiplicity, and other effects. 
    
    \item Perhaps the most pernicious source of contamination, eclipsing binaries blended into the large \TESS pixel can, after dilution by the target starlight, produce periodic variations that closely resemble planetary transits. In some cases, it can be very difficult, using either automated classifiers or human vetting, to distinguish planetary candidates from blended eclipsing binaries (BEB). 
    
    \item Vetting efforts to reduce BEB contamination often rely on pixel-based crowding and centroid analysis that can inadvertently eliminate valid planetary candidates in the Galactic Plane.   
    
    \item Instrumental problems are present in addition to stellar variability in TESS FFI light curves.  These include short-term flux variations from the appearance or disappearance of nearby stars as satellite pointing settles after instrument turn-on, and `ghost' light from bright variable stars contaminating wide areas of the image.  The latter effect is called `ephemeris matching' \citep{Coughlin14}. 
    
    \item Methodological difficulties arise in the analysis of the light curves. These include: the sparsity of observations during brief transits at longer periods that hinders Gaussian-based statistical measures; spurious periodogram peaks from deviations from a regular cadence (e.g. periodic instrument closures due to the 13.7 satellite orbit); and mathematical difficulties in reliably evaluating false alarm probabilities in any periodogram \citep[e.g.][]{VanderPlas18, Delisle20, Koen21}.
\end{enumerate}

It is challenging to design a detection procedure that effectively removes such a variety of different stellar variations, and treats the instrumental and methodological problems, while maintaining the planetary transit signal. BEB contamination, particularly at low Galactic latitudes, can occur more frequently than true planetary signals. The QLP -- and any other transiting detection procedure -- must thus adopt  conservative classification and vetting procedures to reliably identify planetary signals in the midst of these varied sources of contamination and methodological difficulties. As \citet{Barclay18} did not consider these issues, it is not surprising that they overestimated the number of smaller, sub-Neptunian planets that can realistically and reliably be found in \TESS FFI data.   

This situation motivates the search for \TESS FII planets using methodologies different from the QLP used to generate the official FFI TOI list\footnote{\url{https://tess.mit.edu/toi-releases}}.  It is quite possible that different statistical approaches to detrending, periodicity searching, automated classification, and human vetting will find some of the true planetary candidates predicted by \citet{Barclay18} and missing from the TOI FII list. 

Several efforts at \TESS planet detection have been reported using methodologies independent of the TOI pipeline described by \citet{Guerrero21} and \citet{Kunimoto22}.  These include: the DIAmante pipeline \citet{Montalto20} identifying 252 new candidates; \citet{Osborn20} identifying 200 \TESS threshold crossing events as new candidates; \citet{Olmschenk21} finding 181 new candidates; the citizen science Planet Hunters \TESS project finding 90 new candidates \citep{Eisner21}; \citet{Rao21} identifying 38 new candidates; \citet{Nardiello20} finding 33 new candidates in stellar clusters; \citet{Feliz21} finding 24 new candidates around M-dwarfs; and \citet{Dong21} identifying 19 new Warm Jupiter candidates.  Together these efforts add over 700 new planetary candidates. 

One major consideration is the statistical procedure for detrending the lightcurves prior to searching for transit-like periodicities.  The QLP pipeline uses a high pass filter, outlier removal, and spline fits to detrend the light curve \citep{Huang20}.  Other procedures use well-known statistical procedures (such as Principal Component Analysis or Gaussian Processes regression) or more advanced signal processing methods (such as Independent Component Analysis, correntropy, empirical mode decomposition, and Singular Spectrum Analysis). 

Following detrending, most analyses seek transiting signals with Box-Least Squares (BLS) periodograms \citep{Kovacs02}. In the QLP pipeline, BLS peaks with local signal-to-noise ratio $>9$ are flagged as threshold-crossing events.  These are then inputted into a convolutional neural network classifier trained on human-labeled \TESS light curves \citep{Shallue18, Yu19}.  The promising cases are then subject to visual inspection by a vetting team to arrive at the TOI list \citep{Guerrero21}.

\subsection{AutoRegressive Planet Search} \label{sec:arps}

\begin{figure}[tb]
  \includegraphics[width=\textwidth]{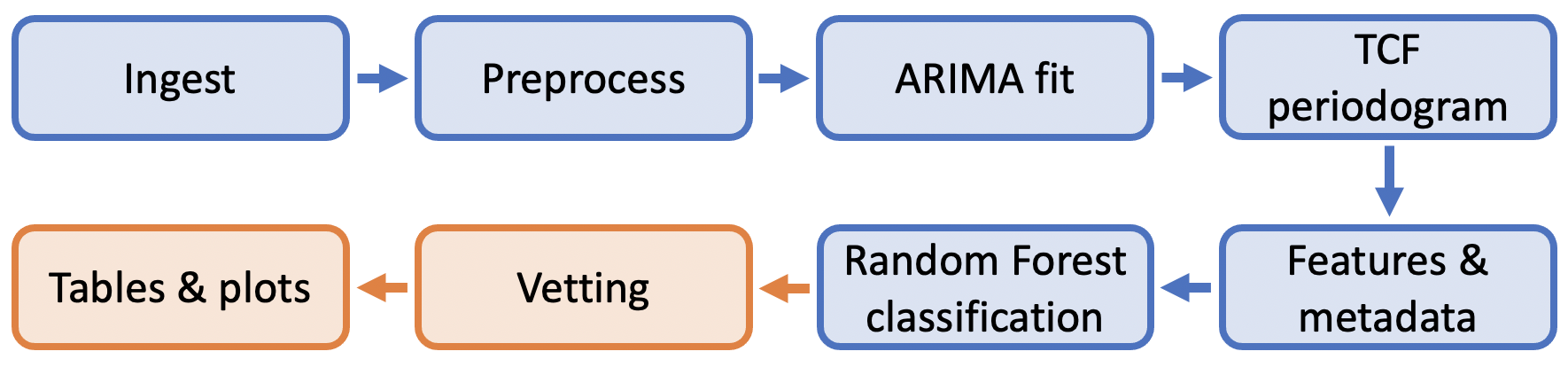}
  \caption{The AutoRegressive Planet Search process. Blue boxes represent steps in the ARPS analysis covered in this paper. Orange boxes represent steps in ARPS covered in Paper II. }
  \label{fig:DTARPS_flow_chart}
\end{figure}

In this paper and its companions, we apply a transit detection procedure shares some similarities to the QLP pipeline but with significant differences.  We adapt the 4-stage AutoRegressive Planet Search (ARPS) developed by \citet{Caceres19b} and applied to the 4-year \Kepler light curves by \citet{Caceres19a}.  The ARPS process is outlined in Figure~\ref{fig:DTARPS_flow_chart}.

The ARPS procedure, presented in detail in \S\ref{sec:methodology}, differs from the QLP pipeline in several respects.  First, non-stationarity in the light curve are removed with a simple nonparametric algorithm called `differencing', rather than a more complicated semi-parametric detrending procedures like spline or Gaussian Processes regression.  Differencing treats both stellar and instrumental variations in a single step, but leaves behind sudden changes such as transit ingress and egress.  Second, we fit low-dimensional parametric autoregressive moving average, or ARMA(p,q), models to remove short-memory autocorrelation in the detrended light curve.  This crucial step is missing in other transit searching pipelines.  Together, these procedures are known as ARIMA or Box-Jenkins analysis that has dominated analysis of stochastic time series analysis since the 1970s in fields such as econometric and engineering signal processing.  The textbook by \citet{Box15} has five edition and over 56,000 citations. 

The differencing operation changes a sharp-edged box-shaped transit into a double-spike representing the planet ingress and egress. This motivated the third innovation of the ARPS procedure, a new sensitive Transit Comb Filter (TCF) periodogram developed by \citet{Caceres19a} to measure the amplitude of periodic double-spike patterns in the ARIMA residuals.  This replaces the traditional BLS method applied to detrended light curves that do not involve differencing. A simulation study by \citet{Gondhalekar23} shows in detail that the ARIMA-TCF sequence outperforms a standard detrender followed by BLS in most \TESS-like light curves.  The TCF periodogram exhibits fewer spurious speaks, reduced heteroscedasticity (noise varying with period) and reduced trend compared to the BLS periodogram.  Combined with ARIMA's removal of short-memory autocorrelation, the ARPS procedure is very sensitive to smaller transiting planets. 

ARPS continues with tuning a Random Forest (RF) classifier based on dozens of features to select possible exoplanet transiting candidates from the vast number of non-transiting systems. Following \citet{Montalto20}, the classifier is trained towards a positive training set of light curves with injected simulated planetary transit signals, and trained away from a negative training set that includes simulated BEB signals. 

The final stage of the ARPS procedure, like the QLP procedure, involves multi-faceted visual visual to reduce False Alarms and False Positives that passed over the Random Forest threshold. This stage is presented in \citet[][Paper II]{Melton22b}.  ARPS is thus a comprehensive planetary transit analysis system, starting with extracted light curves cleaned of most instruemntal effects and ending with a new list of planetary candidate transiting systems. 

When the ARPS method was applied to $\sim 150,000$ \Kepler 4-yr light curves by \citet{Caceres19b}, though without the final vetting step, it recovered 97\% of \Kepler Golden sample providing the \Kepler model had signal-to-noise ratio SNR$>20$, and identified 97 new \Kepler exoplanet transit signals. Most of the new ARPS identified \Kepler planet candidates were (sub)Earths with $P<20$ day periods orbiting faint stars that could not be readily confirmed with follow-up radial velocity spectroscopy. One case, a Mars-size planet orbiting an M star, is discussed by \citet{Canas21}. The application to \TESS data discussed here produces a collection of candidates that is much more accessible to follow-up study than the earlier \Kepler ARPS study.

\subsection{DTARPS-S: Application to \TESS Year 1 Southern Ecliptic Hemisphere} \label{sec:DTARPS}

In this paper, together with Paper II and \citet[][Paper III]{Melton22c}, we combine procedures from two pipelines for \TESS FFI exoplanet detection: the DIAmante pipeline developed by \citet[][henceforth M20]{Montalto20} for light curve extraction and preprocessing, and the ARPS pipeline outlined above for candidate transiting planet detection.  M20 extract $0.9$ million light curves from \TESS Year 1 FFIs covering the southern ecliptic hemisphere.  Using the BLS periodogram and a RF classifier, they proceed to identify 396 exoplanet candidates. The present study is based on the 0.9 million DIAmante extracted and preprocessed light curves, but then diverts the analysis to the ARPS procedure for planetary transit identification. We call the combined effort the  DIAmante \TESS AutoRegressive Planet Search or DTARPS. Application to the \TESS Year 1 southern ecliptic hemisphere data is called DTARPS-S.  The Year 1 full-frame images are available at the \citet{MAST22} and the extracted DIAmante light curves are available at \citet{Montalto20b}. 

The DTARPS-S study of \TESS Year 1 FFIs covering the southern ecliptic hemisphere is presented in three papers. Paper I here (Melton et al. 2024a) describes the application of the ARPS method through the application of the RF classifier to the data (Figure \ref{fig:DTARPS_flow_chart}), producing a list of 7,377 stars that exhibit transit-like behavior.  This stage is roughly comparable to the threshold-crossing events (TCEs) of the QLP TOI analysis \citep{Guerrero21}.  Paper II (Melton et al. 2024b) describes the rigorous multi-faceted vetting procedure applied to the RF classifier results and presents the final list of 463 DTARPS-S candidates.  Paper III (Melton et al. 2024c) provides an initial scientific analysis of these candidates. 

Our papers are purposefully more detailed than most presentations of transiting exoplanet discoveries.  For example, in the present study, we analyze the performance of the RF classifier with respect to planetary injections, astronomically confirmed planets, previously identified planetary candidates and False Positives in the full DIAmante data set. We present a thorough analysis of the performance of the ARPS methodology on injected planetary signals into the DIAmante \TESS FFI light curves and the performance of the RF classifier on these synthetic planetary injections.  

This level of detail has two benefits.  First, it allows us to analyze, and seek to improve, each step of the DTARPS pipeline.  Second, it provides a rigorous foundation for science results such as the first \TESS-based planet occurrence rate calculation in \citet[][Paper III]{Melton22c}.  

The present paper is structured as follows. The ARPS methodology, updated from \citet{Caceres19a}, is presented in \S\S2.1-2.3.  DIAmante extraction and preprocessing is outlined in \S\ref{sec:DIAmante}. Section \ref{sec:ARIMA} discusses the ARIMA modeling and TCF periodogram for detrending and transit search. Sections \ref{sec:rf_training} through \ref{sec:rf_final} describe the RF classifier training set (including creating synthetic injections), the process of RF optimization and the final RF classifier. The performance of the final RF classifier are presented in Section \ref{sec:rf_final} producing a list of 7,377 \TESS stars exhibiting transit-like behavior (\S\ref{sec:DAL}).  Section \ref{sec:TCF_acc} discusses the accuracy of the TCF routine for transit fitting. Section \ref{sec:other} compares the results of the RF classifier with other exoplanet surveys and Section \ref{sec:complete} discusses the completeness of the RF classifier. The principal product is the DTARPS-S Analysis List of 7,377 \TESS light curves in \S\ref{sec:DAL}.  The findings are summarized in \S\ref{sec:summary1} with motivation for the vetting stage described in Paper II.  The Appendix describes external data sets used for comparison and validation of this effort.

\section{DTARPS-S Methodology \label{sec:methodology}}

\subsection{Background: Detrending and Transit Identification} 

Any transiting exoplanet search must try to remove a wide range of stellar variability behaviors and variations due to instrumental effects. 
Detrending procedures used in transit detection outlined in the Appendix include: NASA MIT Quick Look pipeline \citep{Huang20} which uses a high pass filter and fitted splines to detrend the light curve; \texttt{eleanor} pipeline \citep{Feinstein19} that cotrends the extracted light curves using Principal Component Analysis (PCA); photometry extraction with difference image analysis \citep{Oelkers18}; DIAmante pipeline \citep{Montalto20} with difference image analysis, PCA cotrending, and individual star spline fits; NEMISIS \citep{Feliz21} for M-dwarf targets which combines pixel-level decorrelation with an iterative smoother. In the search for transiting planets, common choice for light curve detrending is Gaussian Processes regression \citep[e.g.,][]{Luger16, Angus18} but other methods have been tried such as Independent Component Analysis \citep{Waldmann12}, correntropy \citep{Huijse12}, empirical mode decomposition \citep{Roberts13}, and Singular Spectrum Analysis \citep{Greco16}.

The analysis pipeline used for transit detection of pre-selected \TESS stars observed with rapid cadence is the NASA-Ames Science Processing Operations Center pipeline \citep{Jenkins17b, Guerrero21}. It involves a complex series of operations including autoregressive filling of gaps, removal of some instrumental systemic effects, whitening with power spectral density analysis, removal of multiscale temporal structures with wavelet analysis, and identification of transiting exoplanet signals with adaptive wavelet-based matched filters. A statistical bootstrap test is applied to the light curve and transit detection to determine the probability of the event being a false alarm \citep{Jenkins17c}.

Following detrending in most analyses, transiting signals are sought with Box-Least Squares (BLS) periodograms \citep{Kovacs02}. The BLS method models a transit signal as simple box-shape based on the fraction of duration (duration/period), the transit depth, and the epoch of the transit and relies on the anticipated rigid-shape of the transit light curve to identify transiting exoplanets. For each period being searched for an associated transit signal, BLS utilizes a least squares algorithm to fit the other box-model transit parameters to the folded light curve and returns the signal residue as the `strength-of-fit' measure to create the periodogram. When handling a large number of observations, BLS bins the folded light curve data into small bins with respect to the expected transit duration \citep{Kovacs02}. Variants to BLS include the transit least squares (TLS) algorithm that considers the effect of stellar limb darkening during planetary ingress and egress \citep{Hippke19} and the fast BLS computational algorithm \citep{Shahaf21}.

\subsection{AutoRegressive Planet Search: ARIMA and TCF \label{sec:ARPS_intro}} 

Stellar activity is often classified as an autoregressive behavior, wherein future photometric values depend on current and past values \citep{Caceres19b}. For example, the waiting time between X-ray flares have strong temporal autocorrelation on timescales of hours \citep{Wheatland00, Aschwanden10}. Standard nonparametric detrending procedures with a kernel or  window $-$ such as running medians, spline fitting or Gaussian Processes regression $-$ will not remove short-memory stochastic autocorrelated behaviors. But short-memory autocorrelation can be removed with low-dimensional parametric regressions such as ARMA models that are specifically designed to fit stochastic autocorrelated behaviors. ARMA-type modeling is well-established in many fields of time series analysis with extensive methodology described in textbooks such as \citet{Box15}, \citet{Chatfield19}, and \citet{Hyndman21}. \citet{Feigelson18} argue that these models can be effective for many time domain problems in astronomy. 

For transiting planet identification, stationary low-dimensional linear autoregressive models are combined with a simple differencing operator to remove non-stationary trends arising from stellar and instrumental variations. These ARIMA models are flexible low-dimensional parametric models fit by maximum likelihood estimation without any free parameters. The best ARIMA model is chosen by balancing improvement in likelihood with model complexity. \citet{Caceres19a} show that the shape of a transit is transformed from a box to a double-spike by ARIMA modeling, necessitating a new periodogram $-$ called the Transit Comb Filter (TCF) $-$ to identify periodic sequences of double-spike patterns in the cleaned, whitened ARIMA residual light curves.  A brief overview of the ARPS methodology is given here, the interested reader will find more details on ARIMA and the TCF periodogram in \citet{Caceres19b}.

\subsubsection{Autoregressive Modeling of the Light Curves \label{sec:ARIMAi}} 

The autoregressive moving average(ARMA) model family is very broad and can treat an enormous variety of both stationary and non-stationary time series \citep{Box15}. ARMA-type models are widely used in signal processing, econometrics, and voice recognition, among many other applications.  ARIMA and its extensions like ARFIMA, GARCH and HAR, are able to treat short-memory processes, long-memory processes, volatility, burstiness, nonstationarity in the light curve without being subject to choices like a smoothing kernel function or bandwidth in Gaussian process regression or choices of a basis function and denoising threshold in wavelet analysis. ARMA-type models are fit by maximum likelihood estimation without any free parameters while balancing goodness-of-fit with model complexity. 

The linear ARIMA model has three components: autoregressive (AR), integrated (I), and moving average (MA).  We start our analysis with the `I' (`Integrated') component that treats nonstationarity; typically nonstationarity arises from variations in the mean fluxes of the light curve.  This operation is described by
\begin{equation}
  (1-B)^d x_t = \epsilon_t \,\, \text{where} \,\, B x_t = x_{t-1} \label{eq:arima_ii}
\end{equation}
and $d$ is the order of differencing, $B$ is the backshift operator, $\epsilon$ is a Gaussian noise term, and $t$ is the integer index of an observation in a regularly cadenced light curve. We chose the simplest setting with $d=1$; experimentation shows that higher order differencing is not necessary for most \TESS light curves. This essentially removes the narrowest possible median filter of the time series, equivalent to the transformation
\begin{equation}
  x^{dif}_t = x_t - x_{t-1}. \label{eq:diff_tsi}
\end{equation}
When the ARIMA model is applied to the light curve, we separate the differencing step (equation \ref{eq:diff_tsi}) from the ARMA fit and apply it to the light curve before passing the differenced values to the ARMA function. By separating the difference step from the application of the ARMA model, the transits (if present) are guaranteed to be changed into the double-spike pattern and, thereby, detectable by TCF. \citet{Caceres19b} found that the differencing step greatly reduced the interquartile range (IQR) of light curves with intrinsic stellar variations (such as rotationally modulated starspots) but slightly increases the IQR for light curves without noticeable variation present. 

The autoregressive AR(p) portion of the model represents how the stellar flux responds to recent previous values of the flux according to 
\begin{equation}
  x_t = \phi_1 x_{t-1} + \phi_2 x_{t-2} + \ldots + \phi_p x_{t-p} + \epsilon_t, \label{eq:arima_ari}
\end{equation}
where $x_t$ is the value of the light curve at time $t$, $p$ is the order of the AR component, and $\phi$ is a vector of unknown real-valued coefficients with length $p$. As with most regressions, the error term is assumed to be homoscedastic Gaussian, $\epsilon = N(0, \sigma^2)$ where the variance $\sigma^2$ is another parameter of the model. 

The moving average MA(q) portion of the model represents the effects of random shocks to the light curve by modeling the current flux value as 
\begin{equation}
  x_t = \epsilon_t + \theta_1 \epsilon_{t-1} + \theta_2 \epsilon_{t-2} + \ldots + \theta_q \epsilon_{t-q} \label{arima_mai}
\end{equation}
where $q$ is the order of the MA component, $\epsilon_t$ is the same error as in equation (\ref{eq:arima_ari}), and $\theta$ is a vector of unknown coefficients with length $q$. 

For every possible combination of $p$ and $q$, the $\phi$ and $\theta$ coefficients are computed using a maximum likelihood estimator. In order to reduce computation time for this step, we restricted $p$ and $q$ to have a sum less than or equal to 10. In practice, this restriction has little effect on the solution.  The best ARIMA(p,d,q) model is chosen to balance the accuracy of the ARIMA model compared to the difference light curve data and the overall model complexity using the Akaike Information Criterion \citep{Sakamoto86}, a penalized likelihood measure that balances the model complexity and accuracy of fit in a self-consistent manner. It is similar to, but with different penalty, to the more commonly used Bayesian Information Criterion.

The temporal structure is examined using the nonparametric autocorrelation function for both the original light curve and its ARIMA(p,d,q) residuals.  In practice, we find that most \TESS ARIMA model residuals have little or no autocorrelation and are consistent with white Gaussian noise. 

A critical question is whether the ARIMA model absorbs the planetary signal in addition to stellar and instrumental variations.  Since an exoplanetary transit signal, if present, occurs during only a very small fraction of the observations and the number of time steps between transits is larger than our maximal $p$ and $q$ values, it is mostly ignored by the maximum likelihood estimator. We find a bias does occur in the depth of the deepest transits (e.g. inflated hot Jupiters) as the ARIMA model incorporates some of the transit signal. This bias is corrected in a later stage of analysis (Paper II).

\subsubsection{Transit Comb Filter Periodogram \label{sec:TCFi}} 

\begin{figure}[b]
  \centering
  \includegraphics[width=0.8\textwidth]{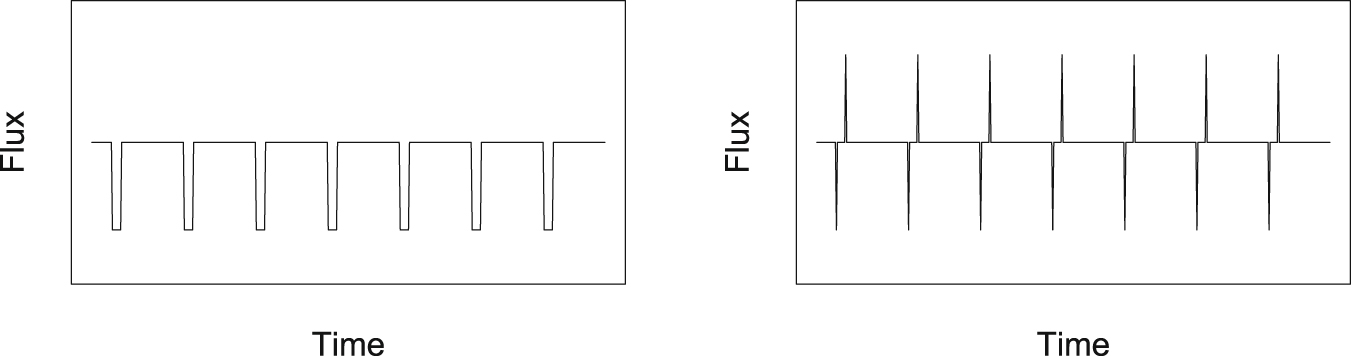}
  \caption{The transformation of the shape of the transit from the original light curve by the differencing step of the ARIMA processing into the double-spike shape \citep{Caceres19b}.}
  \label{fig:diff_trans_shapei}
\end{figure}

The difference step of the ARIMA processing in equation \ref{eq:diff_tsi} changes the shape of a planetary transit from a periodic box pattern to a double-spike pattern (Figure \ref{fig:diff_trans_shapei}).  The period, depth, duration, and phase of the transit are still available in the transformed light curve.   \cite{Caceres19b} developed the Transit Comb Filter (TCF), a matched filter algorithm that searches over a grid of durations and phases to find the strongest periodic double-spike patterns at a chosen trial period.  For a time series of Gaussian white noise, the algorithm is equivalent to the maximum likelihood estimator.  A periodogram is constructed from the strength of the matched filter fit to the ARIMA residuals for each period passed to the TCF.  The code involves the same triple-loop as the traditional Box-Least Squares algorithm \citep{Kovacs02}. \citet{Gondhalekar23} show that the TCF periodogram of ARIMA residuals is typically more sensitive to small planetary transits than the BLS periodogram of residuals from a standard smoother.  

As with the BLS periodogram \citep{Ofir14}, the TCF periodogram can have systematic changes in mean as one passes from short to long periods. We remove this trend with a smoothed locally fitted, robust least squares regression polynomial function $-$  the LOESS algorithm \citep{Cleveland88}. The power of the TCF for a specified period is then measured from the TCF power above the LOESS curve. The peak with the highest signal-to-noise ratio (SNR) in a window around the peak, with respect to the LOESS curve, is chosen as the most likely fit for an exoplanet transit in the light curve \citep{Caceres19b}.

\subsubsection{ARIMAX Model} \label{sec:ARIMAXi}

After the best TCF periodogram peak is found for a light curve, the transit parameters are used to fit a new ARIMA model to the light curve in order to jointly fit the transit and the autocorrelated noise in the light curve. In the parlance of ARMA modeling, this is an ARIMAX model where `X' refers to `exogenous' variables \citep{Hyndman21}. A simple box transit mask is built for the best peak from the TCF periodogram using the transit period, phase, and duration from the periodogram strongest peak. The depth of the box transit mask is left as a free parameter of the exogenous variable. Therefore, when the ARIMAX model uses maximum likelihood estimation to jointly model the transit depth and the autocorrelated noise of the light curve, it also models a transit depth with a confidence interval (error value). Further details about the ARIMAX modeling are provided by \citet{Caceres19b} and \citet{Caceres19a}. 

We found that the ARIMAX depth tended to underestimate the depth of the transit necessitating astrophysical transit models to be fit to the candidates. Like the bias produced by ARIMA modeling (\S \ref{sec:ARIMAi}), this has to be corrected in the later stage of DTARPS-S analysis so reliable estimates of the planet radius can be obtained  (Paper II).

\subsection{Classification and Vetting to Identifying Planet Candidates} 

While a prominent peak in the periodogram is a necessary indicator that a transit-like periodicity is present in a light curve, this alone is not a sufficient criterion. \citet{Caceres19b} compared the results of classifying based on periodogram strength alone and a machine learning classifier based on many features of the light curve and periodogram.  They found that the machine learning classifier performed better.  The principal reason is not lack of sensitivity to planetary signals by the TC periodogram, but the capture of non-planetary signals, particularly BEBs. Classifiers are developed to maximize the number of planets in the identified planet candidates while at the same time minimizing the number of non-planetary objects. But reliance on automated classifiers without human vetting risks statistical False Alarms and a higher percentage of astronomical False Positives in the final data set \citep{Burke15}. 

Two main families of classifiers used in exoplanet transit identification are deep learning classifiers and decision-tree-based classifiers \citep{JaraMaldonado20}, though other types are available.  Deep learning classifiers learn features automatically from training sets of light curves.  While training the neural network, the parameters of the linear combination inputs into each hidden layer feature are tuned using a cost function to minimize the classification prediction error.  Convolutional neural networks have been developed for \Kepler and \TESS planet discovery \citep{Shallue18, Ansdell18, Yu19}; these are used in the QLP pipeline leading to the TOI list$^1$ \citep{Guerrero21}.  Other transit detection groups use decision trees based on extracted features rather than the lightcurves themselves \citep{autovetter15, robovetter17, Armstrong18}.

\subsubsection{AutoRegressive Planet Search: Random Forests and Human Vetting} 

The ARPS method utilizes a Random Forest (RF) decision tree classifier to identify the most promising exoplanet candidates from the full data set with a high recall rate, followed by human vetting to further refine the candidate sample. We carefully test different training sets, features, and classifier settings to optimize its performance. 

RF machine learning classifiers were developed by \citet{Breiman01} as an extension of his earlier Classification and Regression Tree procedure \citep[CART,][]{Breiman84}.  CART classification uses a decision tree that has been grown based on the problem's training set in an iterative procedure.  Each node produces two daughter nodes based on a break in a single data feature that best separates the classes according to some cost function. To reduce over-fitting, the tree is pruned to a predetermined level. The main drawbacks of CART trees are they tend to over-fit the training data and often use only a few of the possible features in classification.  RF overcomes the disadvantages of CART classification by using multiple CART trees with randomized data subsets and  feature subsets at each branching node. This `bagging' strategy avoids over-fitting because each tree in the RF sees a different data set and avoids over-emphasis on just a few features in the classifier.  Whereas a single decision tree produces a `hard' prediction for each object in the test set, a RF gives `soft' or probabilistic predictions arising from votes of many trees.  The RF prediction value is pseudo-probability; higher predictive scores point to more likely exoplanet transiting candidates.

RFs are extremely versatile classifiers that can use data of different types (integer, categories, floating point numbers), units, and scales. RFs have been shown to be robust to imbalanced training set problems, performing well on training sets whose positive class comprises only 2\% of the entire training set \citep{Chen04} and can handle small fractions of mislabeled data in the training set \citep{Mellor15}. The contribution of each feature to the classifier allows for RFs to be partially interpretable, whereas most deep learning classifiers are not readily interpretable. 

Vetting the results of a machine learning classification is necessary to remove lingering False Alarms and False Positives in samples that pass the RF classifier. Although time consuming, every object classified as a potential candidate by the RF is examined by human vetters.  The vetting procedure employed here, described in Paper II, is a mixture of multifaceted automated vetting tests and subjective vetting by humans.

Due to the rarity of transiting planets in a random samle of stars, the RF classifier and the subsequent vetting procedures must strive to reduce the number of False Alarms and False Positives in the final catalog.  For a RF classifier with a False Positive Rate $\gtrsim$ 1\% as measured with a validation set, the number of expected False Positives in sample of a million light curves would overwhelm the planet candidate sample, even if the classifier were to identify every true transiting planet.

\subsection{Light Curve Input from the DIAmante Project \label{sec:DIAmante}} 

DIAmante is a pipeline for extracting and analyzing light curves from the \TESS full frame images (FFIs) developed by M20 and applied to \TESS Year 1 FFIs (M20) and Year 2 FFIs \citep{Montalto23}. M20 used the DIAmante extracted light curves to search for exoplanet transits and identified 396 exoplanet candidates. We applied the DTARPS-S method to the M20 light curves extracted and preprocessed with the DIAmante pipeline. 

M20 defines a sample of 976,814 dwarfs and subgiants with spectral types F5 to M falling in the footprint of \TESS sectors 1$-$13 surveyed during Year 1 with identifications in the TESS Input Catalog \citep[version 8,][]{Stassun19}. FGK stars were restricted to V $<$ 13 magnitude while M stars were restricted to V $\leq$ 16 magnitude and distance D $<$ 600 pc. The sample is further limited to dwarf and sub-giant stars with $\log g \geq 3$. 

The DIAmante extraction was applied to the calibrated FFIs available from the Science Processing Operations Center \citep{Jenkins16, Tenenbaum18}. These calibrated FFIs are reduced CCD images have already been processed with \TESS instrument specific corrections. The DIAmante extraction pipeline is based on Difference Image Analysis \citep{Alard98} that reduces the impact of contaminants on the target photometry through the efficient subtraction of a reference image convolved with a kernel to separate the target and the background flux. Because \TESS FFIs are known to have erratic background variations that depend heavily on the boresight angle between the camera, Sun, and Moon, a flux-conserving delta basis kernel was utilized to create a differential background model using a 20 pixel box smoothing region. After calibration to 250 standard stars in the reference images for each CCD, photometry was extracted from a circular aperture with a radius of one pixel. 

The DIAmante light curves from each CCD for each camera and sector were processed with cotrending to remove systemic variations from the instrument. Principal Component Analysis was applied to the most highly correlated light curves to extract top eigenvectors to cotrend the light curves. After cotrending, individual stellar variations are further detrended with an 8 hour median filter and a B-spline interpolation. Outliers more than twice the interquartile range (IQR) from the median were iteratively removed from the original light curve. The final DIAmante light curves are the averaged values of the B-splines evaluated at each observation time. 

\begin{figure}[tb]
  \includegraphics[width=\textwidth]{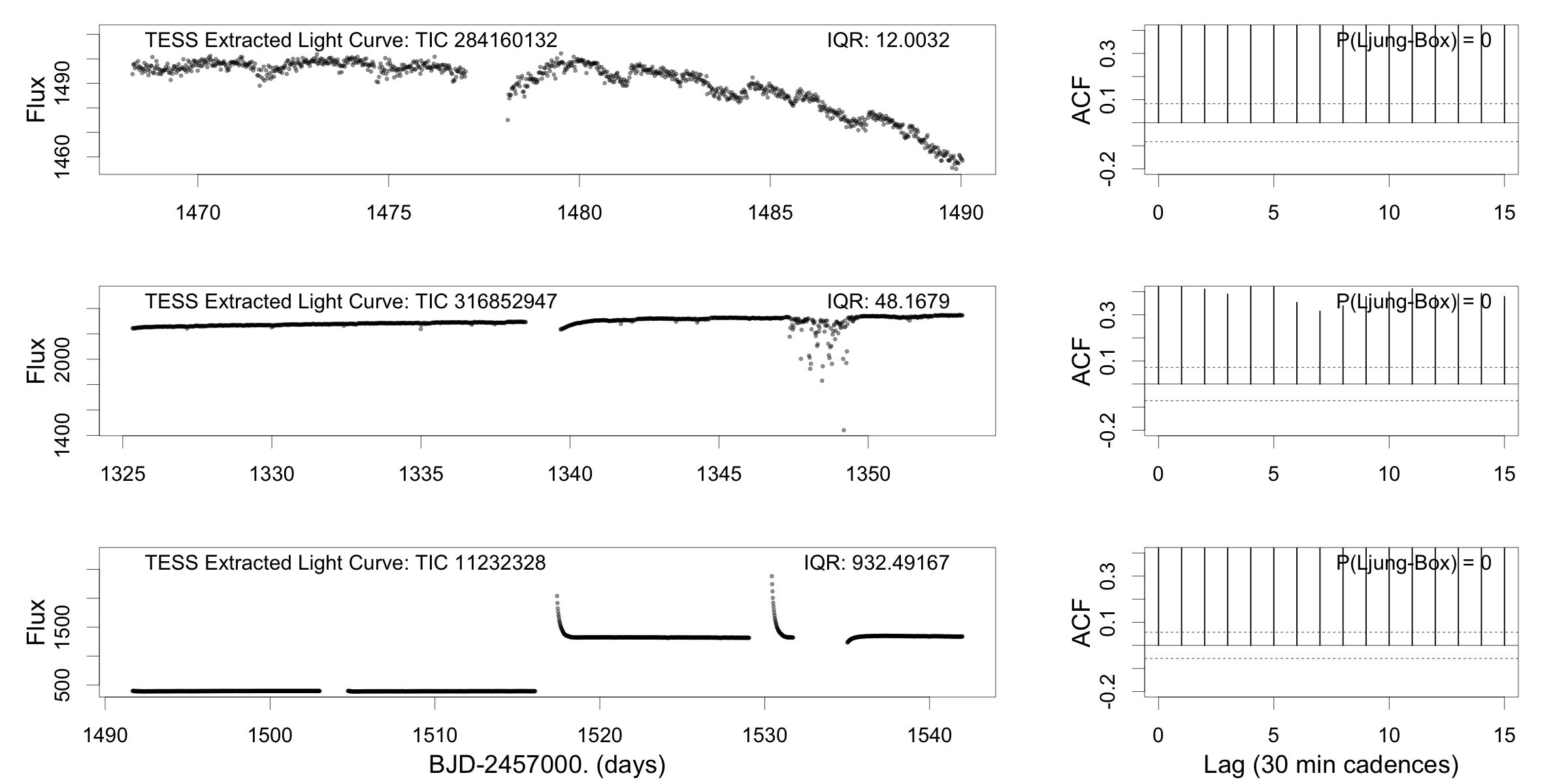}
  \caption{Left: The raw light curves extracted from \TESS FFIs for three example stars in the DIAmante data that were later found to have DTARPS-S candidate transiting planets. Right: Plot of autocorrelation present in the light curve as a function of lag in units of the 30 minute FFI candence. The $p$-value from the Ljung-Box test is 0 for all three light curves indicating that the flux values of the light curves are autocorrelated.}
  \label{fig:extract_lc}
\end{figure}

Figure \ref{fig:extract_lc} shows the raw \TESS extracted light curve for three stars in the DIAmante data set prior to any preprocessing. These examples were extracted using the Python \texttt{Lightkurve} package \citep{lightkurve18}. DTARPS-S identified a new planetary candidate around each of these three examples that had not been identified to date (Paper II). Two of the light curves shown (top and middle panels) are the length of one \TESS sector, as are most of the light curves from the DIAmante data set. The bottom light curve panel is a source observed in two sectors because it lay in the overlap area between sectors.

The three panels to the right of the light curves in Figure \ref{fig:extract_lc} show strong autocorrelation present in the light curves between the 30 minute time steps for the \TESS FFIs. The ARIMA fitting in the ARPS procedure is designed specifically to remove autocorrelation in light curves; the presence of autocorrelation is measured with the Ljung-Box test \citep{Ljung78}. The $p$-value from Ljung-Box test is included in subsequent figures to indicate the presence or absence of autocorrelation in the light curves. A small $p$-value indicates that there is significant autocorrelation present in the light curve and $p$-value $\gtrsim 0.01$ means that the light curve is consistent with white noise without autocorrelation.

\subsection{Additional Ramping and Outlier Removal \label{sec:outlier}}

There is a well known issue in \TESS of erroneous flux variations lasting a few hours introduced to the FFI light curves near the beginning of a sector, end of a sector, and near the beginning and end of the mid-sector gap in the light curve for the data download \citep{Ricker15}. Due to spacecraft jitter before the pointing settles, the target star can land on CCD areas with different quantum efficiency or a neighboring star can enter or leave the field changing dilution levels.  The DIAmante pipeline was able to remove most --- but not all --- of the trends and the jitter effects. There is a weak ramp up effect seen in the top and middle panels of Figure \ref{fig:extract_lc} after the mid-sector gap in the light curves as well as in the second sector of the bottom panel of Figure \ref{fig:extract_lc} after the second gap. The bottom panel light curve in Figure \ref{fig:extract_lc} has two strong ramps during the second sector. Most, but not all, of these ramps are removed by the DIAmante pipeline extracted light curves. 

We therefore add a preprocessing step to reduce remaining ramps around data gaps, flares, and other outliers that may be in the data. The clipping routine is characterized by the outlier threshold and the gap threshold. The outlier threshold defines the maximal distance between the median value of the light curve and a data point; we set the threshold at 5 times the standard deviation of the light curve. The gap threshold defines how large a gap of missing data can be, in time steps, before the clipping routine will examine the points on either side of the gap for evidence of ramping. We set the gap threshold to be 50 time steps, or 25 hours. With this gap threshold, the clipping routine removes erroneous ramping points from the beginning and end of the light curve, and points leading up to and away from a large gap in the light curve. After removal of a data point, the clipping routine is recursively applied to the modified light curve until no more points are removed as outliers.

\begin{figure}[tb]
  \includegraphics[width=\textwidth]{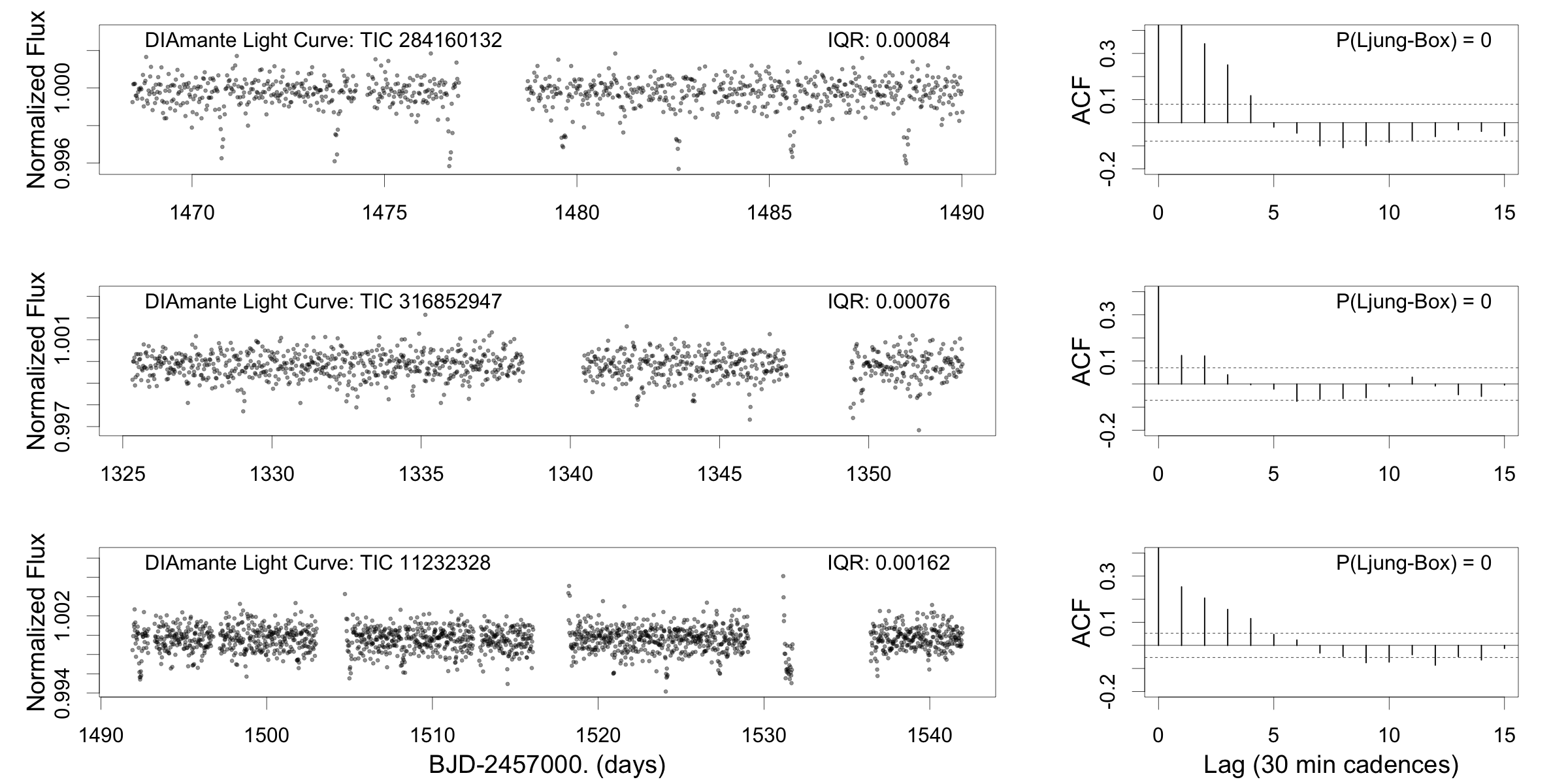}
  \caption{The DIAmante light curves for three example stars in Figure \ref{fig:extract_lc} after the removal of trends, ramping effects and outliers (\S\ref{sec:outlier}). The autocorrelation functions on the right show that autoregressive structure is still present in all three light curves.}
  \label{fig:DIAmante_wo_Outliers_lc}
\end{figure}

Figure \ref{fig:DIAmante_wo_Outliers_lc} shows the DIAmante light curves in Figure \ref{fig:extract_lc} after DIAmante detrending and our additional ramping and outlier reduction procedure. The ramping procedure removed most of the ramping in TIC 11232328 left by the DIAmante extraction pipeline. A weak residual ramping effect around Day 1531 remains (Figure \ref{fig:DIAmante_wo_Outliers_lc}). However, such brief and weak effects will have little effect on our transit search algorithm (\S\ref{sec:TCFi}).  None of these three transiting objects have been identified previously in the literature, despite each of them having strong transit signals in the DIAmante extracted light curves. 

\section{ARIMA Modeling and Periodogram Analysis \label{sec:ARIMA}} 

The right panels of Figure~\ref{fig:DIAmante_wo_Outliers_lc} illustrate that short-memory autocorrelation is often still present in the preprocessed light curves, even though the noise level is often reduced below 0.1\%.  This autocorrelated behavior can increase noise in periodograms such as Box Least Squares (BLS) and thus reduce sensitivity to weaker planetary transits \citep{Gondhalekar23}. Some of autocorrelation variations may be due to planetary transits, but typically it is intrinsic to the star. Transits from large planets can be seen with the unaided eye in Figure~\ref{fig:DIAmante_wo_Outliers_lc} .

\subsection{Autoregressive Modeling of the Light Curves} 

\begin{figure}[tb]
  \includegraphics[width=\textwidth]{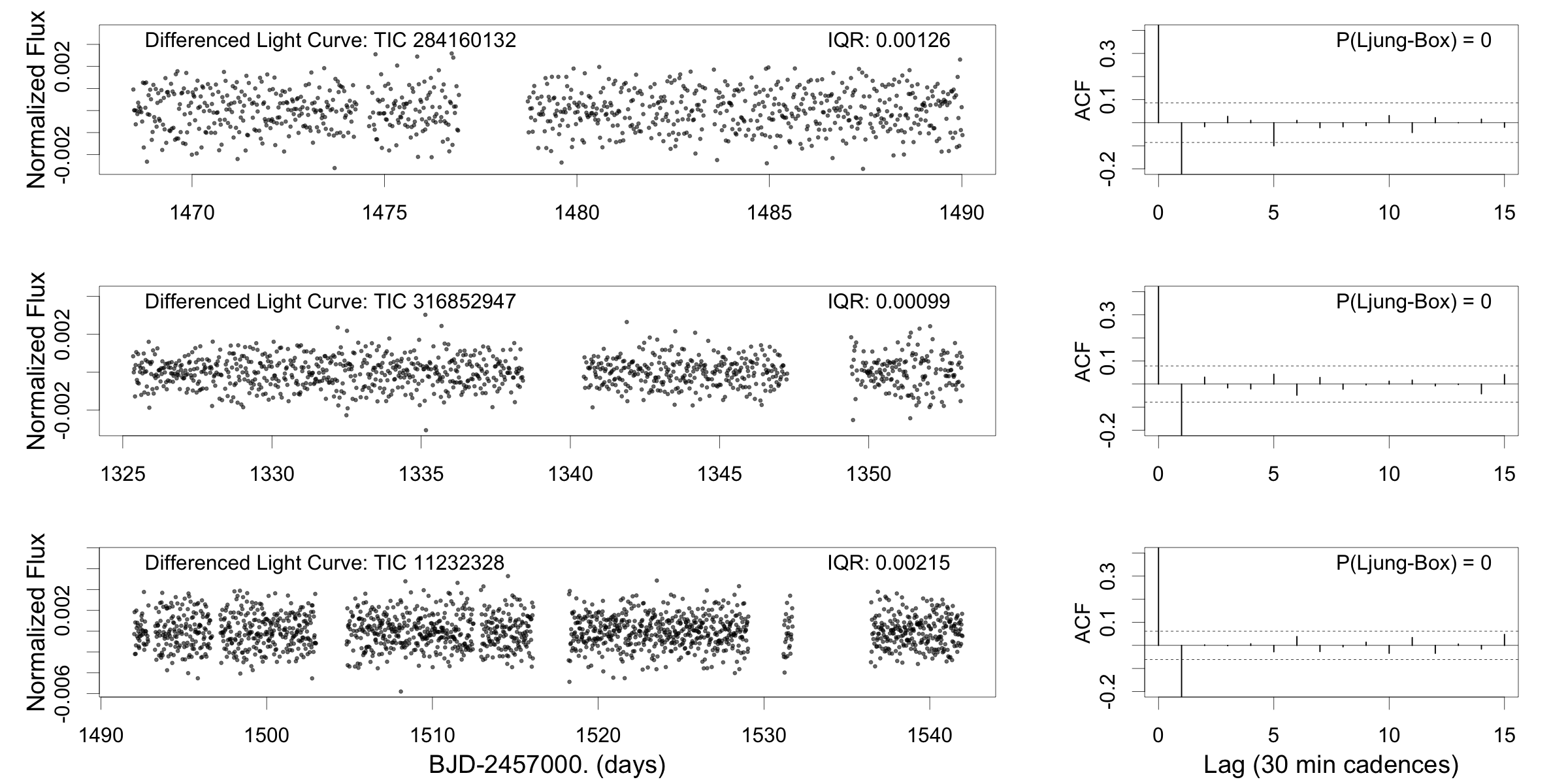}
  \caption{The differenced light curves for three example stars in Figure \ref{fig:extract_lc} with autocorrelation functions. }
  \label{fig:Differenced_lc}
\end{figure}

When the ARIMA model is applied to the light curve, we apply the differencing operation (equation \ref{eq:diff_tsi}) first and then obtain fits from the ARMA model (equations \ref{eq:arima_ari}-\ref{arima_mai}).  This guarantees that any real transits will be changed into the double-spike pattern and thereby be detectable with TCF.  Figure \ref{fig:Differenced_lc} shows the differenced light curves for each of the three example light curves in Figure \ref{fig:DIAmante_wo_Outliers_lc} along with the autocorrelation panels off to the right of the light curves showing the amount of autocorrelation in the light curve as a function of the lag in time steps. The $p$-value from the Ljung-Box test is 0, which is expected due to the negative correlation at lag=1 induced by the differencing operation. Applying the full ARIMA model to the differenced light curve will remove this autocorrelation introduced at lag=1 cadence. 

\citet{Caceres19b} found that the differencing step greatly reduced the IQR of light curves with intrinsic stellar variations, but slightly increases the IQR for other light curves. In our case, the DIAmante pipeline removes most of the light curve trends before we receive them. Therefore, it is not surprising that the IQR of the light curve does not change much over the course of the ARIMA processing.

\begin{figure}[tb]
  \includegraphics[width=\textwidth]{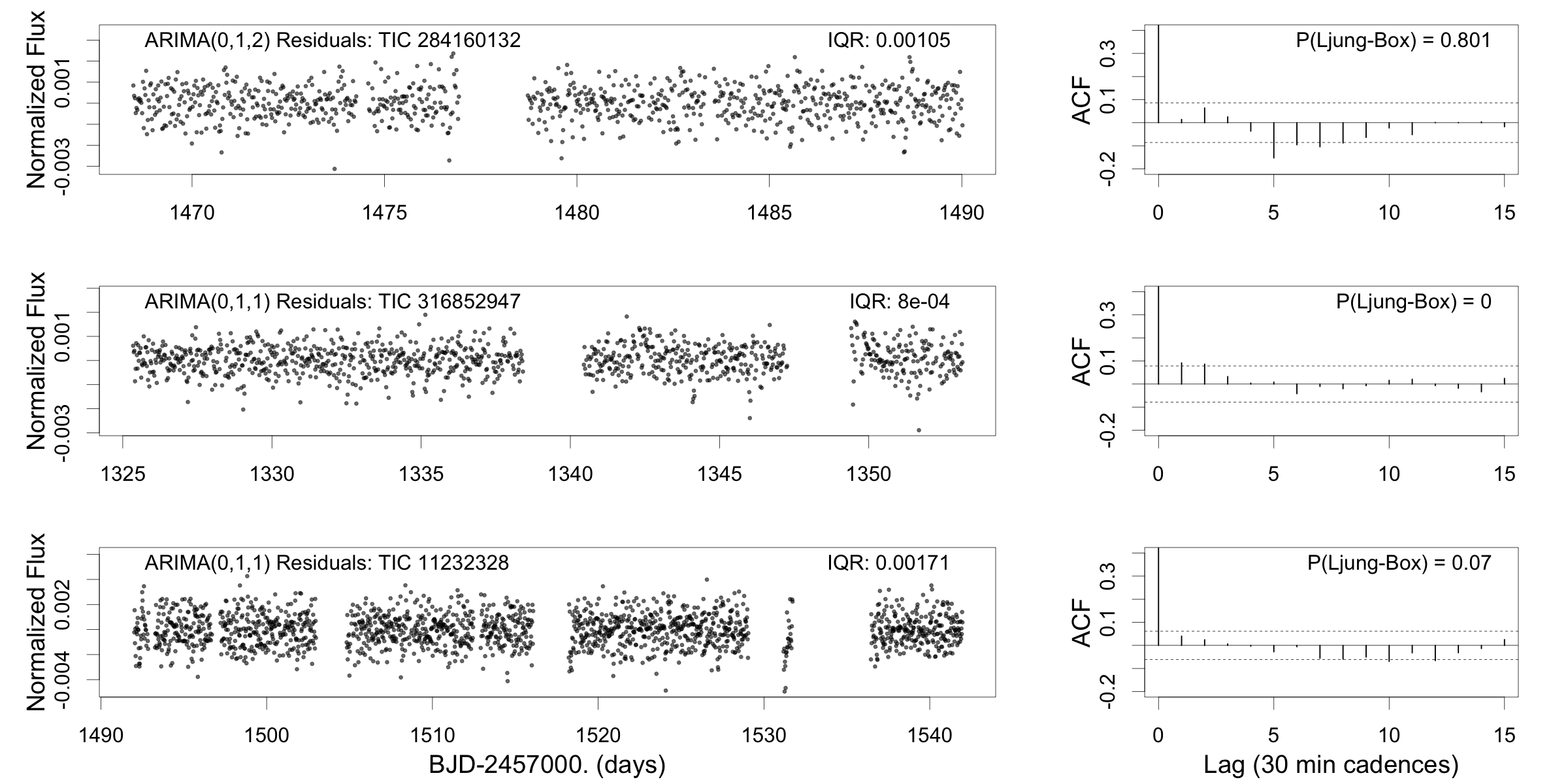}
  \caption{The residuals after the best fit ARIMA model had been subtracted from the differenced light curves for three example stars in Figure \ref{fig:extract_lc}. The three plots on the right show the amount of autocorrelation present in the light curve as a function of time step between points. The Ljung-Box test shows that two of three ARIMA residuals are consistent with Gaussian white noise.}
  \label{fig:ARIMA_resid_lc}
\end{figure}

\begin{figure}[tb]
  \includegraphics[width=0.49\textwidth]{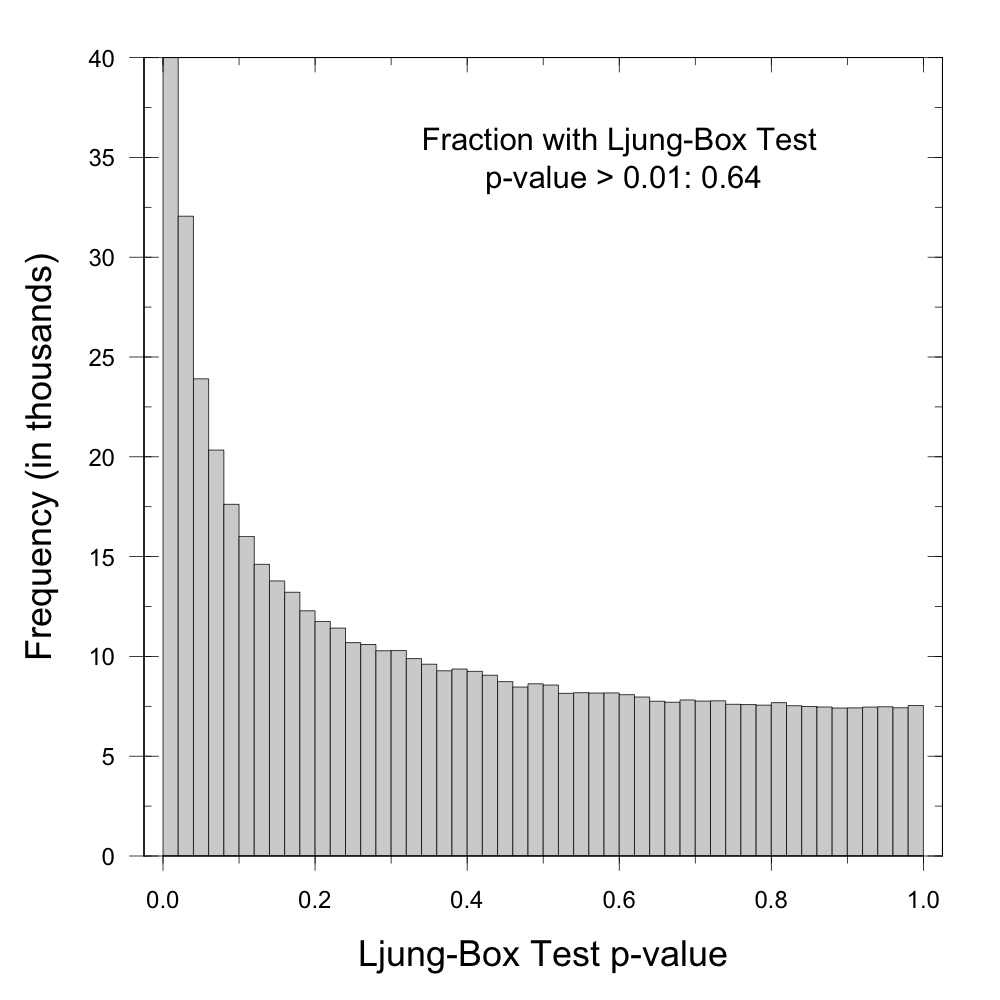}
  \includegraphics[width=0.49\textwidth]{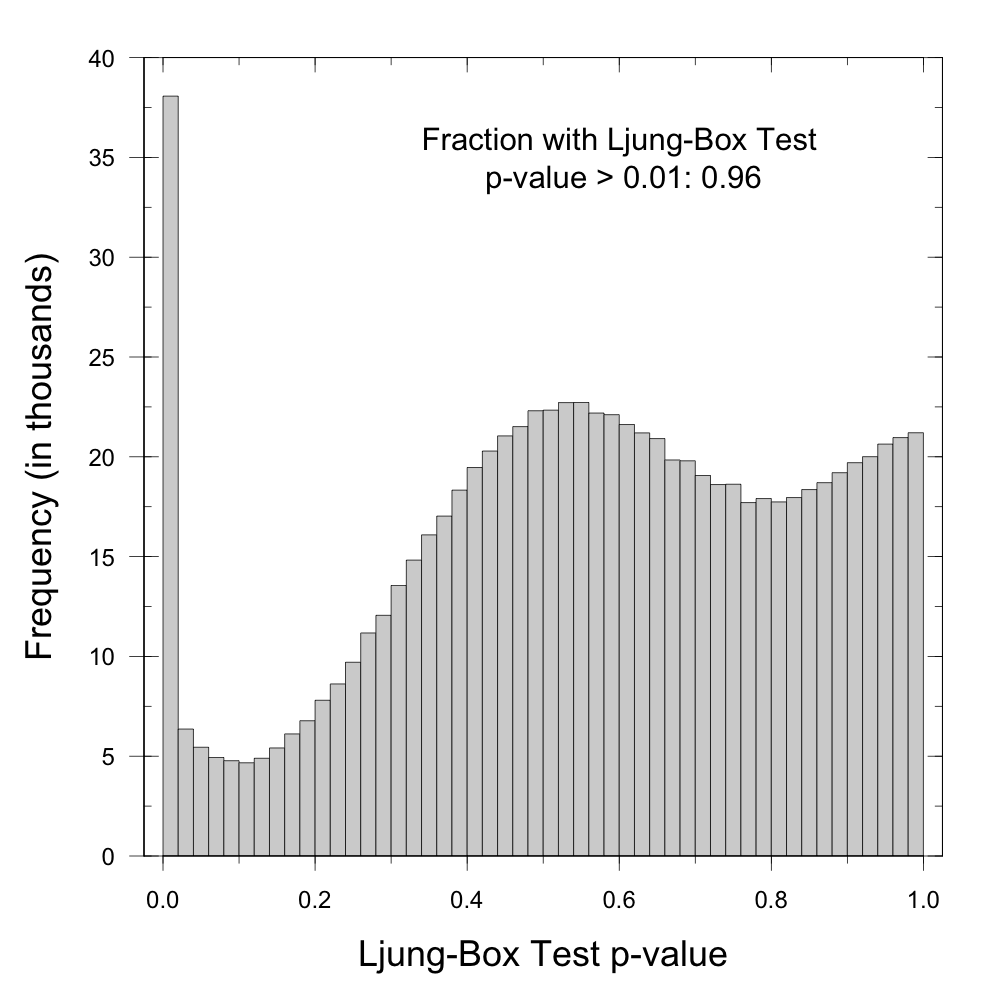}
  \caption{Distribution of $p$-values from the Ljung-Box test for the light curves extracted from the DIAmante data set (on the left) and for the ARIMA residuals (on the right). The first bin in the histogram of Ljung-Box test $p$-values in the DIAmante data (left) has a frequency of 318,227.}
  \label{fig:lb_test_comp}
\end{figure}

We implement the ARIMA model using the \emph{auto.arima} function from the \emph{forecast} package \citep{Hyndman21} in the statistical computing language R \citep{Rcore}. Figure \ref{fig:ARIMA_resid_lc} shows the residuals after the best fit ARMA model has been subtracted from the differenced light curve for the three example light curves in Figure \ref{fig:extract_lc}. The $p$ and $q$ values for the best fit ARIMA models are given in each of the three panels where it states ``ARIMA($p$, $d$, $q$)''. The autocorrelation function of the residuals is often consistent with Gaussian white noise with associated Ljung-Box test $p$-value $ > 0.01$. The improvement in Ljung-Box probabilities for the full DIAmante sample is shown in Figure \ref{fig:lb_test_comp}: 46\% of the light curves from the DIAmante pipeline have significant autocorrelation present in the light curve while only 4\% have autocorrelation after ARIMA modeling. 

The residual light curves in Figure \ref{fig:ARIMA_resid_lc} now exhibit periodic double-spike patterns characteristic of a transiting planet but without stellar or instrumental autocorrelated behavior. In most cases, the DIAmante preprocessing, outlier removal, and ARIMA modeling together successfully remove the structure present in the \TESS light curves except for transits or other brief non-stochastic behaviors.

\subsection{Transit Comb Filter Periodogram \label{sec:TCF}}  

The TCF algorithm described by \cite{Caceres19b} is coded in Fortran for computational efficiency and is called from our R-based DTARPS pipeline.  The Fortran program is available in the Astrophysics Source Code Library \citep{Caceres22}.  The periods tried by TCF in its matched filter algorithm were restricted to periods between 0.2 and 30 days. The TCF algorithm finds the optimal phase, duration, and depth for each period passed to the algorithm, so the periods were restricted so that they covered the length of an entire sector with a 3 day pad so that longer period planets could be identified in multi-sector light curve data. Thirty days is long enough that the TCF periodogram would be able to fully characterize the shape of a high powered peak near 27 days (the length of a single sector) and allow any peak near 27 days to be correctly sorted as either a genuine peak from a transit signal match in the data or noise from a multi-sector light curve being folded by the length of a sector. The lower limit of 0.2 days was chosen to facilitate the search for extreme ultra short period exoplanets. It is just larger than the shortest reported period for a confirmed planet in the NASA Exoplanet Archive (as of March 15, 2022), K2-137 b with a period of 0.179 days. The 354,982 periods passed to the TCF search algorithm were chosen to be evenly distributed in log-space. The durations looped over for each period were limited to a range from a minimum of 15\% of the period to a maximum of 25 hours or 50 time steps \citep{Caceres19b}. 

\begin{figure}[b!]
  \includegraphics[width=\textwidth]{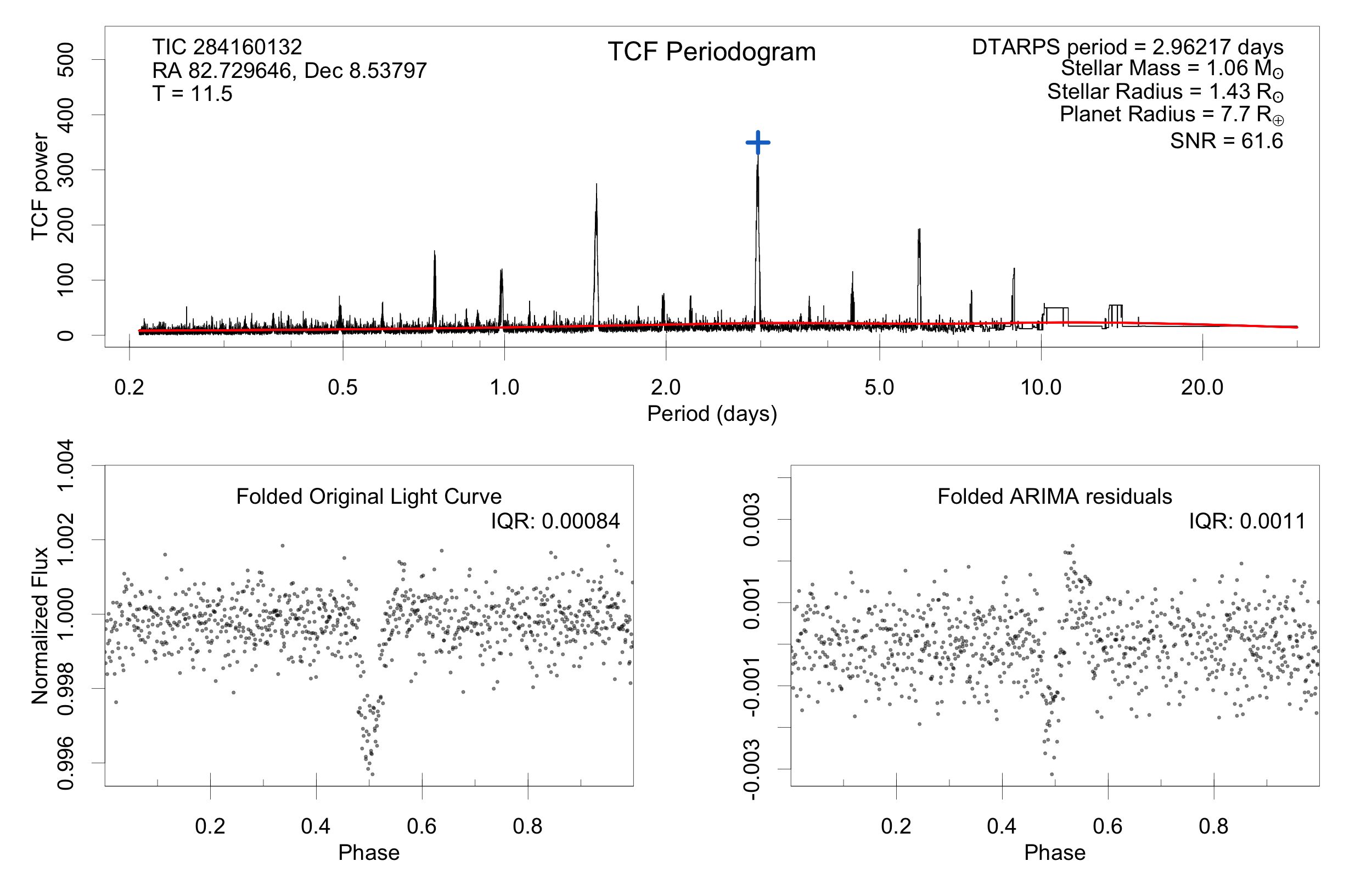}
  \includegraphics[width=\textwidth]{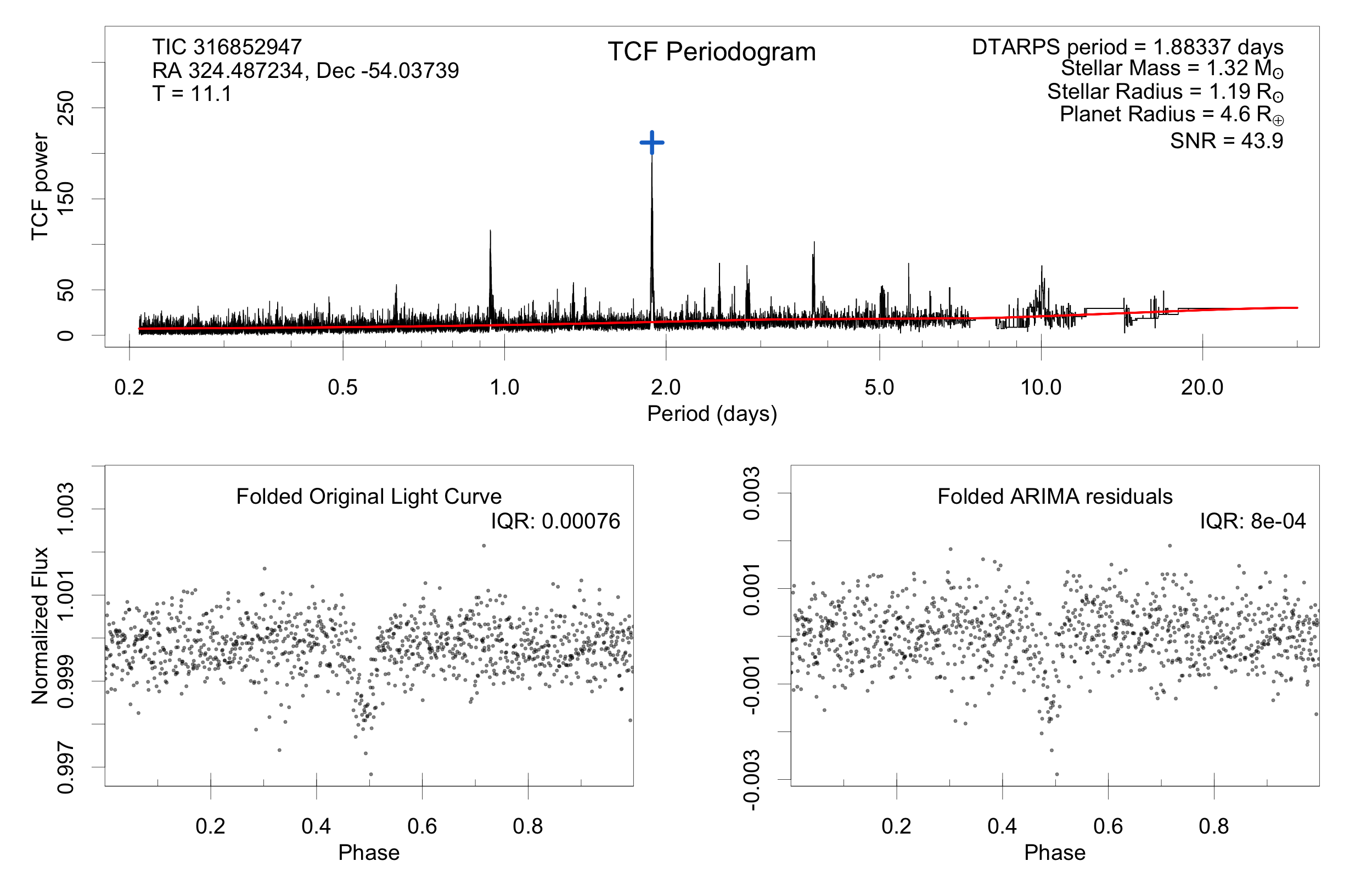}
\end{figure}

\begin{figure}[t]
  \includegraphics[width=\textwidth]{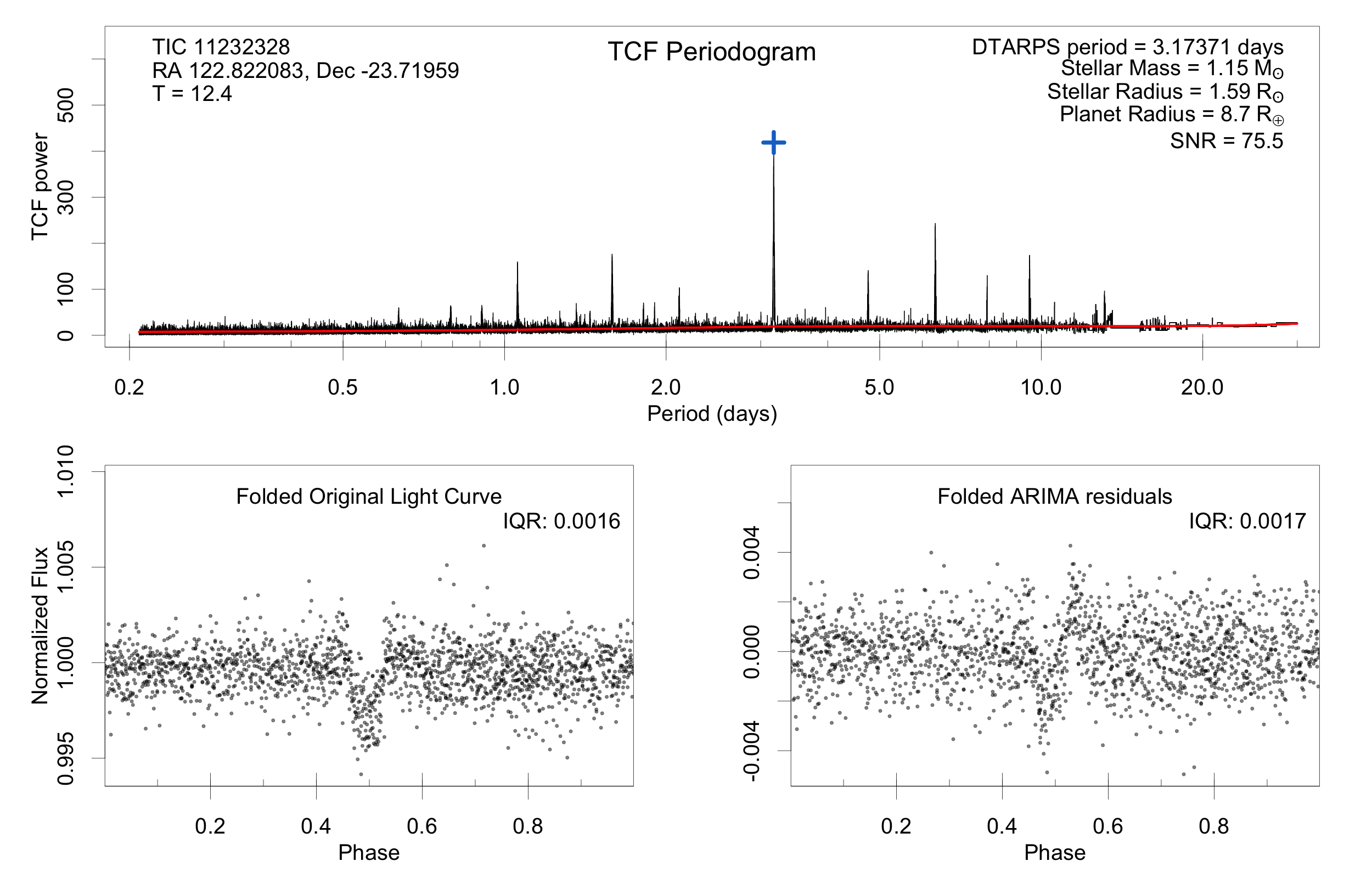}
  \caption{The TCF periodograms and best period phase-folded light curves for the stars in Figure \ref{fig:extract_lc}. The red curve is the LOESS fit to trends in the median of the periodogram. The blue cross indicates the peak with the highest SNR over a window around that peak.  The lower left panel shows the original light curve phase-folded on the best TCF period. The lower right panel shows the residuals after the best fit ARIMA model has been subtracted from the light curve phase-folded on the parameters from the best TCF periodogram peak. Note that the ordinate scales of the two folded light curves differ.  
  \label{fig:tcf_284160132}}
\end{figure}

Figure \ref{fig:tcf_284160132} shows the resulting TCF periodograms and phase folded light curves for the three stars in Figure \ref{fig:extract_lc}. The periodograms in the top panels show the power of the best transit fit in phase, duration, and depth as a function of the period investigated. The LOESS curve is plotted in red to allow removal of any trend in the periodogram noise values. The period that was used to fold the light curve and the ARIMA residuals was chosen by the peak in the periodogram with the best SNR in a window of 10,000 periodogram values to either side of the peak. (The SNR reported in the TCF periodogram refers to the SNR of the periodogram peak, not the SNR of the transit depth as is often seen elsewhere.) The typical best SNR of a TCF peak seen in stars without planetary transit signals is typically between 9 and 13, much lower than the peak SNRs between 43 and 75 in Figure \ref{fig:tcf_284160132}. The potential DTARPS-S candidates identified by the RF classifier have top peak SNRs typically between 17 and 55 and the top peak SNRs for our final DTARPS-S candidates are between 32 and 71. 

The bottom two panels below each of the TCF periodograms in Figure \ref{fig:tcf_284160132} show the original DIAmante light curve and the ARIMA residual fluxes plotted modulo the TCF peak period. The phase is adjusted so the transit is centered at phase 0.5. The phase-folded ARIMA residuals shows the double-spike shape that the transit shape was transformed into due to differencing step of the ARPS processing. The double-spike shape is clearly seen in the folded ARIMA residuals. TCF ran its nested loop matched filter algorithm on the ARIMA residuals light curve shown in the bottom right panel but the transit is more intuitively identified by human vetters as a box shape in the bottom left panel showing the folded DIAmante light curve.

Ancillary information from the \TESS Input Catalog is provided in the TCF periodogram panels. TIC 284160132 (DTARPS-S 548 in Paper II) is an early-G star with a V = 12.2 mag at a distance 384 pc. DTARPS-S identified periodic dips consistent with a gas giant with orbital period 2.96217 days (Paper II). Alias peaks at the double period ($\sim 6$ days) and half period ($\sim 1.5$ days) are easily seen in the TCF periodogram. TIC 316852947 (DTARPS-S 604) is a mid-F star with V = 11.6 at a distance of 341 pc. DTARPS identified a transit consistent with a Neptune-sized object with a period  1.88337 days. TIC 11232328 (DTARPS-S 25) is a fainter star with V = 12.9 mag, a late-F star at 714 pc. DTARPS-S identified a transit consistent with a Saturn-size planet orbiting with period 3.17371 days.

\subsection{ARIMAX Model\label{sec:ARIMAX}} 

The ARIMAX model is run using the \emph{auto.arima} function from the R CRAN package \emph{forecast} \citep{Hyndman21}. We often found that the ARIMAX model underestimated depths for the TESS DIAmante light curves, probably because the ARIMA model incorporated some of the transit signal.  We did, however, use the transit depth error as a feature in the RF classifier (\S\ref{sec:rf_final}).

\section{Random Forest: Training Set\label{sec:rf_training}} 

In human affairs, training sets for machine learning procedures are often well-defined; e.g. photographs of 'dogs' $vs.$ 'cats'.  But in astronomical applications, there is often considerable flexibility in defining the training sets, and these choices can be a dominant contributor to classifier performance.  For binary classification (planet vs. non-planet), a RF classifier requires labels for both positive training examples (light curves with exoplanet transit signals) and negative training examples (light curves without exoplanet transit signals).  Following M20, we introduce simulated `injected' False Positive signals into the negative training sets to steer the classifier away from BEBs. Section \ref{sec:kep_inj} describes the positive training sample with injected exoplanet transit signals and \S\ref{sec:neg_train} explains the negative training sample with injected eclipsing binary (EB) signals. Section \ref{sec:train_test} describes how the two samples were combined into a training set and a validation set for Random Forest classification.

\subsection{Kepler-Based Planet Injections \label{sec:kep_inj}}

The injected transit signals are drawn from the \Kepler 4-year mission exoplanets that can be considered to be an unbiased sample of the true planetary occurrence rate for the shorter period exoplanets  and higher radii that \TESS is designed to identify during its prime mission. The \Kepler 4-year mission exoplanet sample was acquired from the NASA Exoplanet Archive (accessed March 14, 2021).  We confine injections to stars with a \Kepler identifier that have a confirmed planet with $P < 13.5$ days (allowing at least two transits during a 27-day \TESS sector exposure).  Following the finding of \citet{Caceres19b} that roughly half of the `confirmed' Kepler Objects of Interest (KOIs) with low \Kepler $Model.SNR$ were not recovered with ARPS analysis, we  removed KOIs with \Kepler $Model.SNR < 20$.  Of the 2,356 \Kepler confirmed planets, 949 are suitable for injections.

\begin{figure}[tb!]
  \includegraphics[width=0.99\textwidth]{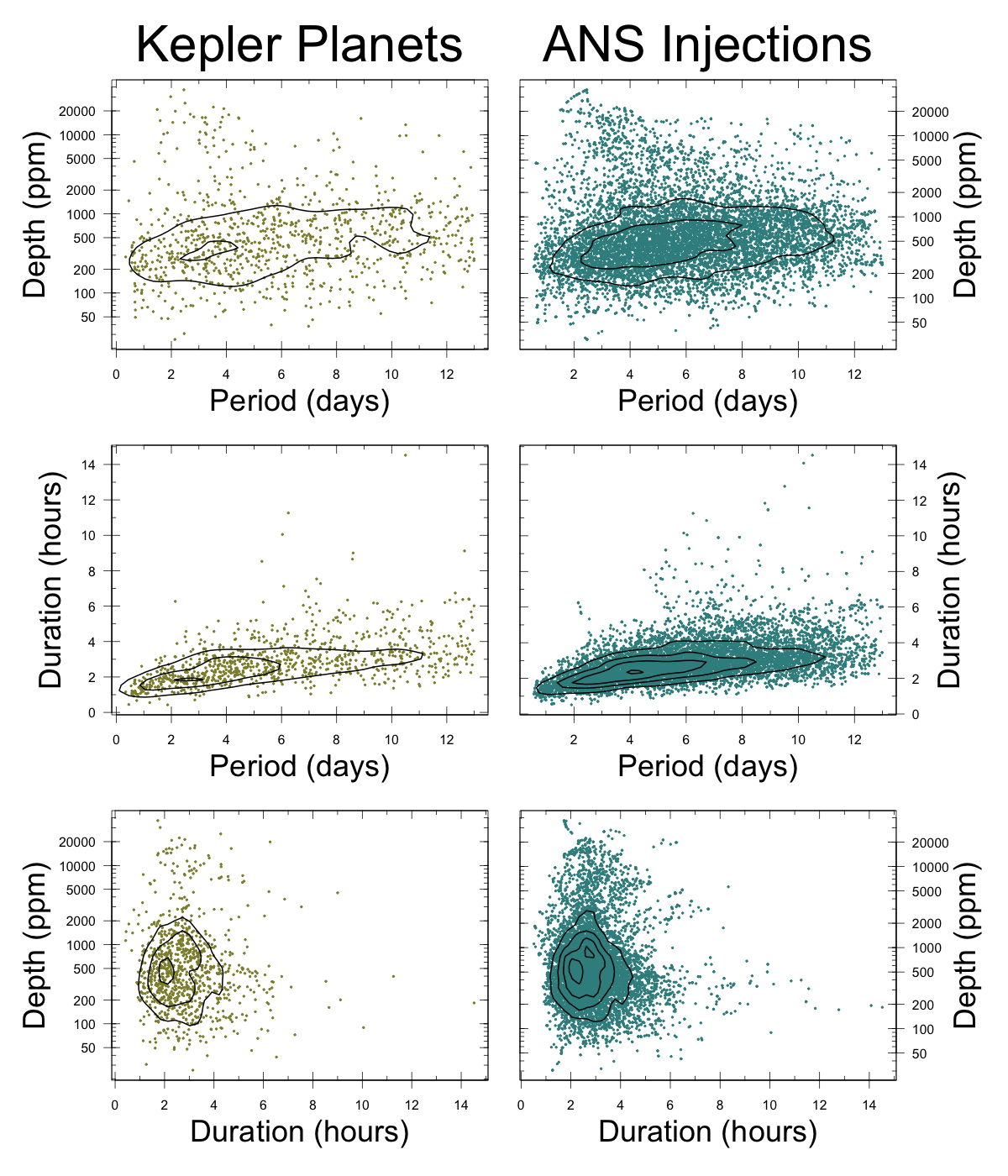}
  \caption{Scatter plots of the transit parameters of the 949 confirmed \Kepler planets from the NASA Exoplanet Archive \citep{NEA-CP} accessed March 14, 2021 in the left column plotted in olive green. The distribution of the synthetically created exoplanet transit parameters created using Adaptive Neighbor SMOTE (ANS) on the 949 \Kepler planets in the right column plotted in teal. Smoothed contours have been included to better illustrate the distributions.}
  \label{fig:kep_v_ans}
\end{figure}

The left panels of Figure \ref{fig:kep_v_ans} show the distribution of transit parameters period, duration and depth for these 949 confirmed \Kepler planets in the left column.  Most of the transit signal depths from the Kepler sample are below 1,000 parts per million (ppm) corresponding to a planetary radius of $\sim 3.5 R_{\oplus}$ for a Sun-like star
. Only 62 gas giant planets are among the 949 \Kepler planet sample with planetary radii $> 8 R_{\oplus}$. 

If only \TESS-detectable planets are considered, this training sample is too small for viable training of a RF classifier for an imbalanced classification problem.  We therefor augmented the sample of 949 \Kepler planets with a synthetic exoplanets sharing the same distribution of transit parameters. This allowed us to inject thousands of DIAmante light curves with planetary transit signals without reusing any set of transit parameters, preventing our training set from being biased by a few \Kepler planets. A modified version of the widely utilized Synthetic Minority Oversampling Technique \citep[SMOTE,][]{Chawla02} was applied to the 949 \Kepler planets in order to create synthetic exoplanet transit sample. Specifically we used Adaptive Neighbor SMOTE (ANS) described in \citet{Siriseriwan17} with code implementation in CRAN package {\it smotefamily} within the R statistical software environment \citep{Siriseriwan19}.
  
The SMOTE algorithm selects a random instance in the minority class and finds the $k$-nearest minority class neighbors in feature space. The parameter $k$ is set by the user with a common default value of five. SMOTE then randomly selects one of the $k$-nearest neighbors and generates a synthetic minority class instance by randomly choosing a point along the line between the original minority class instance and the neighbor minority class instance in feature space \citep{Chawla02}. ANS modifies SMOTE by removing the need for a user-set parameter $k$ and finds an optimal $k$ for each instance of the minority class based on the maximum distance between the minority class points and their nearest minority class neighbor in feature space, $\eta$. For each instance of the minority class, $p_i$, $k_i$ is set the number of minority class instances within $\eta$ of $p_i$. The regular SMOTE method of generating synthetic minority class members is then executed using the assigned $k_i$ value for each minority point until the desired number of synthetic minority class instances have been generated \citep{Siriseriwan17}.

We used the ANS SMOTE procedure to create a sample of periods, durations, and transit depths for 10,850 synthetic exoplanets shown in the right panels of Figure \ref{fig:kep_v_ans}. The ANS algorithm preserves the distribution of transit parameters and does not extend the sample of points beyond the extreme values of the distribution. For instance, like the \Kepler planet sample, $\sim 6\%$ of the synthetic exoplanet transits had depths consistent with gas giant planets.

A well-sampled exoplanet transit signal will also have an ingress and egress duration as a transit parameter. We assume the ingress/egress time for the exoplanets ranges from $\sim$ 5 minutes for smaller exoplanets (Earths to sub-Neptunes) up to $\sim$ 30 minutes for hot Jupiter exoplanets. Our injection model is a trapezoidal shape with straight-line ingress, duration, and egress. Given the \TESS Year 1 FFI cadence is 30 minutes, the ingress and egress is instantaneous in most cases. 

We injected the 10,850 planetary signals into 6,506 random light curves from the DIAmante light curve sample. Unlike other studies that inject planetary transit signals into the pixel data from the instrument \citep[e.g.][]{Christiansen20}, we inject the transit signal into the light curve data after DIAmante preprocessing but prior to any DTARPS-S processing. This focuses the analysis on the efficiency of DTARPS-S to identify planetary transit signals in the light curve data set without including instrumental and light curve extraction effects in our analysis.

The light curves that received planetary injections were selected randomly from the sample of $\sim$0.9 million DIAmante stars.  The following cuts were made using the TIC stellar radii and effective temperatures to avoid subgiants: $T_{eff} < 4750 K$ and $R < 1 R_{\sun}$, $4750 K \leq T_{eff} < 5250 K$ and $R < 1.125 R_{\sun}$, $5250 K \leq T_{eff} < 5600 K$ and $R < 1.325 R_{\sun}$, $5600 K \leq T_{eff} < 5900 K$ and $R < 1.45 R_{\sun}$, $5900 K \leq T_{eff} < 6200 K$ and $R < 1.55 R_{\sun}$, $6200 K \leq T_{eff} < 6500 K$ and $R < 1.65 R_{\sun}$, and $6500 K \leq T_{eff}$ and $R < 1.7 R_{\sun}$. 

\begin{figure}[tb]
  \includegraphics[width=\textwidth]{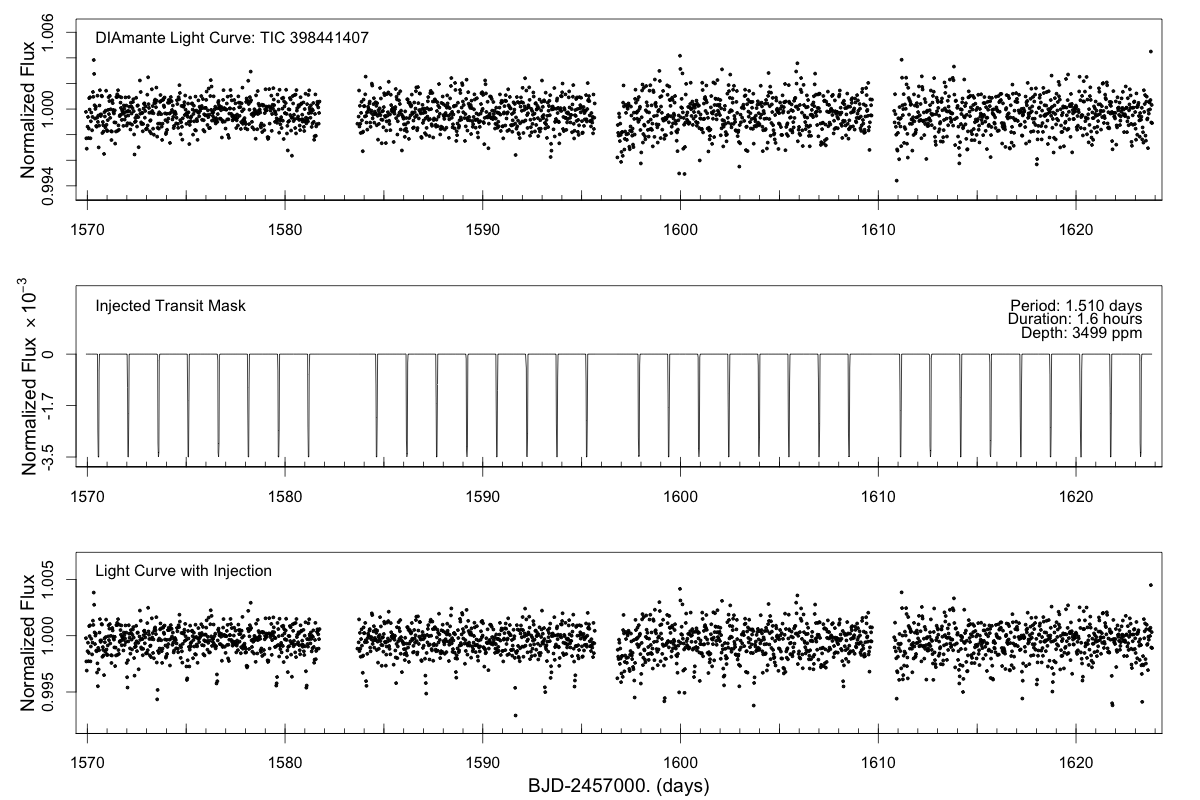}
  \caption{The steps for injecting a synthetic transit signal into a random DIAmante light curve. The top panel shows the original DIAmante light curve. The middle panel shows the transit mask created for a planetary injection. The bottom panel shows final injected light curve.}
  \label{fig:inj_example}
\end{figure}

Figure \ref{fig:inj_example} illustrates the injection process for a planetary injection. The top panel shows the DIAmante extracted \TESS FFI light curve for TIC 398441407, a V=12.1 G2V star radius of 1.1 $R_{\sun}$. The middle panel shows the modeled transit that was injected into each light curve characterized by the transit period, depth, and duration. The phase of each transit was chosen randomly. Sometimes the depth of the injection signals may appear to vary due to the transit jittering with respect to the 30 minute time step cadence. The third panel shows the resulting light curve after the flux values associated with the modeled transit were added point-wise to the flux values of the DIAmante light curve. 

After the injected light curves had been modeled with a best-fit ARIMA model and analyzed by TCF, the resulting TCF periodograms were vetted by visual inspection  to identify injected transit signals that had been successfully recovered by TCF. The injected transit signal was only considered recovered if the orbital period of the peak TCF period matched the injected orbital period (or an integer ratio of the injected orbital period) within $1\%$ and a transit dip in the folded light curve was visible. Smaller injected planets were more likely rejected by human vetting due to the lack of a visibly discernible transit. Of the 10,850 injections, 1,327 synthetic injections were recovered for use as the positive training set for the RF classifier.

\subsection{Negative Training Set \label{sec:neg_train}}

Identifying transiting exoplanets is inherently a highly imbalanced classification problem, as most planets have unsuitably inclined orbits or are too small for transit detection. However, the RF technique is well-adapted to this situation; \citet{Chen04} showed that RF classifiers can perform well with a positive training set that is as small as $2\%$ of the entire training set. We chose to a $\sim$20:1 ratio of negative to positive training set sizes with 26,953 lightcurves without injected planetary signals compared to 1,327 light curves with planets. 

The negative training set of the RF classifier should be made up of light curves with no transiting exoplanet signals. However, it is infeasible to manually vet and remove transits from this large negative training set. From the Kepler survey, \citet{Howard12} found that the expected planet occurrence rate of exoplanets with radii $2-32 R_{\oplus}$ and periods less than 10 days for GK dwarfs is $0.034\pm0.003$. Since most of these have inclined orbits without transits, an unvetted random sample of \TESS light curves thus suffers negligible contamination from transiting exoplanets. In any case, RF classifiers have been shown to perform well when a small fraction of their training set has been mislabeled \citep{Mellor15}. 

One of the largest sources of expected False Positives (FPs) in an exoplanet transit survey are BEBs. In order to push the classifier away from labeling eclipsing binary (EB) transit signals as an exoplanet transit signal, we follow the procedure of M20 by supplementing the negative training with the injected FP light curves. The injected FP light curves in M20 are made up of injected EB transit signals corresponding to secondaries with radii larger than 2.5~R$_J$ in both circular orbits and eccentric orbits. Short period sinusoidal signals mimicking rotationally modulated spotted single stars are also injected.  The entire set of injected FP was separated evenly between EBs with a circular orbit, eccentric EBs, and sinusoidal variables. The injected FP signals were not vetted to see if the FP signal was recovered after the ARIMA processing and the TCF analysis because the classification and characterization of FPs is not the goal of the classifier. Therefore all of the injected FP signals were included in the negative training set. We used 11,342 injected FP light curves and 15,611 random light curves in our negative training set.

\subsection{Training and Validation Sets \label{sec:train_test}}

The full set of labeled objects used for training is split into a training set for the RF classifier and a validation set to measure the performance of the RF classifier on a set of labeled data.   We reserved $20\%$ for the validation set, randomly chosen, leaving a training set of 1,048 injected exoplanet signals, 9,095 injected FP signals and 12,475 random light curves. The validation set holds 279 injected exoplanet signals, 2,247 injected FP signals and 3,136 random light curves.

\section{Random Forest: Optimization \label{sec:rf_optimize}}

\subsection{Training RF Classifiers}

In order to optimize the final RF classifier, we trained thousands of trial RF classifiers with different parameters.  Trials examined different combinations of feature selection, feature weights, and algorithmic options to maximize the performance of the RF classifier on the validation set.   The number of features to try at each node was left at the default value of 7, but instead of testing a set number of splits in the data for the node features, the optimal split for each node was found.  Since the training set is highly imbalanced (\S \ref{sec:train_test}), the balanced Random Forest option is used \citep{Chen04, Ishwaran22}. Balanced RF compensate for an imbalanced training set by undersampling the majority class for each tree in the RF classifier so that each tree is grown using a balanced subsample of the full training set.  The number of trees in the forest was varied from 500 to 1000.  The RF analysis was performed using CRAN package \emph{randomForestSRC} \citep{Ishwaran22} implemented by public domain R statistical software environment \citep{Rcore}.

The feature selection and weights were the focus of our tuning parameters to build the best possible performing RF classifier. Over 100 features were gathered from every stage of the ARIMA and TCF analysis as in \citet{Caceres19b}. Features describing the light curve were extracted from the light curve, the differenced light curve, and the residuals of the light curve after the best fit ARIMA model had been subtracted. Features from the TCF analysis included the features from the top 100 peaks of the TCF periodogram as well as features from the peak with the greatest peak signal-to-noise ratio (SNR). Features were also created for the light curve folded according to the parameters from the best TCF periodogram peak. Stellar metadata from the TIC v8 \citep{Stassun19} and the \emph{Gaia} DR2 \citep{GaiaCollab_Mission, GaiaCollab_DR2} were gathered for each light curve. Finally, two features that were of high RF feature importance in M20 were calculated for all of the light curves. We examined a wide variety of features because RF classifiers are data driven classifiers, not physically motivated classifiers. The features that are useful to the RF classifier may not have physical interpretations \citep{Genuer10}. 

The optimal RF classifier was found by creating a large number of test RF classifiers, each with 500 trees, to test different combinations of features. The RF classifiers whose Area Under Curve (AUC) for the Receiver Operating Characteristic (ROC) curve was greater than 0.9 and the AUC for the precision-recall curve was greater than 0.85 were kept for further consideration. The features from the top performing RF classifiers were then combined with different feature weights and a new batch of RF classifiers were grown. Altogether approximately 20,000 classifiers were considered. The RF classifiers with the same criteria for the AUCs of the ROC and the Precision-Recall curves were retained and the optimization process was repeated three more times, creating a smaller number of RF classifiers each time to narrow down the feature and feature weight choices to find the optimal combination. In the final two rounds of optimization, the number of trees for each RF classifier was raised to 1,000 trees. 

In the last round of optimization, we added 133 random candidates from M20 to the validation set and used the True Positive Rate (TPR) of the M20 candidates to help make the final classifier decision. The final RF classifier that we chose had the highest AUC of the ROC and AUC of the precision-recall curve along with the highest recall rate of the 133 random M20 candidates at the threshold for the maximal Youden's J index.

\subsection{Classification Metrics \label{sec:rf_class_metric}}

To evaluate classifier performance, \citet{Akosa17} describes several classification metrics appropriate for machine learning problems with an imbalanced training set. Criteria for selecting the best classifier are based on scalar classification metrics and the AUC for the ROC curve and the Precision-Recall curve.

The output of a RF applied a new data point is a prediction value between 0 and 1. The prediction value is not a probability value, but can be considered a pseudo-probability that the input to the RF belongs to the positive class (in this case, a light curve with an exoplanet transit signal).  After the classifier is applied to a validation set, we must set a classifier threshold to convert this `soft' classification pseudo-probability to a `hard' classification to produce a confusion matrix of True Positives, False Positives, True Negatives, and False Negatives. The threshold of the classifier can be placed anywhere between 0 and 1.  

There is no rule for guiding the choice of the threshold for a classifier other than {\it ex post facto} performance metrics like the Matthew’s Correlation Coefficient (MCC), Youden's J Index, and adjusted F-score \citep{Powers11}.   These are defined as follows: 
\begin{eqnarray}
  MCC ~&=&~ \frac{TP \times TN - FP \times FN}{\sqrt{(TP + FP)(TP+FN)(TN+FP)(TN+FN)}} \\
  Adjusted\, F\textrm{-}Score ~&=&~ 5 \sqrt{\frac{TP \times TN}{\left(5\, TP + 4\, FP + FN\right) \times \left(5\, TN + 4\, FP + FN\right)}} \\
  Youden's\, J\, Index ~&=&~ \frac{TP}{TP+FN} - \frac{FP}{TN+FP}
\end{eqnarray}
where TP is the number of true positives (exoplanet injections above the RF threshold), TN is the number of true negatives (negative validation set objects below the RF threshold), FP is the number of False Positives (negative validation set objects above the threshold), and FN is the number of false negatives (exoplanet injections below the RF threshold). 

MCC is the correlation coefficient between the labeled test data set and the predicted labels for the validation set. It can have values between -1 and 1, with 1 corresponding to a perfect classifier, 0 indicating a random classifier, and -1 corresponding to the worst possible classifier. The adjusted F-score, ranging from 0 to 1, is an improvement to the normal F-score that balances classifier recall and precision for imbalanced classes. It gives a higher weight to the correctly classified positive instances in the test data set and a stronger weight against FPs than the traditional F-score. In Youden's J, also ranging from 0 to 1, the first term is the classifier recall rate or True Positive Rate (TPR) and the second term is the False Positive Rate (FPR).  When evaluating a trial classifier, the threshold corresponding to the maximum Youden's J index was used. This is not the final threshold used for the final RF classifier, but a standard choice for performance comparison.

The ROC plots the TPR as a function of the FPR for every possible threshold value of the classifier. We used CRAN package \emph{ROCR} \citep{Sing05} implemented with the R software \citep{Rcore} to calculate the ROCs. The AUC for the ROC is a measure of classifier performance that does not depend on a single threshold choice. An AUC of the ROC of 1 indicates a perfect classifier, 0.5 indicates a classifier that performs no better than assigning random labels, and 0 indicates the worst possible classifier. 

Closely related to the ROC, the Precision-Recall curve is used for imbalanced classification problems because both precision and recall focus on the correct classification of the positive class. The AUC for the Precision-Recall curve is another measure of classifier performance that does not depend on a threshold choice.

\subsection{Feature Selection} \label{sec:features}

Table \ref{tab:rf_features} lists the 37 features and feature weights in the final optimized RF classifier.  The weights indicate the probability of the feature being among the randomly chosen features for each node calculation. The table organizes the features by the stage of DTARPS-S analysis. The feature groups are described here:

\begin{deluxetable}{rcl}[tb!]
    \centering
    \caption{Scalar features used in the optimized Random Forest classifier \label{tab:rf_features}}
\tablehead{
    \colhead{Feature} & \colhead{Weight} & \colhead{Description} 
}
\startdata
\multicolumn{3}{c}{\textbf{Stellar Properties}}\\
    tic\_Radius & 0.027 & Radius of star [$R_{\sun}$]\\ 
    logg & 0.027 & \emph{Gaia} DR2 log(g) measurement \\
    Gaia.color & 0.027 & \emph{Gaia} DR2 color, $G_{BP}$ $-$ $G_{RP}$\\
    Luminosity & 0.027 & \emph{Gaia} DR2 luminosity of the star \\
    \hline
    \multicolumn{3}{c}{\textbf{DIAmante Light Curve Properties}}\\
    median.lc & 0.027 & Median value of the DIAmante light curve\\
    Redchisq.lc & 0.027 & Reduced $\chi^2$ value for a flat model \\ 
    tail\_range.lc & 0.027 & 98th quantile - 2nd quantile divided by the IQR\\
    POM.lc & 0.027 & Greatest positive outlier measure\\
    skew.lc & 0.027 & Measure of Non-Gaussian asymmetry\\ 
    kurt.lc & 0.027 & Measure Non-Gaussian outlier strength\\ 
    P\_autocor.lc & 0.027 & $p$-value from the Ljung-Box test for autocorrelation \\ 
    \hline
    \multicolumn{3}{c}{\textbf{Differenced Light Curve Properties}} \\
    quantiles.diff.01 & 0.027 & 1st quantile\\
    quantiles.diff.90 & 0.027 & 90th quantile\\
    POM.diff & 0.027 & Greatest positive outlier measure\\ 
    \hline
    \multicolumn{3}{c}{\textbf{ARIMA Residuals Properties}} \\
    quantiles.resid.10 & 0.027 & 10th quantile \\
    quantiles.resid.99 & 0.027 & 99th quantile \\
    POM.resid & 0.027 & Greatest positive outlier measure \\
    IQR.resid & 0.027 & Interquartile range \\
    Redchisq.resid & 0.027 & Reduced $\chi^2$ value for a flat model \\
    Prob\_norm.resid & 0.027 & $p$-value from the Anderson-Darling test for normality \\
    Prob\_autocor.resid & 0.027 & $p$-value from the Ljung-Box test for autocorrelation \\
    IQR.improv & 0.108 & Ratio of the IQR for the ARIMA residuals to the \\
    & & DIAmante light curve\\
    Redchisq.improv & 0.108 & Ratio of the reduced $\chi^2$ value of the ARIMA residuals to the\\
    & & DIAmante light curve \\
    \hline
    \multicolumn{3}{c}{\textbf{TCF Periodogram Properties}} \\
    LOESS\_mnsnr & 0.081 & Mean SNR of top 100 periodogram peaks with respect to \\
    & & LOESS fit \\ 
    TCFpeaks\_mean & 0.081 & Mean raw power of the top 100 TCF peaks\\
    LOESSpeaks\_mean & 0.081 & Mean power above the LOESS fit of the top 100 TCF peaks\\
    \hline
    \multicolumn{3}{c}{\textbf{Best TCF Transit Properties}} \\
    TCF\_power & 0.135 & Strength of best peak in the TCF periodogram \\ 
    TCF\_period & 0.054 & Period of the best TCF periodogram peak \\
    TCF\_depthSNR & 0.135 & SNR of the best TCF periodogram peak\\ 
    TCF\_shape & 0.135 & Mean of the in-transit folded light curve divided by the MAD\\  
    & & out-of-transit folded light curve \\ 
    Folded\_AD & 0.135 & Anderson-Darling test on the phases of the folded light curve\\ 
    arbox\_deperr & 0.135 & Transit depth error from ARIMAX model \\ 
    even.odd.p\_value & 0.135 & $t$-test $p$-value for even and odd transit depths \\ 
    trans.p\_value & 0.135 & $t$-test $p$-value for in-transit and out-of-transit depths\\ 
    snr.transit & 0.135 &  $\delta / \sigma \left( 1 / \sqrt{N_{in}} + 1 / \sqrt{N_{out}} \right)$ (See \S 10.1.1 M20) \\ 
    frac\_dur & 0.135 & Fractional transit duration (See \S 10.1.8 M20)\\ 
    planet\_rad\_tcf & 0.054 & Planet radius calculated from the TCF transit depth\\\hline
    \enddata
    \tablecomments {All features for the light curve have been calculated after the removal of outliers and ramping problems from spacecraft jitter.}
\end{deluxetable}

\begin{list}{}{\setlength\itemindent{-\leftmargin}}
  \item[] \emph{Stellar properties}: Stellar metadata from the TIC v8 \citep{Stassun19} and from the \emph{Gaia} DR 2 catalog \citep{GaiaCollab_DR2}. Stellar properties tested include the effective temperature, mass, \TESS T magnitude, \emph{Gaia} parallax, \emph{Gaia} G magnitude, G magnitude SNR, and others. The classifier optimization found that stellar radius, surface gravity, luminosity, and \emph{Gaia} $G_{BP}$ - $G_{RP}$ color index played significant roles in classification. 
  
  \item[] \emph{DIAmante light curve properties}:  The reduced $\chi^2$ measures the goodness-of-fit of the light curve to a constant median brightness. The tail range compares the range of the middle 96 percent of the light curve flux values with the range of the middle 50 percent of the light curve flux values. The Positive Outlier Measure (POM) measures the most extreme positive outlier in the light curve with respect to the median. As discussed by \citet{Caceres19a}, the POM helps identify stars with strong flares that may cause spurious peaks in the TCF Periodogram. Skewness, the third standardized moment of the distribution, is a measure of the asymmetry of the distribution of light curve flux values around the mean. Kurtosis, the fourth standardized moment of a distribution, is helpful to measure the strength of outliers with respect to a Gaussian distribution. The Ljung-Box test for autocorrelation applied to the light curve has a null hypothesis that the flux values are independently distributed and tests the alternative hypothesis that the flux values show correlation. A $p$-value $\gtrsim 0.01$ indicates that the light curve is consistent with white noise.
  
  \item[] \emph{Differenced light curve properties}: Statistics of the distribution of flux values for the differenced light curve including the 1st and 90th quantiles of the distribution. The POM, again, measures the most extreme positive outlier in the differenced light curve with respect to the median of the differenced light curve, which would identify a sharp transit brightening. 
  
  \item[] \emph{ARIMA residual properties}: These features include: four statistics describing the residuals after the best fit ARIMA model had been subtracted from the differenced light curve (10\% and 90\% quantiles, POM, IQR); three statistical tests applied to the residual light curve ($\chi^2$, Anderson-Darling and Ljung-Box); and two measures of the importance of the ARIMA fitting (IQR and $\chi^2$ improvements). 

These last two measures, with high feature weights, raised classifier performance more than most features; this demonstrates the importance of ARIMA modeling for planet detection. The ratio of the IQR of the ARIMA residuals to the IQR of the DIAmante light curve measures the improvement in the noise of the time series with a smaller ratio indicating a greater effect of the ARIMA fit. The ratio of the reduced $\chi^2$ values for a constant flux model of the ARIMA residuals to the DIAmante light curve measure how well the ARIMA model removed variation and trend from the light curve. The Ljung-Box test applied to the ARIMA residuals indicates how well the ARIMA model did at removing short-memory autocorrelation from the time series. The Anderson-Darling normality test is used here to determine if the ARIMA residuals fluxes follow a Gaussian distribution. A $p$-value $\gtrsim 0.01$ indicates that ARIMA residuals are consistent with Gaussian noise. 
  
  \item[] \emph{TCF periodogram properties}: \citet{Caceres19a} found that the collective top periodogram peak properties, not just the results from the strongest TCF periodogram peak, improved the performance of the classifier for \Kepler light curves. We find the same behavior for the \TESS classifier. The mean SNR of the top 100 TCF peaks is used to identify periodograms with many noisy peaks in the periodogram, suggesting that no strong periodicity was identified. The mean power of the top 100 TCF peaks calculated from the raw TCF output and from the LOESS regression line are used to pick out periodograms with just a single or a few strong peaks. A low mean power of the top 100 peaks but a strong power for the best peak indicates that the best exoplanet transit peak is highly significant and does not arise from noise in the ARIMA residuals. 
  
  \item[] \emph{Best TCF transit properties}: The eleven properties based on the best TCF period have high feature weights. These include: three properties of the highest TCF periodogram peak; six measures of the time series folded modulo the best period; and the significance of the depth derived from the parametric ARIMAX model. 
  
  The three properties of the strongest TCF peak are the period, TCF power, and SNR of the strongest peak in the TCF periodogram within a small window after subtracting the LOESS smoother. \citet{Caceres19a} refrained from including the transit period and corresponding planetary radius from their RF classifier to avoid biasing their candidate results. We include the transit period because it reduces the significance of spurious peaks with periods $13-15$ days due to the \TESS satellite orbit (Figure \ref{fig:rf_pred}).
  
  The shape parameter of the folded DIAmante light curve compares the mean value of the transit with the median absolute deviation (MAD) value for the other mean values of the non-transit sections of the phase-folded light curve. This measures the transit's flux difference compared with the rest of the light curve, and provides a subtle distinction between planets and EBs. 
  
  The Anderson-Darling test is applied to the distribution of phases for the observations in the phase-folded light curve to test if the phases are consistent with an underlying uniform distribution. If not, then the TCF may have found a spurious periodic signal by aligning gaps in the light curve rather than identifying a true exoplanet transit.  This is a crucial feature for identifying and reducing spurious periodogram peaks arising due to periodicities in cadence gaps.  
  
  The $t$-test, designed to quantify the difference in means of two Gaussian distributions, is applied to the even and odd transit light curve flux values and the in-transit and out-of-transit light curve flux values. A larger $p$-value is desired for the even-odd $t$-test to distinguish planet transits from EBs, while a smaller $p$-value is desired comparing the in-transit and out-of-transit fluxes states to show that the transit represents a statistically significant dip in flux. These tests rely on a transit mask to label points in the light curve that are in-transit and out-of-transit, as well as to label even and odd transits. 
  
  The SNR of the primary transit feature describes how well the transit signal with the period and phase from TCF add up over the phase-folded light curve. It is described in both M20 and \citet{Kovacs02}.
  
  The fractional transit duration is the ratio of the transit duration to the transit period.  This feature was used by M20.

\item[] \emph{Inferred planet radius}:
  The radius of an exoplanet is calculated from the depth of the transit from the best TCF peak and the stellar radius from the TIC.  We include the planetary radius because the injected FP signals included in our negative training set (\S \ref{sec:neg_train}) allows us to train the RF classifier away from likely astrophysical FPs with very deep transit depths. Note however that this may reduce the DTARPS-S classifier sensitivity to very large, inflated gaseous exoplanets. 
\end{list}

\section{The DTARPS-S Final Classifier} \label{sec:rf_final}

\begin{figure}[tb]
  \includegraphics[width=0.49\textwidth]{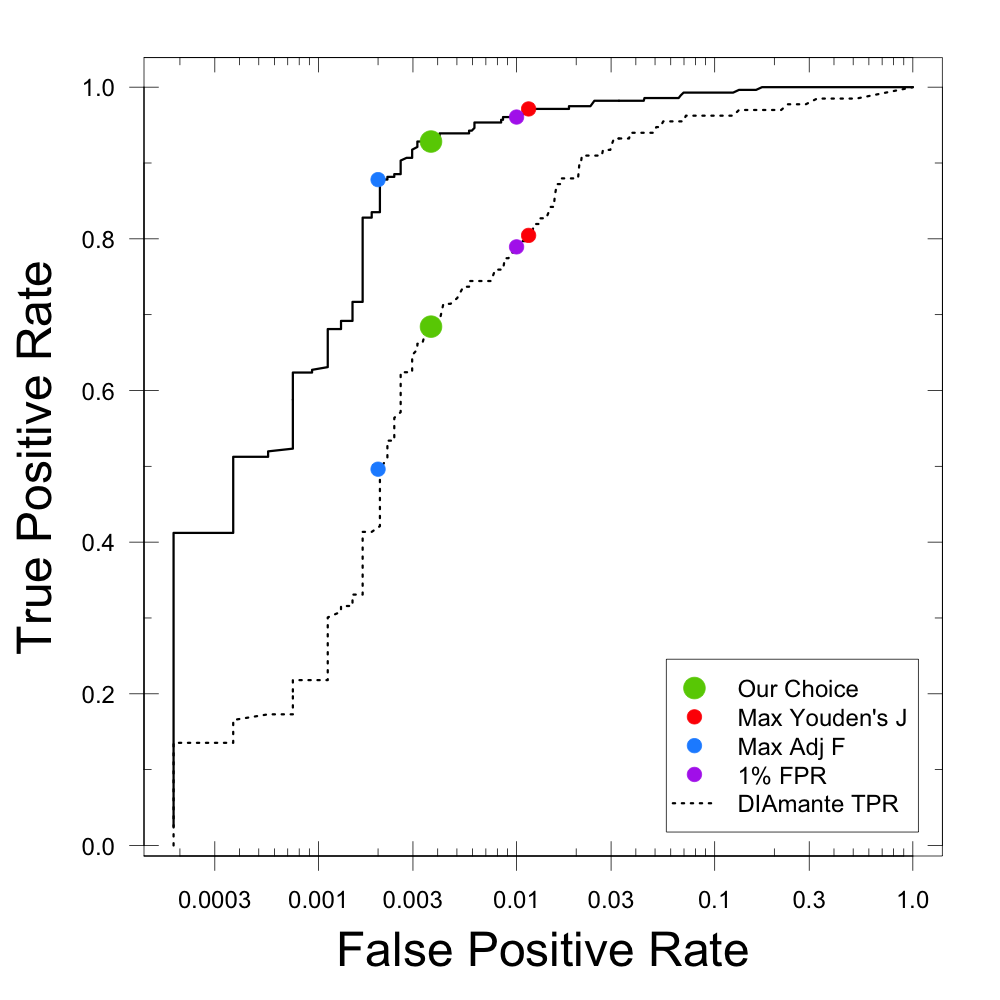}
  \includegraphics[width=0.49\textwidth]{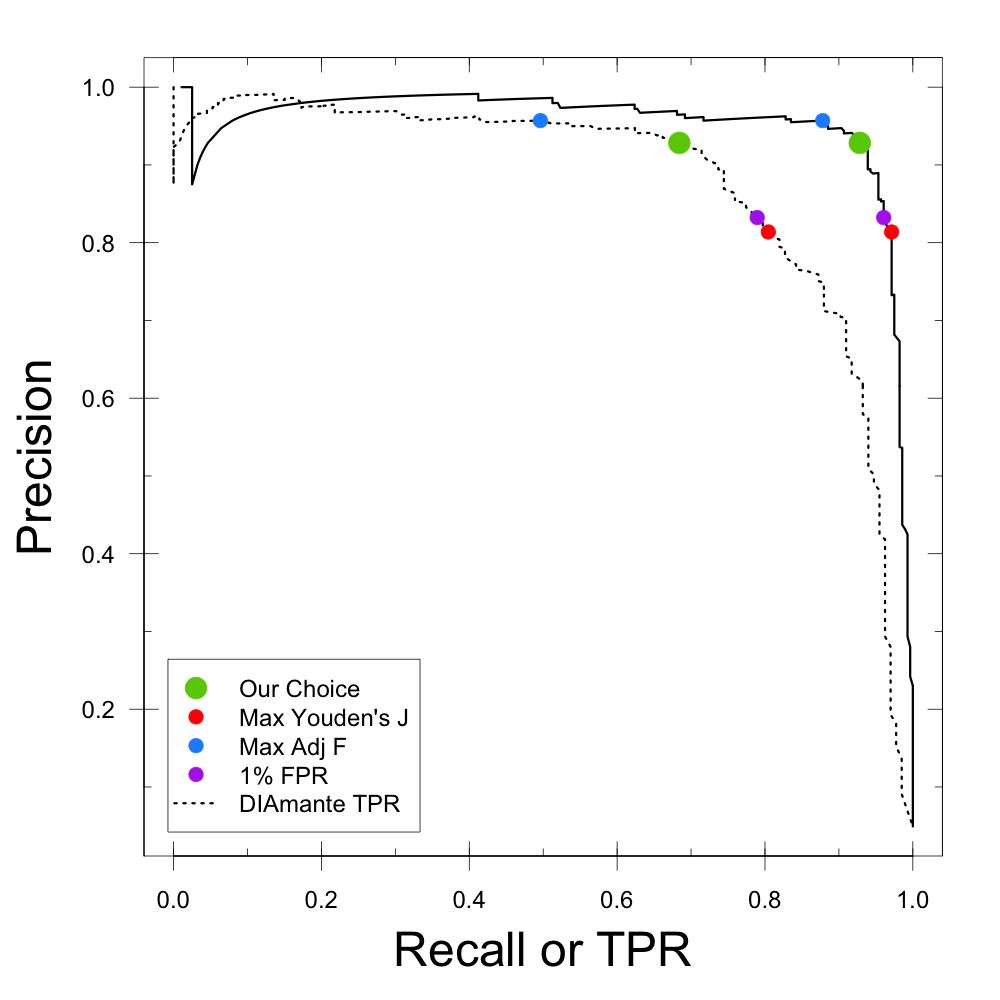}
  \caption{Performance of the final Random Forest classifier for every possible threshold choice shown with the Receiver Operator Characteristic (left) and Precision-Recall (right) curves. The solid lines derive from application to the validation set, and the dashed lines are the recall from 133 planet candidates of M20.  Our choice of threshold $P_{RF} = 0.300$ (shown in green) is compared with other possible threshold choices, including the 1\% FPR choice used by M20 (purple). \label{fig:rf_ROC_prec_recall}}
\end{figure}

Figure \ref{fig:rf_ROC_prec_recall} shows the ROC and Precision-Recall curves for the final RF classifier. Note that the abscissa is logarithmically transformed to highlight small values of FPR needed for reliable transit discovery. The solid lines give the TPR and the FPR (or the TPR and the precision) for every possible threshold value between 0 and 1. The dashed lines give the recall rate for random sample of 133 planet candidates from the M20 DIAmante study. 

\begin{table}[b!]
    \centering
    \caption{Classification metrics for the validation set} \label{tab:rf_metrics}
    \begin{tabular}{lcccccc}
        \\\hline \hline
          & $P_{RF}$ &  &  &  & Adjusted & Youden's\\
          &  threshold &  TPR & FPR & MCC & F-score & J Index \\\hline
         Max Youden's J & 0.155 & 0.9713 & 0.0115 & 0.8830  & 0.8869  & 0.9598 \\
        Max Adj F &  0.481 & 0.8781 & 0.0020 & 0.9127  & 0.9682 & 0.8761 \\
        1\% FPR & 0.174 & 0.9606 & 0.0100 & 0.8884 & 0.8927 & 0.9505 \\
        {\bf Our Choice} & {\bf 0.300} & {\bf 0.9283} & {\bf 0.0037} & {\bf 0.9246} & {\bf 0.9283} & {\bf 0.9246} \\\hline
    \end{tabular}
\end{table}

Our threshold choice of Random Forest probability $P_{RF} = 0.300$ is shown with the larger green points. We chose this threshold to minimize the FPR as much as possible while maintaining high DIAmante survey recall and the TPR. Three other threshold choices are shown for comparison; M20 chose a threshold that gave a FPR of $1\%$ that lies very close to the maximum Youden's J threshold. The final TPR and FPR values for our chosen threshold, and comparison thresholds, are listed in Table \ref{tab:rf_metrics} along with classification metrics described in \S \ref{sec:rf_optimize}. 

\begin{figure}[tb]
  \centering
  \includegraphics[width=0.7\textwidth]{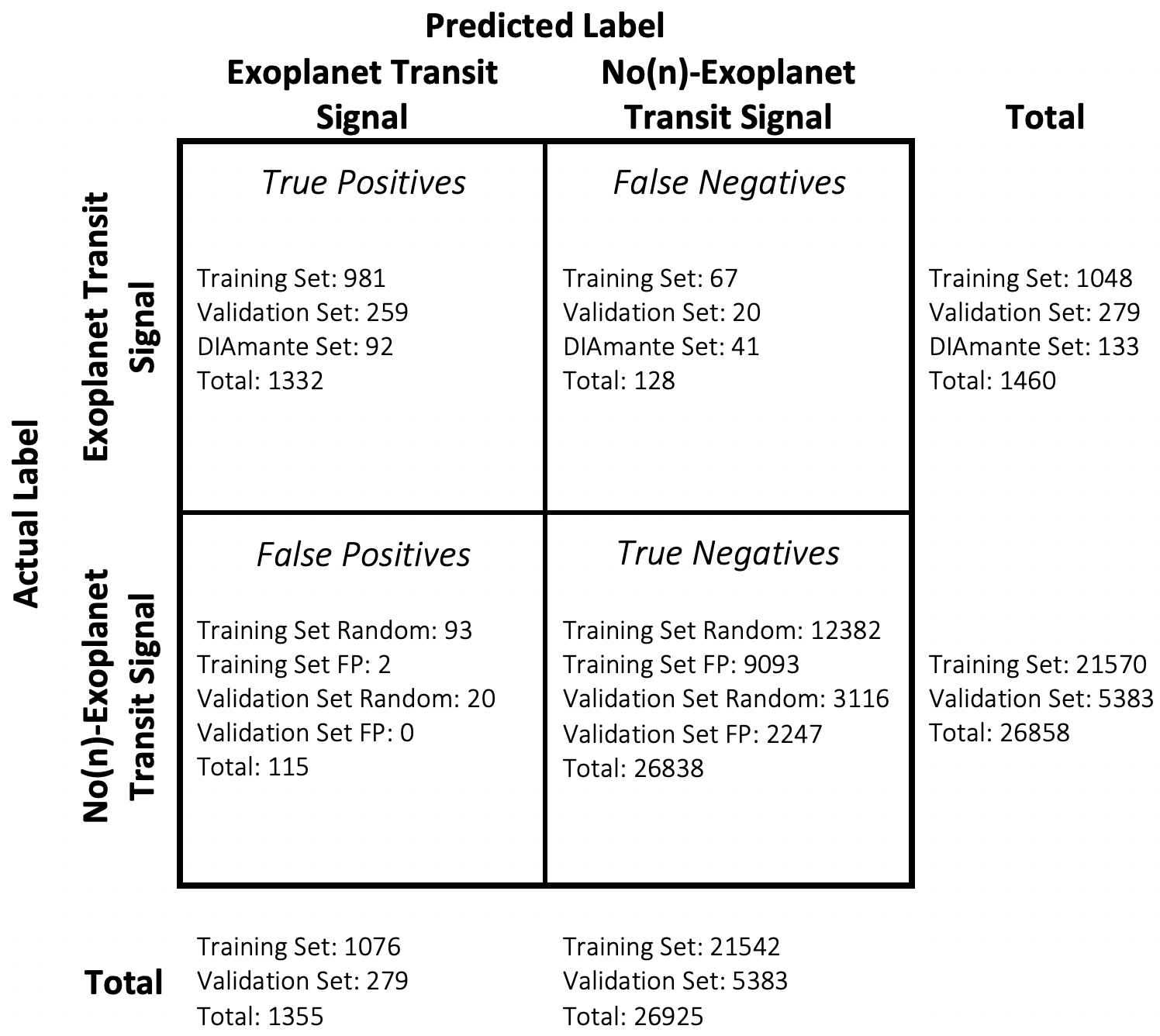}
  \caption{Confusion matrix for the final RF classifier with threshold $P_{RF}=0.300$. Values are based on the validation set and OOB predictions for the training set. \label{fig:rf_conf_matrix}}
\end{figure}

Further detail is given in Figure \ref{fig:rf_conf_matrix} showing the confusion matrix for the final RF classifier based on our chosen threshold $P_{RF}=0.300$. The confusion matrix shows how well the predicted labels from the RF classifier line up with the actual labels of the data in the training and validation sets. For the training set, we used the out-of-bag (OOB) RF prediction value to determine the predicted label. Each tree in the RF classifier uses a bootstrapped sample of the training set for construction, called bagging. OOB means that a data case from the training set was classified only by trees in the classifier that the data case was not used to `grow' or train the tree. OOB prediction values are calculated using only decision trees in the RF ensemble that were not grown using that training data case \citep{Breiman01}. We include the predicted labels for the random sample of 133 DIAmante candidates (M20) used to create the dashed TPR line in the ROC curve (see Figure \ref{fig:rf_ROC_prec_recall}, left plot). The $P_{RF} = 0.300$ threshold for the RF classifier gave us a DIAmante TPR of 69\% for the 133 randomly selected candidates.

For the data whose actual label is non-exoplanet transit signal, we separately counted the negative data sets of randomly selected light curves and injected FPs. Perhaps the most important result here is the extraordinary effectiveness of the final RF classifier with respect to injected False Positive EBs: only labeled 2 out of the 11,342 (0.02\%) injected FPs used in the labeled data sets are falsely labeled as an exoplanet transit signal.  {\it Overall, the optimized Random Forest classifier attains a 92.5\% True Positive recovery rate with 0.4\% False Positive contamination with respect to injected exoplanet transits and simulated variable stars.} 

\begin{figure}[tb!]
  \centering
  \includegraphics[width=0.99\textwidth]{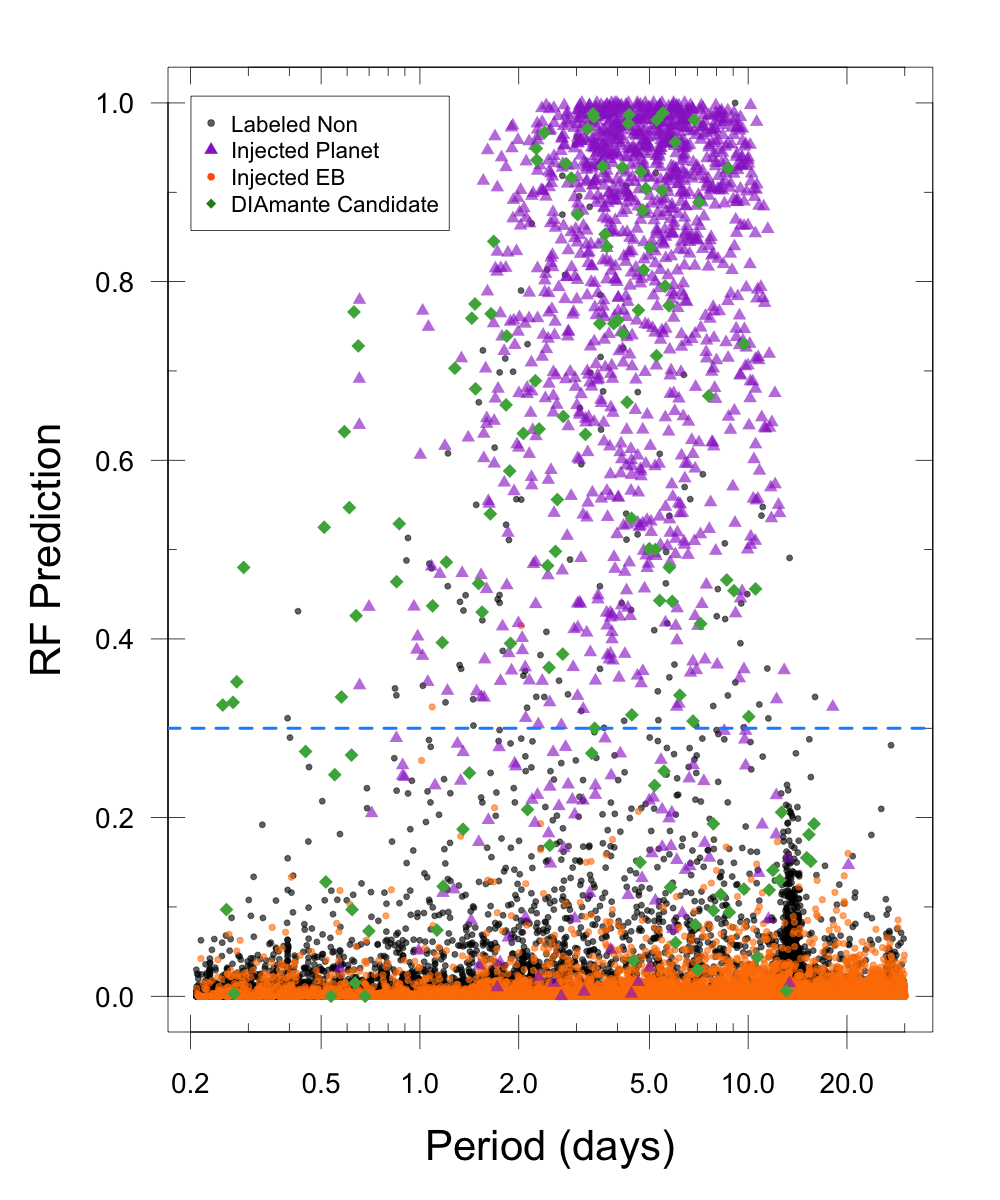}
  \caption{Distribution of the Random Forest pseudo-probability $P_{RF}$ for the optimized classifier on the training and validation sets. The purple triangles represent light curves with injected transiting exoplanet signals that passed human vetting and were utilized in the positive training set (\S \ref{sec:kep_inj}). The green diamonds represent the 133 candidates identified in M20, that assisted in optimizing the classifier. The black points indicate random light curves that were given negative class labels; most have $P_{RF}$ values near zero.  The orange points represent light curves with injected False Positive signals (\S \ref{sec:neg_train}). Our chosen threshold for the final classifier is shown as the blue dashed line at $P_{RF} = 0.300$. The strong performance of the classifier is directly seen: purple and green points lie mostly above the threshold while the black and orange points like mostly below the threshold. \label{fig:rf_pred}}
\end{figure}

The performance of the final optimized classifier is shown visually in Figure \ref{fig:rf_pred} where the classification results are plotted as a function of period from the best TCF peak (\S \ref{sec:TCF}) for the entire labeled data set (\S \ref{sec:train_test}) and 133 randomly selected candidates identified in M20. The RF prediction values for labeled data set objects in the training set is the OOB prediction value.

The plot of $P_{RF}$ against TCF best period in Figure \ref{fig:rf_pred} provides valuable insights into the classifier performance that are not revealed in the confusion matrix. The vast majority of injected planets are recovered with periods $\sim 0.7$ to 11 days, and nearly all labeled negatives are rejected from 0.2 to 30 days. The small number of False Positives (with respect to the threshold) do not have preferred periods. 

A strong spike of negative label points lying below the $P_{RF} =  0.300$ threshold is present at periods $13-15$ days. This arises from the \TESS satellite 13.7 day lunar-synchronous orbital period with a large gap in the middle of the FFI light curve in each sector. This leads the Transit Comb Filter algorithm, in the absence of a strong transit signal, to fold the data in half to line up the gaps in the data. Other period search procedures applied to \TESS light curves are similarly affected \citep[e.g. M20,][]{Chakraborty20}. Many of the trial RF classifiers were less successful than the final classifier in pushing down the $P_{RF}$ values for these spurious periodicities. However, in the final classifier, this spurious spike in Figure \ref{fig:rf_pred} has the indirect effect of causing a sharp drop in the RF prediction value of all objects with periods longer than $\sim 11$ days.  As a result, our classifier is insensitive to true exoplanet transits at longer periods.  This might have been alleviated if our injected exoplanet training set extended to longer periods $\sim 15-25$ days.  

In contrast, although the injected exoplanet periods do not go shorter than 0.625 days (because the injections were based on \Kepler planets based on a transit search truncated below 0.5 days), the optimized RF classifier does not appear strongly biased against short period transit signals. This is seen by the recovery of several DIAmante candidates in the $0.2-0.6$ day regime.  

\begin{figure}[tb!]
   \centering  \includegraphics[width=0.7\textwidth]{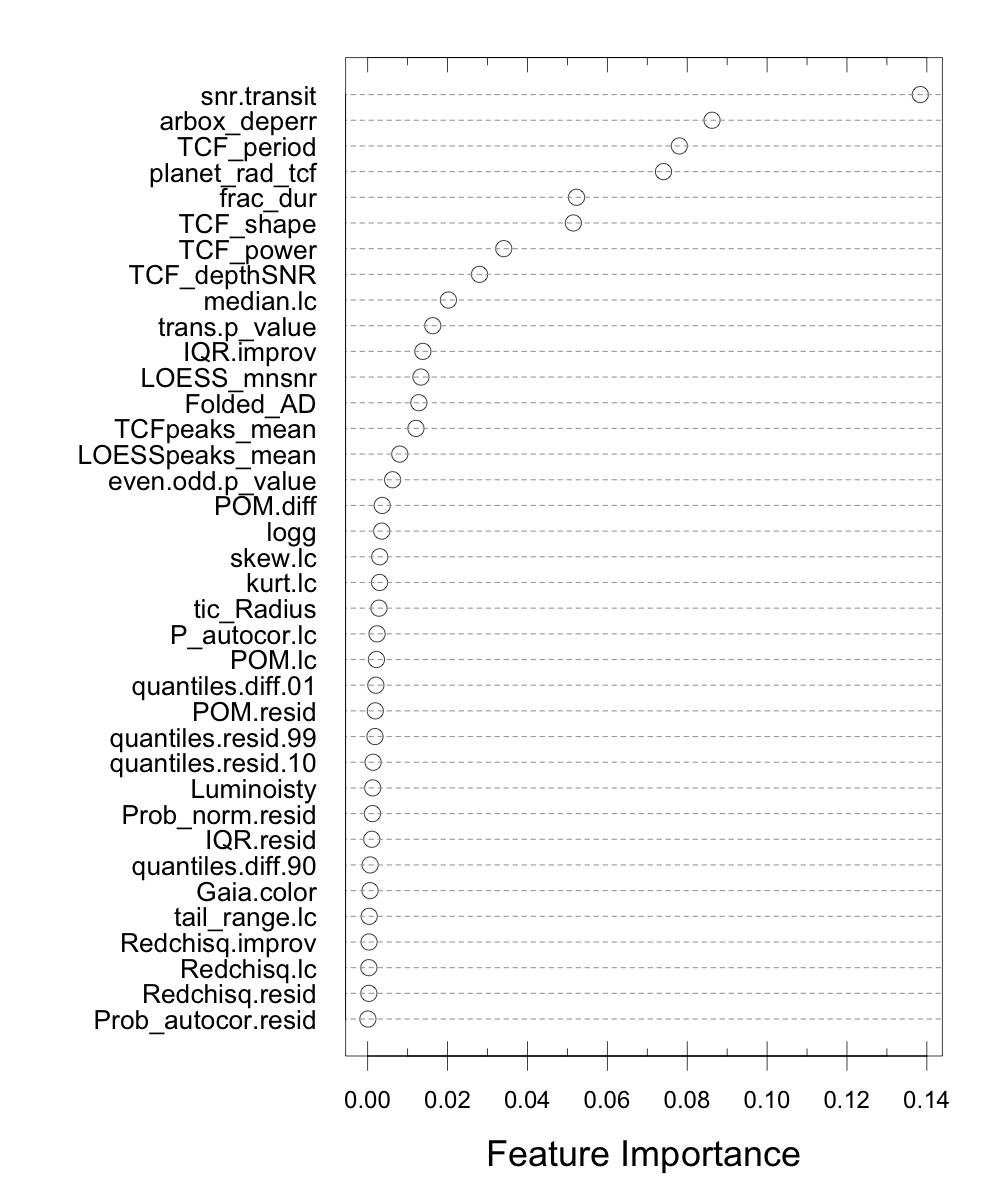}
  \caption{Feature importance for the final Random Forest classifier ordered by importance. Descriptions of the 37 features appear in Table \ref{tab:rf_features}. \label{fig:rf_feature}}
\end{figure}

Figure \ref{fig:rf_feature} shows the feature importance plot associated with the final RF classifier where input features are ordered by their importance to the classification. Feature importance is calculated by comparing the training set label accuracy from a perturbed OOB forest ensemble with the unperturbed OOB forest ensemble \citep{Ishwaran22}. For each feature, the label from the perturbed OOB forest is found by classifying each data case normally on the OOB trees in the forest for that data case, but whenever a node is encountered that is split using the feature for which the importance is being calculated, the opposite daughter node is used for classification. Therefore, the feature importance shows the improvement of the accuracy of the entire RF classifier when the correct classification path in the trees is used for a feature rather than the opposite classification path for that feature. The feature importance is calculated from the predictive success of the feature and often cannot be interpreted physically \citep{Genuer10}.

The signal-to-noise ratio of the transit in the folded light curve is the most important feature, followed by the error on the ARIMAX fitted transit depth, the period, and the planet radius. Some features serve as positive discriminators of planet transits (such as snr.transit and arbox\_deperr) while others serve to push away spurious effects (TCF\_period helps remove $13-15$ day periodogram peaks, and planet\_rad\_tcf helps remove deep EB eclipses). Of the top five most important features, three are also among the most important features in the DIAmante classifier derived by M20 (snr.transit, planet\_rad\_tcf, and frac\_dur). 

\section{The DTARPS-S Analysis List} \label{sec:DAL}

The final RF classifier was applied to a test set of 823,099 DIAmante light curves. This is the full DIAmante collection of \TESS Year 1 light curves minus those with missing features (Table~\ref{tab:rf_features}).  Random Forest classifiers require full characterization of each object, and we did not attempt imputation of missing data.  The classifier threshold of $P_{RF} = 0.300$ was then applied.  The result is 7,377 DTARPS-S processed DIAmante light curves had a RF prediction value above the threshold. We call this the {\it DTARPS-S Analysis List} of \TESS stars.  This DTARPS-S Analysis List represents 0.9\% of input light curves selected by uniform statistical procedures to have periodic transit-like features.  It is roughly similar to the Threshold-Crossing Event (TCE) list by the \TESS official pipeline as a step toward producing their \TESS Object of Interest (TOI) list \citep{Guerrero21}, although their processing steps are quite different from the DTARPS-S analysis. 

A small portion of the DTARPS-S Analysis List is shown in Table~\ref{tab:DTARPS_DAL} with the full list available in machine readable format from the electronic edition of this paper.

{\it We emphasize that, while this DTARPS-S Analysis List of 7,377 \TESS stars has captured many transiting planets, it is still dominated by False Alarm and False Positive objects.} The False Positive Rate of 0.0037 estimated from the combined labeled training and validation sets (Table~\ref{tab:rf_metrics}) predicts that at least $\sim 3,000$ of the 7,377 objects are not valid transiting planets. A rigorous vetting process to remove as many falsely labeled objects as possible is therefore needed to give a smaller catalog with much higher reliability (\S\ref{sec:vetting}). In the parlance of machine learning classification, the 7,377 stars represents the maximum recall of the DTARPS-S analysis but with low sensitivity. The catalogs of 772 stars after several vetting procedures presented in Paper II represents the subset with high sensitivity but reduced recall rate. 
    
\begin{deluxetable}{rccccccrrrrr}[b]
\tablecaption{DTARPS-S Analysis List}   \label{tab:DTARPS_DAL}
\tabletypesize{\small}
\tablewidth{0pt}
\centering
\tablehead{
\multicolumn{8}{c}{Designations} && \multicolumn{3}{c}{Star} \\ \cline{1-8} \cline{10-12} 
\colhead{TIC} & \colhead{DTARPS-S} & \colhead{NEA$_{name}$} & \colhead{NEA$_{disp}$} & \colhead{TOI} & \colhead{TOI$_{disp}$} & \colhead{DIAm} & \colhead{Refs} && 
\colhead{R.A.} & \colhead{Dec} &  \colhead{T} \\ 
\colhead{(1)} & \colhead{(2)} & \colhead{(3)} & \colhead{(4)} & \colhead{(5)} & \colhead{(6)} & \colhead{(7)} & \colhead{(8)} && \colhead{(9)} &
\colhead{(10)} & \colhead{(11)}  }
\startdata
   4711 & \nodata &  \nodata & PC     & 3129.01 & \nodata & F & \nodata && 218.91878 & -26.17928  & 10.9 \\ 
   17361 & 1   &  \nodata & APC   & 3127.01 & \nodata & T & \nodata && 219.33632 & -24.95848 & 11.3 \\  
   44870 & \nodata & \nodata & \nodata & \nodata & PC & F & \nodata && 220.19607 & -29.25022 & 11.4 \\ 
  113636 & \nodata &  \nodata & \nodata & \nodata & \nodata  & F & \nodata && 222.35362 & -28.42375 & 11.6 \\  
  153687 & \nodata & \nodata & \nodata & \nodata & \nodata  & F & \nodata && 223.58956 & -27.16942  & 11.6 \\  
\enddata
\tablecomments{The full table of 7,377 light curves exceeding the DTARPS-S Random Forest threshold is available in the electronic version of the paper. Column descriptions: \\
(1) TIC: TESS Input Catalog identifier.\\
(2) DTARPS-S: DTARPS-S identifier from Paper II.  Sequence number $1-467$ from the DTARPS-S Candidate Planet catalog (Paper II, vetting Levels 1 and 2). $GP$ indicator for Galactic Plane DTARPS-S list with reduced vetting (Paper II, vetting Level 3) \\
(3) NEA$_{name}$: Name, NASA Exoplanet Archive \citep{NEA-CP} \\
(4) NEA$_{disp}$: Disposition, NASA Exoplanet Archive (combined Confirmed Planets and TOI$^1$ list) \citep{NEA-CP}. CP includes CP (Confirmed Planet and KP (Known Planet); PC includes APC (Ambiguous Planet Candidate) and PC (Planet Candidates); FP includes EB (Eclipsing Binary); FA (False Alarm); FP (False Positive). ... = previously unidentified by NEA (accessed March 15, 2022). \\
(5) TOI: \TESS Object of Interest$^1$ (accessed February 2022). \\
(6) TOI$_{disp}$: TOI Disposition$^1$. \\
(7) DIAm: Flag for DIAmante planet candidate (M20) \\
(8) Refs: Identified as a planet candidate or other object of interest by other studies: 
  Co = \citet{Collins18}; Dr = \citet{Dressing19}; Do = \citet{Dong21}; Ei = \citet{Eisner21}; 
  Fe = \citet{Feinstein19}; Ko = \citet{Kostov19}; Kr = \citet{Kruse19}; 
  Ma = \citet{Mayo18}; Ol = \citet{Olmschenk21}; Sc = \citet{Schanche19}; 
  Tu =\citet{Tu20}; vB = \citet{vonBoetticher19}; Yu = \citet{Yu19} \\
(9-10) R.A., Dec = Right Ascension and Declination  from Gaia DR2 catalog.\\
(11) T = \TESS band magnitude 
}
\end{deluxetable}
 
\begin{deluxetable}{rrrrrrrrrrr}
\renewcommand\thetable{3}
\tabletypesize{\small}
\tablewidth{0pt}
\centering
\tablecaption{{\bf (continued)} }
\tablehead{
\multicolumn{5}{c}{Light curve and ARIMA residuals} && \multicolumn{3}{c}{Transit Comb Filter} && \colhead{Classifier} \\ \cline{1-5} \cline{7-9} \cline{11-11}
\colhead{N$_{lc}$} & \colhead{IQR$_{lc}$} & \colhead{P$_{LB,lc}$} & \colhead{IQR$_{AR}$} & \colhead{P$_{LB,AR}$} && 
   \colhead{Period} & \colhead{Depth} & \colhead{SNR} && \colhead{P$_{RF}$} \\
\colhead{(12)} & \colhead{(13)} & \colhead{(14)} & \colhead{(15)} & \colhead{(16)} && \colhead{(17)} & \colhead{(18)} & \colhead{(19)} && \colhead{(20)} }
\startdata
 849~ & 0.0020 & -6.0~~~ & 0.0022 & -0.1~~~  && 2.33034 & 0.0045  & 33~~ && 0.73~~~~ \\
 849~ & 0.0009 & -6.0~~~ & 0.0012 & -0.1~~~  && 3.60412 & 0.0072  & 83~~ && 0.94~~~~ \\
 922~ & 0.0014 & -6.0~~~ & 0.0014 & -0.3~~~  && 3.44101 & 0.0017  & 12~~ && 0.33~~~~ \\
 922~ & 0.0010 & -6.0~~~ & 0.0010 & -0.2~~~ && 10.75002 & 0.0029  & 40~~ && 0.42~~~~ \\
 922~ & 0.0047 & -6.0~~~ & 0.0032 & 0.0~~~  && 0.34148 & 0.0104  & 324~~ && 0.31~~~~ \\
\enddata
\tablecomments{Column descriptions: \\
(12) N$_{lc}$: Number of measurements in TESS FFI light curve.\\
(13) IQR$_{lc}$: InterQuartile Range of normalized light curve fluxes.\\
(14) P$_{LB,lc}$: Log probability of autocorrelation in light curve from the Ljung-Box test. Values above -2 are consistent with white noise while values below -4 are not. \\
(15) IQR$_{AR}$: InterQuartile Range of ARIMA residuals.\\
(16) P$_{LB,AR}$: Log probability of autocorrelation of ARIMA residuals from the Ljung-Box test.\\
(17) Period: Transit period (day) from TCF periodogram.\\
(18) Depth: Transit depth from TCF. \\
(19) SNR: Signal-to-noise of peak power in TCF periodogram. \\
(20) P$_{RF}$: Pseudo-probability of planet classification from Random Forest classifier.
}
 \end{deluxetable}

\vspace*{0.1in}

\section{Effect of ARIMA Modeling on Injected Signals   \label{sec:TCF_acc}}

As the injected signals are the basis of the RF classifier's ability to identify true planetary transit signals,  it is important to understand how the ARIMA modeling affects the injected signals, TCF periodogram, and features that drive the classifier. The ARPS procedure described in \S\ref{sec:ARIMAi}$-$\S\ref{sec:ARIMAXi} is designed for sensitive and reliable $detection$ of planetary transits and may not make accurate $characterization$ of planetary properties such as orbital period and planet radius. In this section, we examine the limitations and biases present in the recovery of injected objects and their properties. We use the injected populations to investigate the effect of DTARPS-S processing on the physical parameters of the injected signals.  The low recovery of the injected planetary transit signals (\S \ref{sec:kep_inj}) is also explained.  We return to the injected population in \S\ref{sec:complete} for a third issue: estimating the completeness of the DTARPS-S Analysis List.

\subsection{Recovered Planet Properties} \label{sec:recovered_prop}

The results of the RF classifier depend heavily on the orbital parameters obtained from the TCF algorithm for the best period. The top eight most important features of the RF classifier (Figure \ref{fig:rf_feature}) are either extracted from the best TCF peak and TCF periodogram or are computed on the light curve phase-folded at the period identified by the best TCF transit model. However, of the 10,850 synthetic planet injections only 1,327 (12\%) had TCF orbital periods that were close enough to the injected synthetic period (or an integer ratio) to be recovered with human vetting.

\begin{figure}[tb!]
  \gridline{\fig{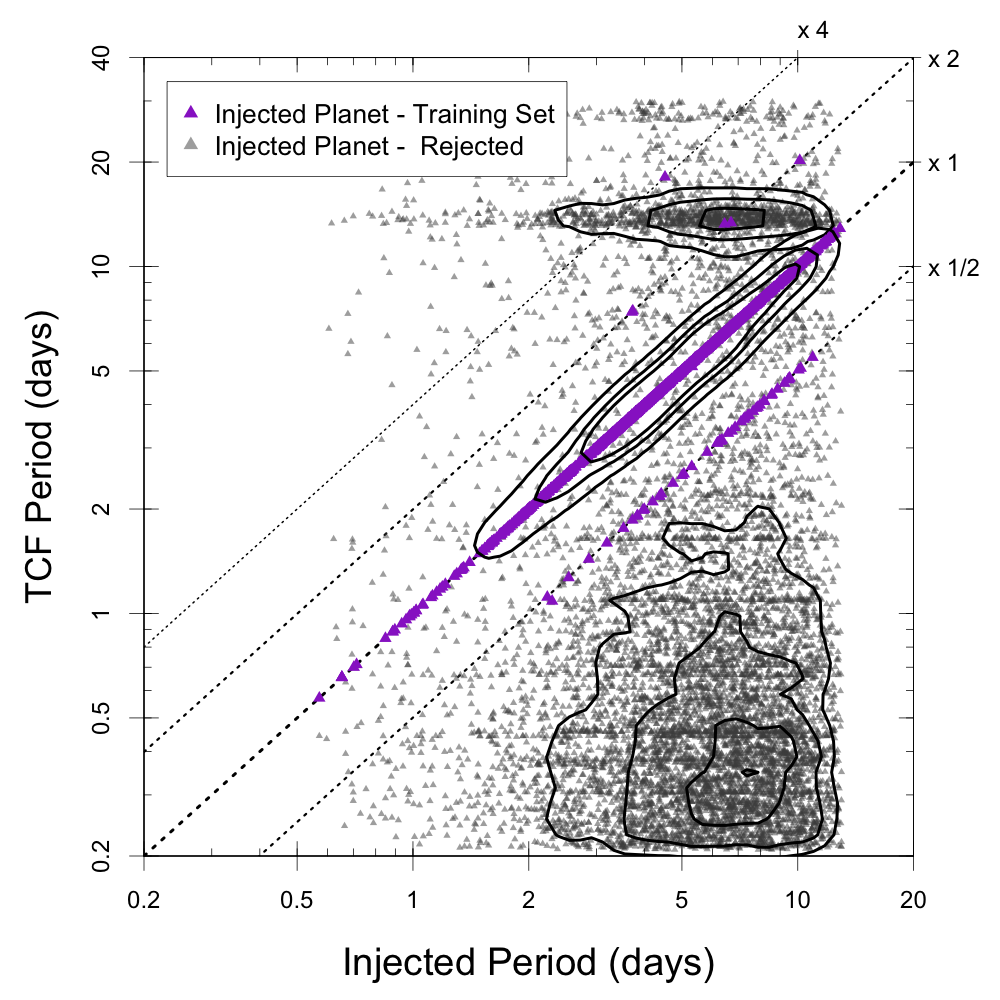}{0.49\textwidth}{(a)}
  \fig{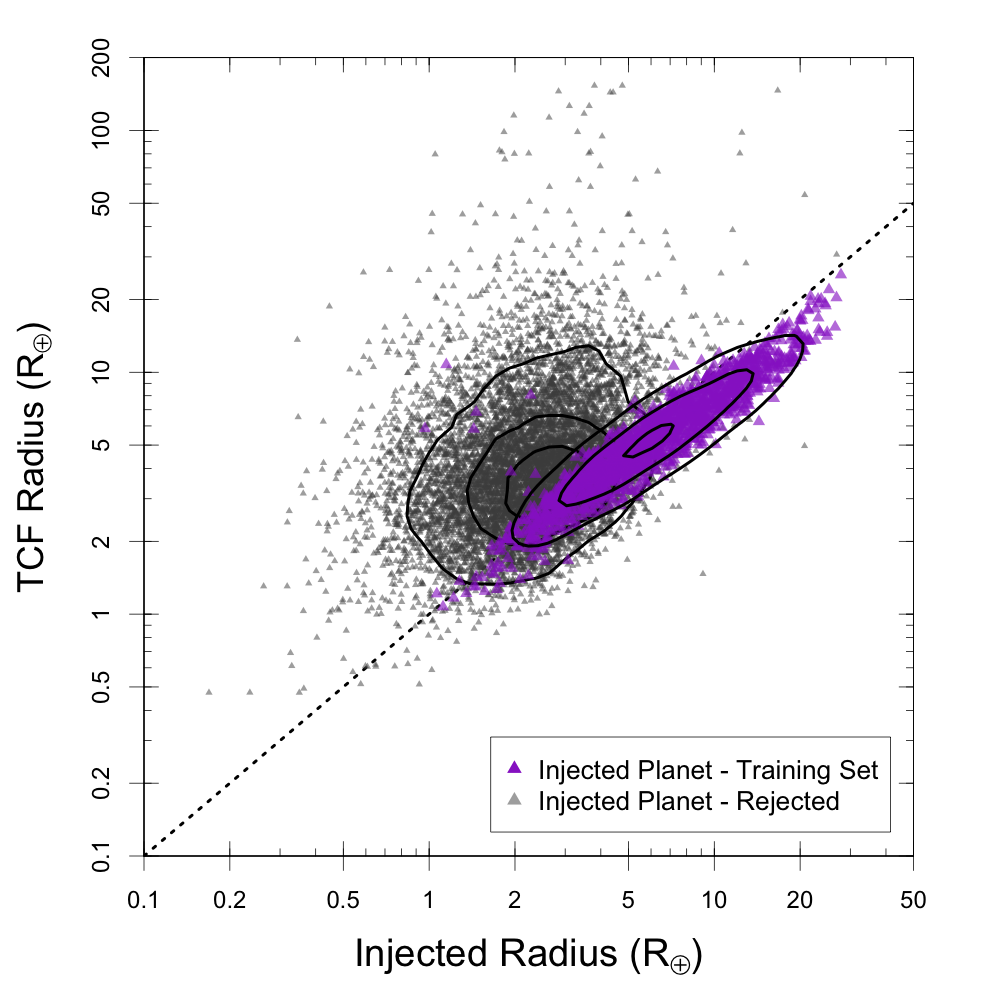}{0.49\textwidth}{(b)}}
  \gridline{\fig{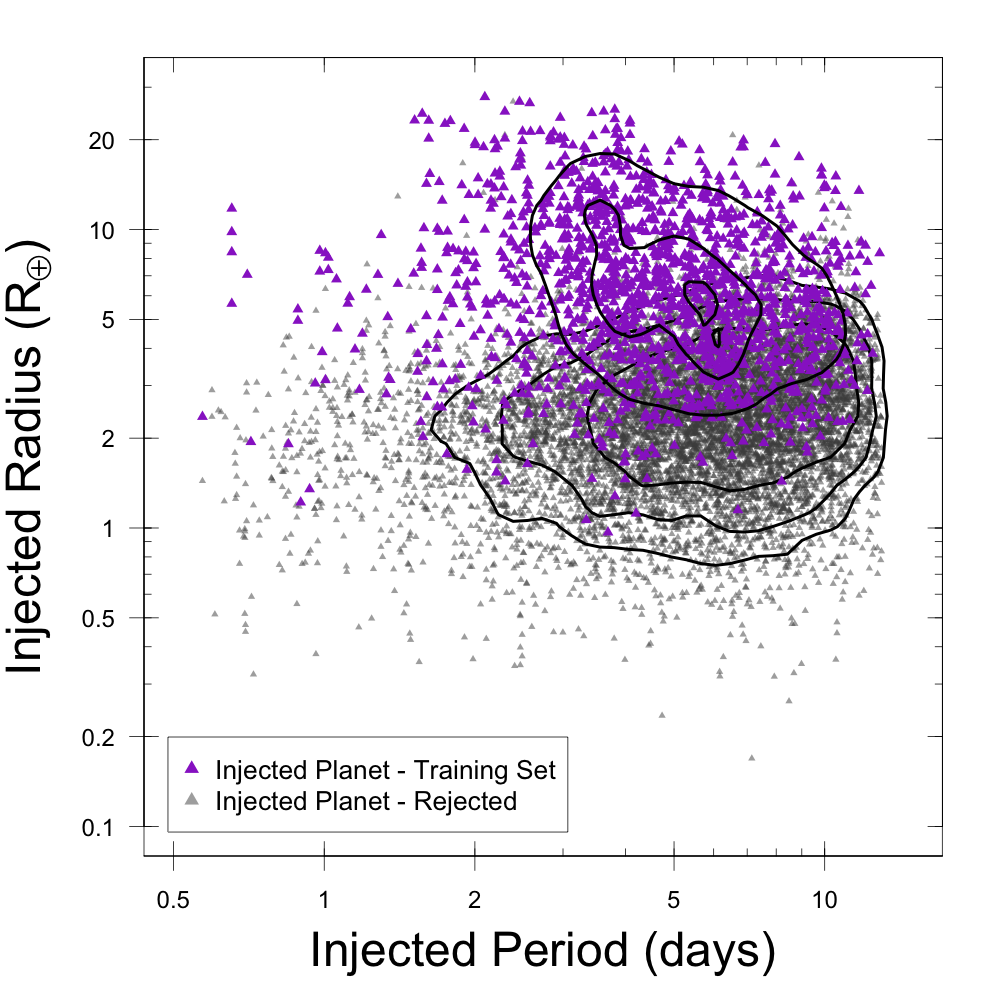}{0.5\textwidth}{(c)}}
  \caption{Comparison of the orbital parameters found from the best TCF peak with the injected orbital parameters for the synthetic injected planets. Recovered injections are shown as purple triangles. (a) Injected and recovered orbital periods. Dotted lines show integer ratios between the injected and TCF periods. (b) Injected and recovered planet radii. (c) Distribution of the injected planets in period-radius space.  \label{fig:inj_param_diff}}
\vspace*{0.5in}
\end{figure}

Panels a and b of Figure \ref{fig:inj_param_diff} compare the injected orbital parameters with the orbital parameters from the TCF analysis for the full set of synthetic planetary injections. The synthetic planetary signals whose best TCF peak orbital period matched the injected orbital period (or an integer ratio) are shown as purple triangles. Spurious periodicities with periods of 13-15 days arise from the 13.7 day orbital period of the \TESS satellite. It is not surprising that TCF would align the two halves of the light curve and find spurious double-spikes associated with ramping problems that escape our data cleaning procedure (\S\ref{sec:outlier}). The pile-up of identified TCF periods between 13 and 15 days and near the extreme limit of the TCF period search of 27 days are both expected and easily removed by the RF classifier and by vetting.

The tendency of TCF to identify much shorter periods than the injected period for injected planetary signals was not expected. This the cloud of gray points in the lower-right of Figure \ref{fig:inj_param_diff}. Just over half of the 10,850 synthetic injected planetary transit signals were assigned periods $<1$ day by the TCF algorithm, while only 3\% of the injected FP signals had a spurious period found by TCF shorter than one day. This difference can be attributed to the injected transit depth: injected FP signals radii exceeded 38 $R_{\oplus}$ while injected planet radii were less than 28 $R_{\oplus}$ and often much smaller. This suggests that the DTARPS-S procedure will have difficulty identifying shallow transit signals, especially at shorter periods. 

This issue is further elucidated in Figure \ref{fig:inj_param_diff}b that compares the injected radius and the radius from the best TCF peak. The injected planets included in the positive training set shown in purple are concentrated along a locus that falls just below the desired 1:1 line. When TCF identifies the correct period for a planetary signal, it gives a slightly smaller radius than the radius of the underlying signal; the effect is more pronounced for radii $\gtrsim 10$~$R_{\oplus}$. This bias has multiple causes. First, the ARIMA model incorporates some of the transit signal with the stellar variability \citep{Caceres19b}. This effect can also occur with other detrending statistical procedures such as wavelet analysis and Gaussian Processes regression. Second, for longer periods, the ARIMA residuals have only a few points in the ingress and egress spikes and the TCF matched filter has difficulty correctly fitting the extreme values of the spike shape. This partially accounts for the paucity of recovered large $10-20$~$R_{\oplus}$ planets at long periods in Figure \ref{fig:inj_param_diff}c. Third, the ingress and egress will often be split between two \TESS cadence slots so neither capture the full height of the spike in the ARIMA residuals. Stellar limb darkening may further slow ingress and egress, weakening the spike. We mitigate this radius bias in Paper II, both by fitting likelihood-based astrophysical models to the stronger transits and by visual correction of transit depths for weaker transits.

The gray points in Figure \ref{fig:inj_param_diff}b reveal another bias: when TCF fails to identify the correct orbital period for a planet, it also tends to overestimate the planet radius. This overestimation of the TCF radii is strongest for smaller planets. For injected planets rejected from the positive training set, the TCF radius is more than twice the true radius for 43\% of injections with R $<$ 4 $R_{\oplus}$ compared to 14\% of injections with R $>$ 4 $R_{\oplus}$. 

The distribution of the injected planets in Figure \ref{fig:inj_param_diff}c shows that the ability of TCF to recover the injected orbital parameters does not depend stringently on the injected period, but is moderately conditional on the injected planet radius. Both distributions cover the same range of injected periods from 0.5 to 13.5 days. The distribution of recovered injected planets is centered at a higher injected radius than the distribution of rejected planetary injections indicating that ARPS has an easier time identifying planetary signals with R $\gtrsim~5~R_{\oplus}$.  This effect will be quantified in Paper II with  completeness curves.

\subsection{Injected False Positives}

\begin{figure}[tbh!]
    \centering
  \includegraphics[width=0.49\textwidth]{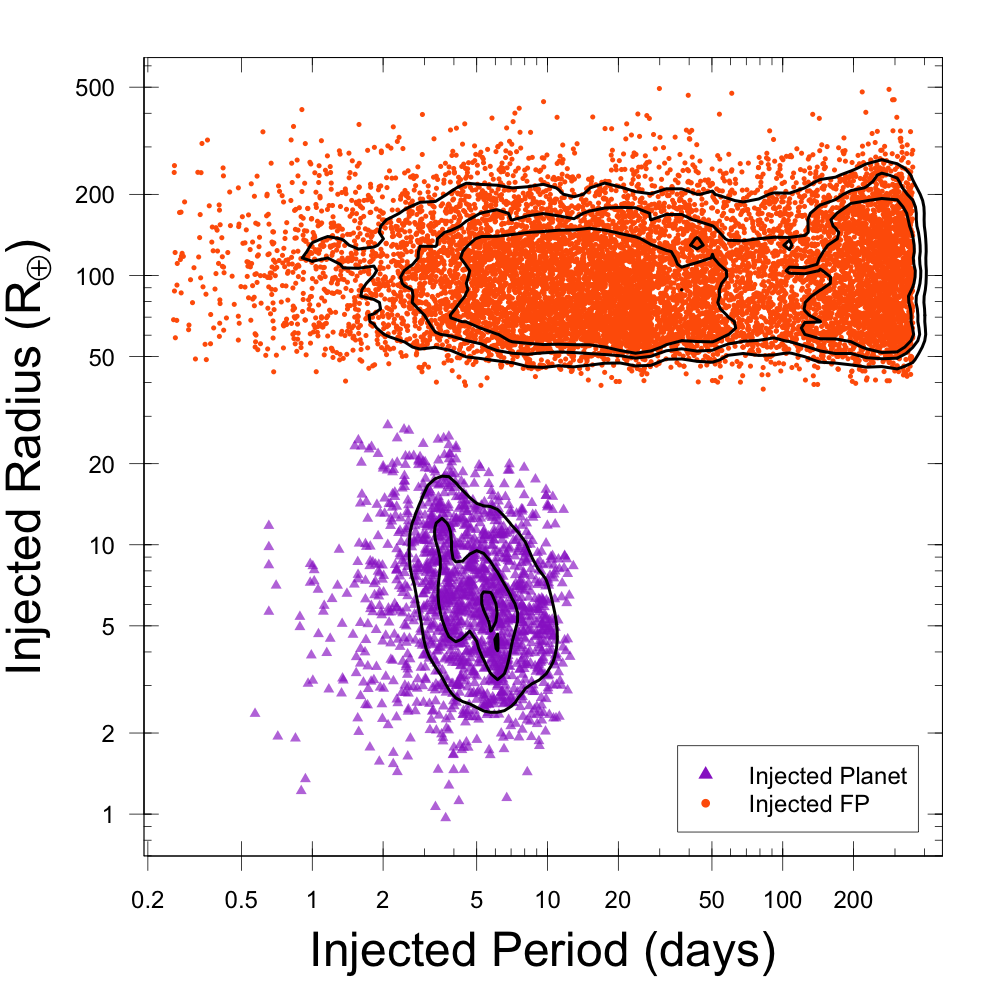}
  \includegraphics[width=0.49\textwidth]{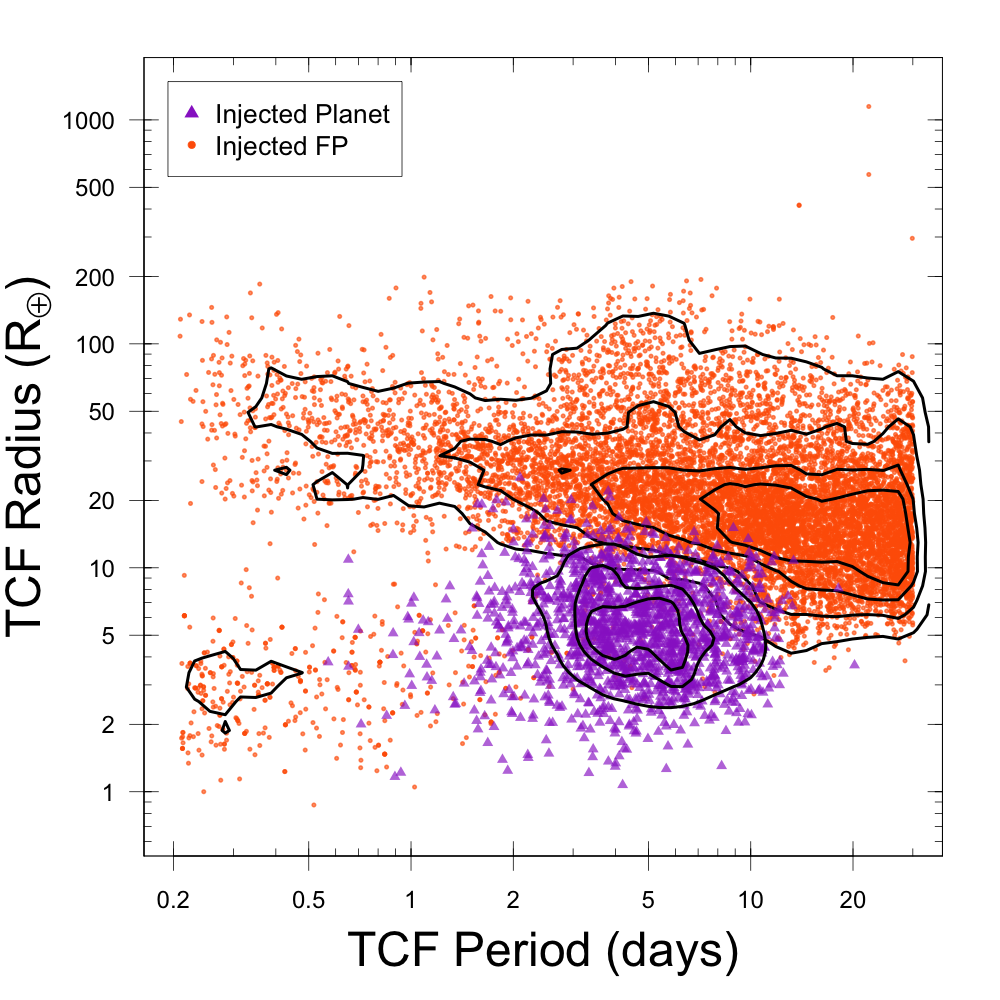}
  \caption{Comparison of the synthetic orbital parameters that were injected into random light curves for the positive (purple triangles) and negative (orange points) training sets with the orbital parameters from the best TCF transit model. Injected FPs were used supplement the negative training set and only represent about half of the negative training set (see \S \ref{sec:neg_train}). The top panel shows the distribution of injected orbital parameters and the bottom panel shows the distribution of the best orbital parameters from TCF analysis. Smooth contours are plotted on both panels to aid the eye in seeing the underlying distribution. \label{fig:inj_orbital_params}}
\end{figure}

Figure \ref{fig:inj_orbital_params} shows the injected and TCF recovered radius-period distribution for both injected planets and injected False Positive (FP) signals. The injected FPs had periods from 0.2 to 357 days, the length of the longest extracted \TESS FFI light curve in the DIAmante data set. Since DTARPS-S does not seek to correctly characterize the orbital parameters of FPs, we chose to include in the negative training set the full range of FP signals that may be present in the DIAmante data set.
\

The tendency of DTARPS-S to underestimate the depth of the transit signal for large injected radii signals (Figure \ref{fig:inj_orbital_params}b) causes DTARPS-S to greatly reduce the radii of injected FPs; compare the orange points in the two panels of Figure \ref{fig:inj_orbital_params}. This undoubtedly is due to incorporation of deep EB transits into the ARIMA model.  The center of the distribution of injected FP radii is reduced from $\sim$100 $R_{\oplus}$ to $\sim$20 $R_{\oplus}$. The FP and planet injection distributions overlap in the TCF parameters while they are fully separated in the injection parameters. The group of injected FPs with radii $\lesssim$ 10 $R_{\oplus}$ and periods $<$ 1 day are not injected EB signals, but the injected sinusoidal signals simulating rotationally modulated variable stars.

The predilection for TCF to report ultra short periods when it fails to find the correct period, combined with the much smaller TCF radius value, created a sample of $\sim 400$ injected FP signals with TCF periods $<$ 2 days and TCF radii $<$ 10 $R_{\oplus}$. Although the injected FP signals include short period sinusoidal variable signals (\S \ref{sec:neg_train}), none of them have a TCF radius smaller than 15 $R_{\oplus}$.  Only injected FP signals with spuriously identified TCF periods $<$ 2 days have TCF radii $<  2$ $R_{\oplus}$.  There are only $\sim 100$ injected planetary signals in the positive training set in that region. The erroneously characterized FPs completely dominate the shortest TCF periods with $0.2-1$~day. This means, even though TCF tends to identify spurious short periods when it cannot correctly identify a transit signal (Figure \ref{fig:inj_param_diff}), the RF classifier will be unlikely to identify them as potential DTARPS-S Candidates because that region of TCF period-radius space is dominated by injections from the negative training set. The classifier is also less likely to recover true planetary signals with periods less than 2 days. The RF classifier is more complicated than drawing boxes in the regions of TCF period-radius space, as it includes influence of over 30 other variables. But since the TCF period and radius are the 3rd and 4th most important features in the RF classifier (Figure \ref{fig:rf_feature}), this mischaracterization of injected FP signals will affect the final classification. 

In addition to the population of injected FPs with short TCF periods and TCF radii consistent with planetary objects, there are a large number of injected FPs with TCF radii consistent with a Jovian planet ($\sim 10-20$ $R_{\oplus}$), particularly at longer periods $\gtrsim 10$~days. The presence of this population of FPs in the negative training sample may cause DTARPS-S to be less sensitive to Jovian planets and long period planets. The large number of injected FPs with TCF radii consistent with Jovian planets irrespective of orbital period may make it more difficult to find Jovian planets despite TCF's ability to better recover the correct orbital period for larger planetary signals.

\subsection{Conclusions from Injected Populations} 

Only 12\% of the injected planets are reliably recovered by the DTARPS-S procedure. This low fraction is not a surprise, as the distribution of injected radii is drawn from the more sensitive \Kepler mission dominated by planets too small for \TESS detection (Figure \ref{fig:kep_v_ans}, top-right panel). The rate of capture of planetary injections will be examined in our completeness analysis below (\S\ref{sec:complete}). 

For the correctly identified injected planets, the TCF periods are mostly accurately recovered from $0.5-13$ days. For a small fraction, the 1/2-period harmonic is preferred by the TCF. We use harmonic, here and later in the dissertation, with the definition used in time series analysis: a frequency (period) is an integer ratio of another frequency (period) in the time series (light curve). TCF-derived radii, on the other hand, are underestimated for correctly identified injected planets, particularly for large-radius injections. This bias is understood as a combination of ARIMA and TCF behaviors. We correct this bias for astronomically interesting candidates by manual intervention or astrophysical modeling in Paper II.

The DTARPS-S analysis also recovers a small fraction of injected false positive signals as potential planetary candidates. The response to astronomical False Positives is complicated and is discussed in \S\ref{sec:recall_FP}. False Positive (and False Alarm) contamination motivates the strictness of our vetting procedures in Paper II. In that study, we remove nearly 90\% of the objects in DTARPS-S Analysis List when creating the DTARPS-S Planet Candidate catalog. This reduces the completeness (in statistical parlance, the `recall') of the listing in this study but greatly improves the reliability (`precision') of the DTARPS-S planet candidates.

\section{Random Forest  Classifier Performance for Planet Injections \label{sec:complete}}

The recall rate of the injected planetary signals across the range of the injected period and radii can quantitatively measure the ability of the RF classifier to recover planets in the DIAmante data set. It is important to understand how the classifier performs across the planetary radius-period distribution to evaluate the completeness of our intermediate DTARPS-S Analysis List with 7,377 objects and the smaller DTARPS-S Planet Candidate catalog produced in Paper II. The completeness (or recall rate) of the RF classifier for different bins in planetary period-radius space is measured using the full set of synthetic planetary injections based on the \Kepler planet sample (\S \ref{sec:kep_inj}). The analysis is based on 7,751 of the 10,850 synthetic planetary injections that were processed by the RF classifier; the remaining objects were omitted due to missing features. 

Analysis of recall rates for synthetic injections gives a more reliable view of the completeness of the DTARPS-S analysis for a physical population of planets than the comparison with other surveys (\S\ref{sec:other}). The latter are subject to the different, and often poorly known, limitations and biases of \TESS light curve analysis and telescopic follow-up by many research groups. 

Comparison with injections by \citet[][and earlier studies]{Christiansen20} has proved invaluable for deriving true planetary occurrence rates from the Kepler 4-year survey. {\it However, we emphasize that planetary occurrence rates cannot be reliably estimated from the analysis at this stage because the DTARPS-S Analysis List is dominated by non-planetary signals. Occurrence rates will be estimated in Paper III after vetting has greatly improved the `sensitivity' of the sample, though with reduced `recall' rates.}

\subsection{Survey Completeness} 

\begin{figure}[tb]
  \centering
  \includegraphics[width=\textwidth]{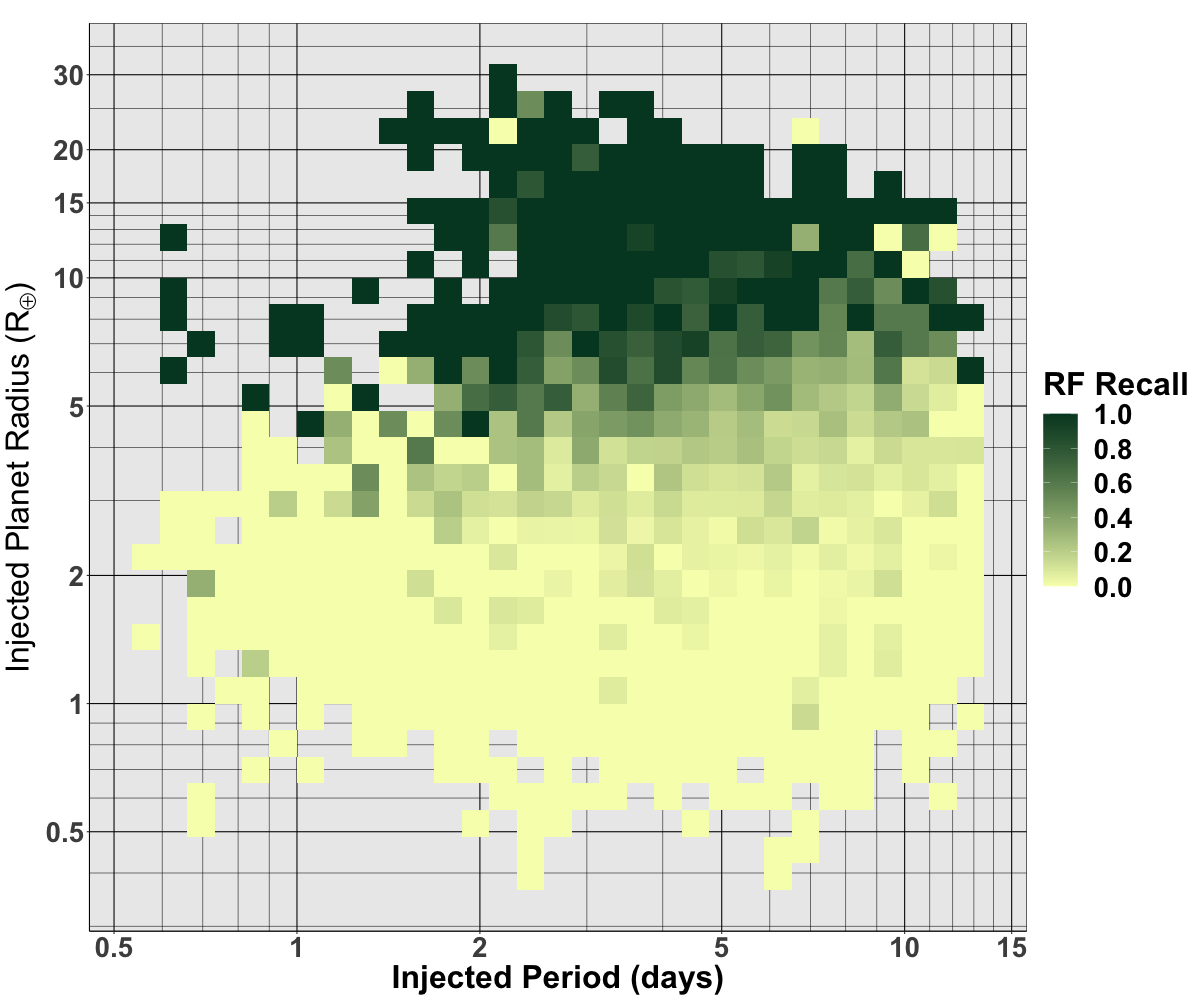}
  \caption{Heat map of recall rates of the Random Forest classifier for the synthetic planetary injections as a function of injected period and radius bins. }
  \label{fig:inj_recall_inj}
\end{figure}

Figure \ref{fig:inj_recall_inj} shows the completeness for the 7,751 synthetic planet injections. The completeness for each period-radius bin is the number of injected planets in the bin with a RF prediction value greater than the $P_{RF} = 0.300$ threshold divided by the total number of injected planets in the bin. The bins are distributed evenly in log-space for the injected planetary radius and the injected orbital period. We chose to analyze the completeness of the RF in the injected orbital parameter space and not in the TCF identified period-radius space because the planetary injections were created to have a realistic distribution in the injected period-radius space which is not preserved in the TCF period-radius space (Figure \ref{fig:inj_param_diff}).

The distribution of the underlying points in Figure \ref{fig:inj_recall_inj} roughly follows the distribution of points in Figure \ref{fig:inj_param_diff}c; Figure \ref{fig:inj_param_diff}c contains all of the injected planetary signals, whereas Figure \ref{fig:inj_recall_inj} only uses injected planetary signals that were classified by the RF classifier.  The completeness in Figure \ref{fig:inj_recall_inj} appears to correspond with the recovered planetary injection signals in Figure \ref{fig:inj_param_diff}c, but only because we trained a highly effective classifier with an OOB recall rate of 92.5\% for the training set of recovered planetary injections.  Not all of the recovered planetary injections in Figure \ref{fig:inj_orbital_params}c had an OOB RF prediction value above the threshold and not all of the rejected planetary injections were rejected by the RF classifier.  There is a subtle, but important distinction that Figure \ref{fig:inj_orbital_params}c shows the distribution of the recovered injected planetary signals used in the positive training set of the classifier and Figure \ref{fig:inj_recall_inj} shows the completeness of the DTARPS-S methodology on the whole set of injected planetary signals.  

Figure \ref{fig:inj_recall_inj} shows poor completeness ($< 10\%$) for radii less than 2 $R_{\oplus}$ across all periods.  There is also poor completeness for periods less than 1 day and radii $<5$~R$_\oplus$. In these regions, the classifier fails to recover enough planets to make meaningful statements about the exoplanet population. The classifier has low completeness (10\% - 25\%) for planets with radii between 2 and 4 $R_{\oplus}$, and high completeness (70\% - 100\%) for periods between 0.6 and 13 days for planetary radii between 8 and 30 $R_{\oplus}$. In the latter region, the classifier essentially captures the full exoplanet population. At a given planet radius, the DTARPS-S classifier achieves somewhat higher recall rates for periods around $2-4$~day than around $7-13$~day, producing a tilt in the heat map.  The outlying bins of the distribution in Figure \ref{fig:inj_recall_inj} are sparsely populated bins where recall rates are uncertain. For example, the pale yellow bins with radii larger than 10 days do not represent an inability of DTARPS-S to recover planets, as only a single injected object is present in those bins.

\subsection{Weakest Signal Recovery \label{sec:sig_recall}} 

\begin{figure}[tb]
  \centering
  \includegraphics[width=\textwidth]{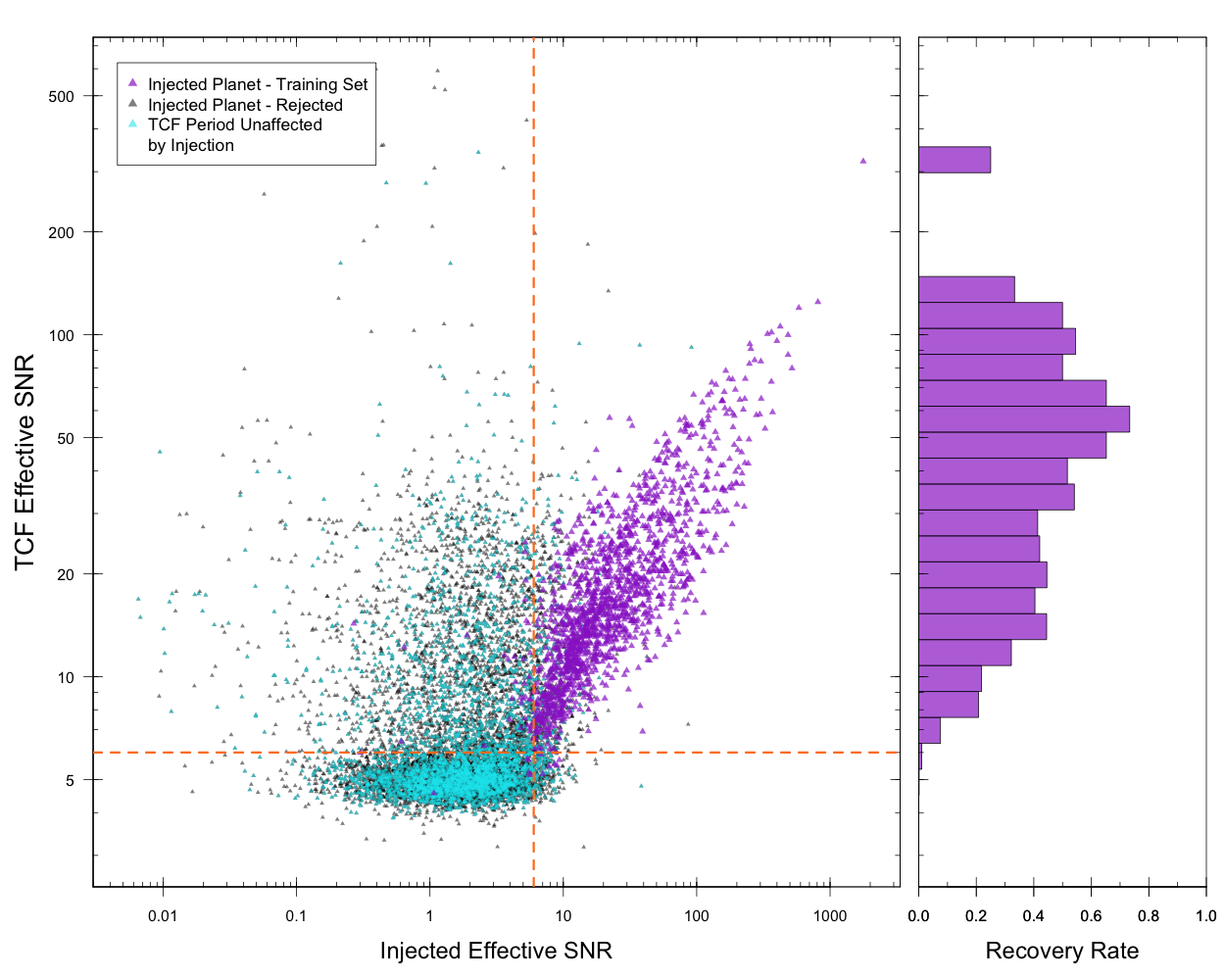}
  \caption{Effective SNR of the injected planetary signals. (Left) Comparison of the effective SNR (based on IQR rather than standard deviation) of the injected planetary transit to the effective SNR of the best transit from TCF. The injected planetary signals that were used in the training set are plotted as purple triangles while injected signals whose period was not recovered by the TCF are gray triangles. Light curves whose TCF peak periods were unaffected by the injected planetary signal are plotted in turquoise. The dashed orange lines show approximate lower bounds of injected planets used in the training set. (Right) Fraction of injected planetary signals that were recovered given the spread of effective SNR of the best transit from TCF. }
  \label{fig:eff_SNR_comp}
\end{figure}

While Figure \ref{fig:inj_recall_inj} shows the gradual deterioration of DTARPS-S recovery of smaller planets, the sensitivity of a transit survey can also be examined in terms of signal strength. A common measure of the ability to recover a synthetically injected transit signal is the {\it effective signal-to-noise ratio (SNR$^{eff}$)} of a transit signal \citep{Kovacs02, Howard12, Christiansen13, Christiansen16}. The SNR$^{eff}$ of a transit signal is the depth of the transit divided by the standard deviation of the average of all measurements in the transit. 

Using features already calculated for use in the RF classifier, we calculate an SNR$^{eff}$ for the injected transit signal and the periodic signal associated with the best peak in the TCF periodogram based on equation 1 in \citet{Howard12}.  Here, the effective SNR of the transit is
\begin{equation}
    \textrm{SNR}^{eff} = \frac{\delta}{\textrm{IQR}} \sqrt{\frac{n_{pts} \, T_{dur}}{P}} \label{eq:eff_SNR}
\end{equation}
where $\delta$ is the depth of transit, $\textrm{IQR}$ is the InterQuartile Range of the light curve from which the transit depth is being measured, $n_{pts}$ is the number of points in the light curve, and $T_{dur}/P$ is the fractional duty cycle of the transit. We substitute the IQR of the original light curve (for the injected signal) and the IQR of the ARIMA residuals (for the best TCF transit) instead of standard deviations. Both the IQR and the standard deviation of a distribution are measures of the spread of the distribution, but the IQR is more robust against non-Gaussianity.

Figure \ref{fig:eff_SNR_comp} compares the SNR$^{eff}$ for the injected planet signal and the SNR$^{eff}$ for the strongest periodic signal in the ARIMA residuals from the TCF periodogram. This is shown for two subsamples: the injections that are successfully recovered by DTARPS-S processing (purple), injections that were rejected (gray). The orange dashed lines in Figure \ref{fig:eff_SNR_comp} show the approximate lower boundaries for the SNR$^{eff}$ for the recovered injected planet signals used in the positive training set. The boundaries were set at $SNR^{eff}=6$ for both the SNR$^{eff}$ of planetary signals injected into the DIAmante extracted light curve, and for the SNR$^{eff}$ of the best TCF periods obtained from ARIMA residuals. 

About 58\% of the injected planetary signals that were rejected from the positive training had a TCF SNR$^{eff}$ less than 6 and 93\% of the rejected injected planetary signals lie to the left or below the boundaries in Figure \ref{fig:eff_SNR_comp}. The recovery rate of the injected planetary signals with effective SNRs above the $SNR^{eff}=6$ boundaries is 71\%, much larger than the overall recovery rate of the injected planetary signals (\S \ref{sec:kep_inj}). The histogram on the right side of Figure \ref{fig:eff_SNR_comp} shows the fraction of injected planetary signals used in the training set (the TCF periodogram peak recovered the injected transit period) for the SNR$^{eff}$ of the TCF periodic signal in even logarithmically-spaced bins.  The decline in the recovery rate of the injected planetary signals above a TCF SNR$^{eff}$ of $\sim60$ is due to the rejected planetary injections whose injected transit SNR$^{eff}$ was overwhelmed by noise or an inherent periodic signals in the light curve.

The low injected SNR$^{eff}$ values often arises from very shallow or short period injections. Here the paucity of \TESS observations $-$ only $\sim 1,000$ points in single sector observation $-$ hinders detection of small planets with small periods. Similar difficulties probably affect other \TESS transit analysis systems: only 6\% of TOI candidates have periods $<$1 day and only 5\% have radii $<$2 $R_{\oplus}$.
 
We can compare this situation to the earlier application of ARPS methodology to the \Kepler 4-year mission where \citet{Caceres19a} identified 97 new mostly Earth- and Mars-size candidates. \citet{Gondhalekar23} finds that application of ARIMA detrending of light curves and TCF periodograms is usually more sensitive to small planets than traditional detrending methods and BLS periodograms. However the DTARPS-S completeness is low for small planets with shallow transits (Figure \ref{fig:inj_recall_inj} and the histogram in Figure \ref{fig:eff_SNR_comp}). This apparent discrepancy stems from the enormous difference in durations of \Kepler and \TESS FFI lightcurves. While both have time steps $\sim$ 30 minutes, the longer baseline of the \Kepler four-year light curves meant that the \Kepler light curves had $\sim$ 77,000 flux measurements while the \TESS FFIs have on average $\sim$ 1,000 flux measurements.  Even in the continuous viewing zone for the Year 1 \TESS data, the stitched together light curve has a duration of only one year as opposed to four years from the \Kepler mission. The increased number of points in the \Kepler light curves  increases the effective signal-to-noise ratio (SNR$^{eff}$) of the transit (equation \ref{eq:eff_SNR}).

\begin{figure}[bh!]
     \includegraphics[width=\textwidth]{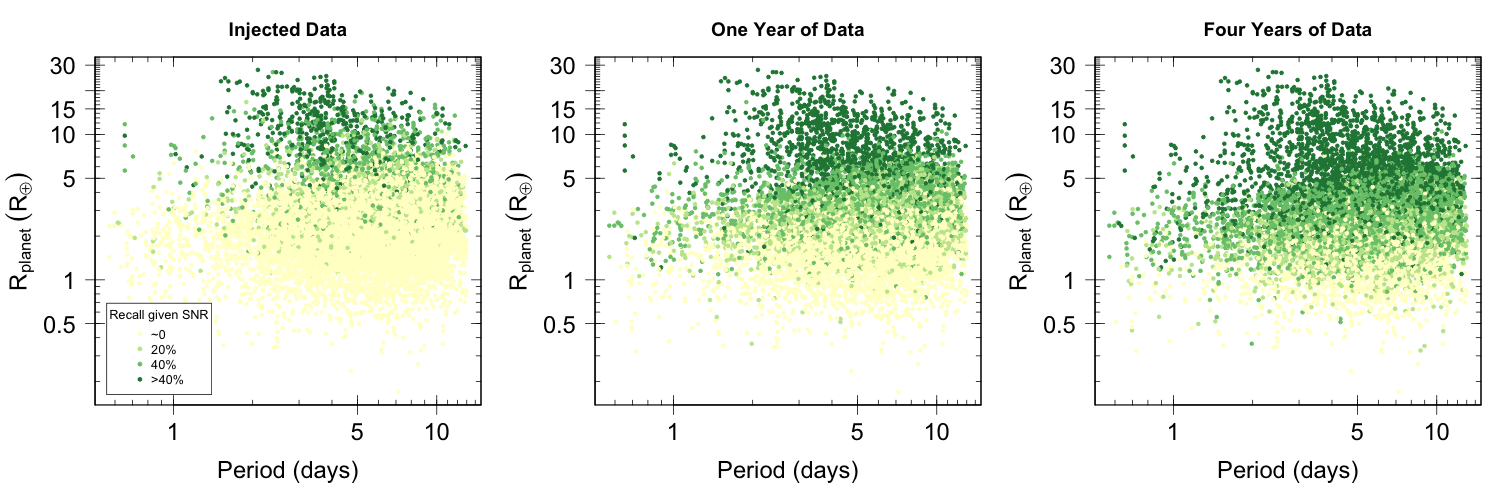}
    \caption{Radius-Period plots of the injected planets (\S \ref{sec:kep_inj}) colored by the recall rate for the injected planet given only the injected transit TCF signal SNR$^{eff}$ for \TESS with three different hypothetical observational durations: one month, one year, and four years. Green indicates a high recall rate for the injected planet and yellow indicates a low recall rate for the injected planet.}
    \label{fig:proj_snr}
\end{figure}

The three panels of Figure \ref{fig:proj_snr} show a simulation of the recall rate of the injected planet signal given the SNR$^{eff}$ from the histogram in Figure \ref{fig:eff_SNR_comp} for the planet injections for \TESS-like surveys with three hypothetical durations: one month (left), one year (middle), and four years (right). It shows that \TESS could detect substantial fractions of Earth-size planets were four continuous years of observation available.  This sensitivity could be approached by the \TESS extended missions near the ecliptic poles. 

\section{Relationship to Other Surveys} \label{sec:other}

In addition to considering recall rates for the synthetic planetary injection sample, we can compare the RF classifier performance to previous studies. The main difficulty in making inferences is that other surveys may suffer incompleteness and erroneous (False Alarm or False Positive) exoplanet identifications. The `Confirmed Planets' from the NASA Exoplanet Archive \citep{NEA-CP} (NEA, accessed March 15, 2022) is likely to have the fewest errors, though its listings are culled from a heterogeneous collection of studies. The \TESS Objects of Interest (TOI) list, the community TOI (cTOI) list \citep{ExoFop-TOI} (both accessed March 15, 2022), and the M20 DIAmante analysis candidates are specifically derived from \TESS data, but their reliability is unknown. The NEA and TOI efforts also list `False Positives' that are useful for comparison with DTARPS-S results. 

We matched the DIAmante data set and the potential candidate transits in the DTARPS-S Analysis List with lists from 15 previous studies on exoplanet surveys or False Positives such as low mass eclipsing binaries, flare stars, and stellar rotation. The exoplanet survey studies utilized here are \citet{Mayo18}, \citet{Dressing19}, \citet{Feinstein19}, \citet{Kostov19}, \citet{Kruse19}, \citet{Yu19}, \citet[][M20]{Montalto20}, \citet{Dong21}, \citet{Eisner21}, and \citet{Olmschenk21}. The False Positive studies are \citet{Affer12}, \citet{Collins18}, \citet{Schanche19}, \citet{vonBoetticher19}, and \citet{Tu20}. Appendix \ref{app:surveys} gives brief description of each of the external surveys and their corresponding entries in the DTARPS-S Analysis List. Where TIC numbers were not available for matching DIAmante light curves with reported objects in these external studies, we used the best match between the right ascension and declination coordinates of the objects with a search radius of 5$\arcsec$. The periods reported in the external surveys (when available) are compared with the period from the best TCF peak.  We consider the period matched when the TCF peak period is within a 1\% fractional difference of the reported period (or a harmonic of the reported period).

\subsection{M20 DIAmante Candidates \label{sec:recall_DIA}}

\begin{figure}[tb!]
   \centering
   \includegraphics[width=0.75\textwidth]{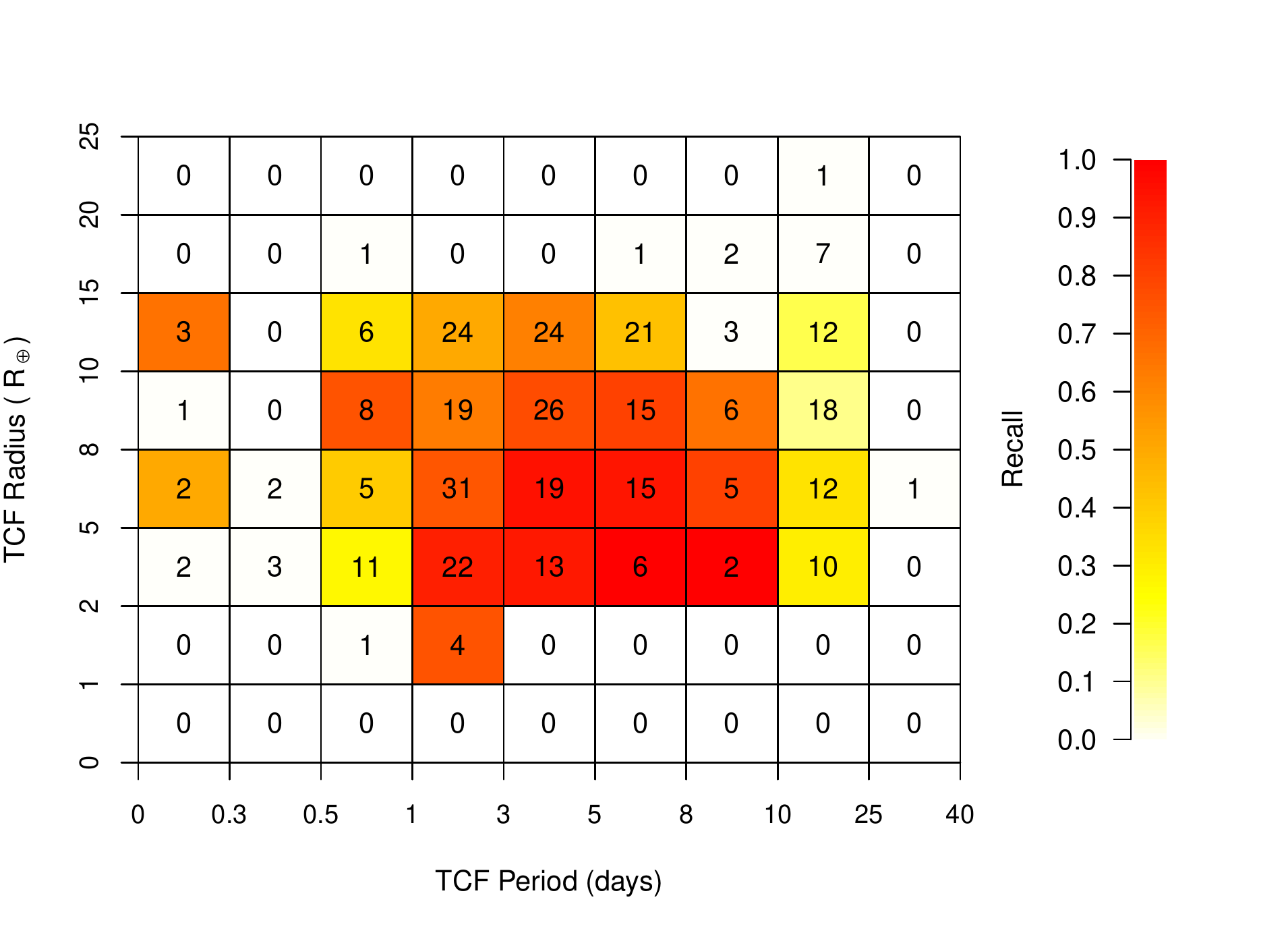}\\
   \includegraphics[width=0.75\textwidth]{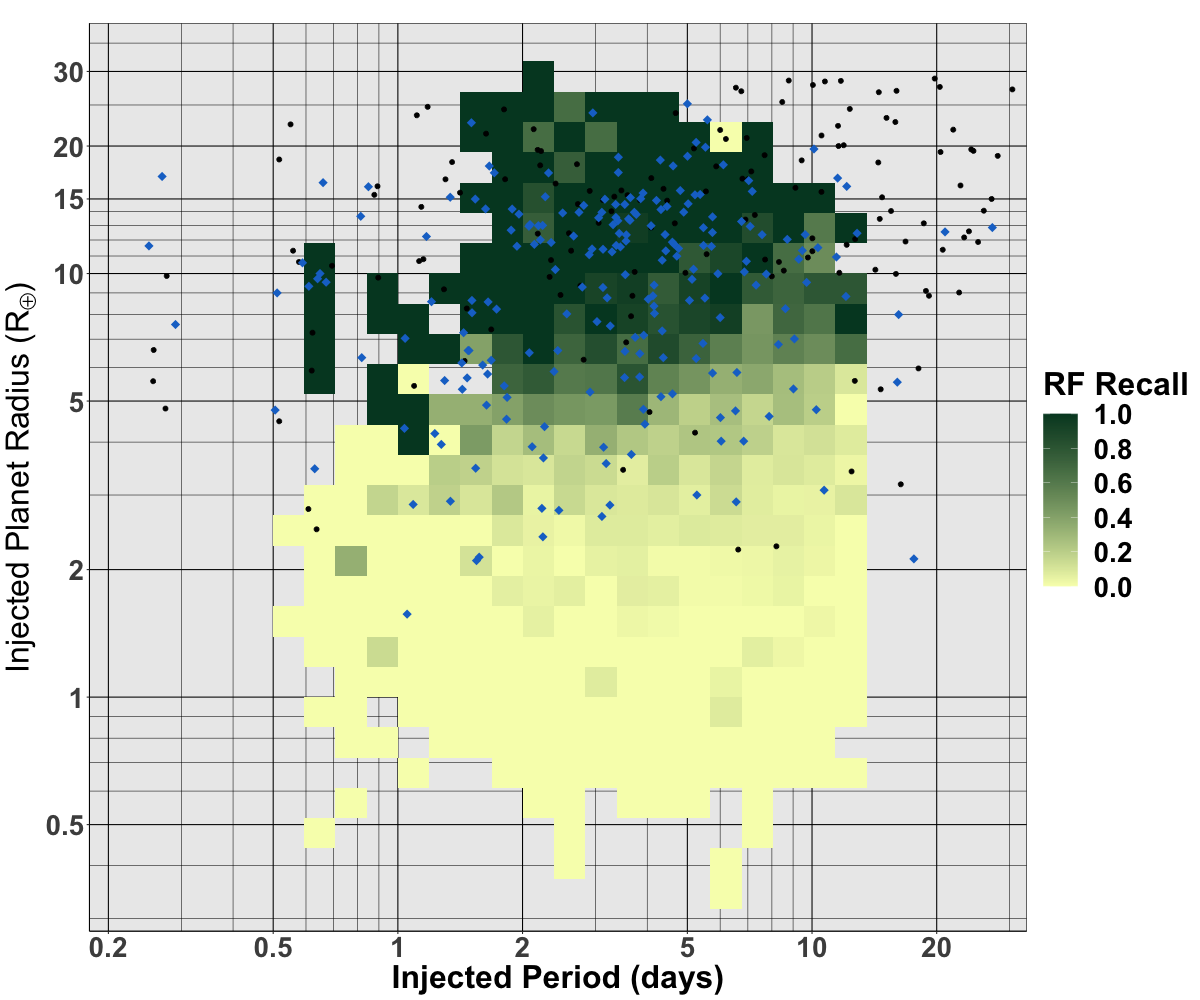}
   \caption{Recall rates by the DTARPS-S RF classifier for 364 M20 DIAmante  candidates. (Top) Heat map as a function of planet radius and orbital period found with TCF. The colors give the recall rate and the numbers are the number of M20 candidates in each period-radius bin. (Bottom) Candidate planets from M20 superposed on the DTARPS-S heat map for injected planets from Figure \ref{fig:inj_recall_inj}. Blue diamonds are recovered objects and black points are not recovered. }
   \label{fig:recall_DIAmante}
\end{figure}

Of the 394 planet candidates identified by the M20 DIAmante analysis (some of which were later confirmed as planets or found to be False Positives), 364 were processed through the entire DTARPS-S procedure. The best DTARPS-S period matched the M20 reported period for 333 (91\%) of the M20 candidates.  The TCF periodogram and BLS periodogram thus emerge with identical results for  nearly all DIAmante candidate planets. Of the 31 M20 candidates where TCF failed to recover the correct period, 17 had M20 periods greater than 13.5 days and were thus outside the range of our injected planet training set. TCF often identified the 1/2 or 1/3 harmonic period of long-period M20 candidates; it is not clear which period is correct in these cases. 

When the $P_{RF} = 0.300$ threshold to the RF classifier is applied (DTARPS-S Analysis List), 213 of the 364 M20 candidates are captured giving a 59\% recall rate for the M20 study. The main difference between DTARPS-S and DIAmante results is thus attributable to the classification stage.  All of these recovered M20 candidates had a TCF period matching the reported M20 period. 
 
Figure \ref{fig:recall_DIAmante} (top panel) shows the recall rate of the RF classifier for different bins in radius-period space from the best TCF peak. The RF classifier has the strongest recall rates for the candidates whose planetary radius from TCF is between 2 and 10 $R_{\oplus}$ and whose TCF orbital period is between 1 and 10 days. The RF classifier only has a 20\% recall rate for M20 candidates with TCF periods greater than 10 days and a 47\% recall rate for TCF periods less than 1 day. The recall rate for the M20 candidates also falls off at larger TCF planet radius likely due to the many injected False Positive  signals in the negative training set with planetary-consistent radii. 

In the bottom panel of Figure \ref{fig:recall_DIAmante}, the colored points show the recovered M20 candidates and the black points show the unrecovered M20 candidates with the RF classifier. The distribution of M20 candidates closely follow the recall distribution of synthetic planet injections.

The candidates reported in M20 mostly have radii $>$7 $R_{\oplus}$, with moderate coverage around $3-7$ $R_{\oplus}$. Neither the DTARPS-S nor the M20 samples cover well the region with small radii; only 16 candidates have radii $<$3 $R_{\oplus}$. Despite this, DTARPS-S does a good job at recovering the M20 candidates with radii less than 5 $R_{\oplus}$ (77\%). 

But for periods shorter than $0.5$ day, DTARPS-S has only moderate recovery of DIAmante candidates. This is likely due to concentration of injected False Positives with TCF periods in this region and TCF radii consistent with planetary objects (Figure \ref{fig:inj_orbital_params}) that bias the classifier against short period planet transit signals. DTARPS-S also has poor recovery of the M20 candidates with periods $>$10 days and radii $>$10 $R_{\oplus}$ for reasons explained in \S \ref{sec:recovered_prop}. 

\subsection{NASA Exoplanet Archive Confirmed Planets \label{sec:recall_KP}}

\begin{figure}[tb!]
    \centering
   \includegraphics[width=0.75\textwidth]{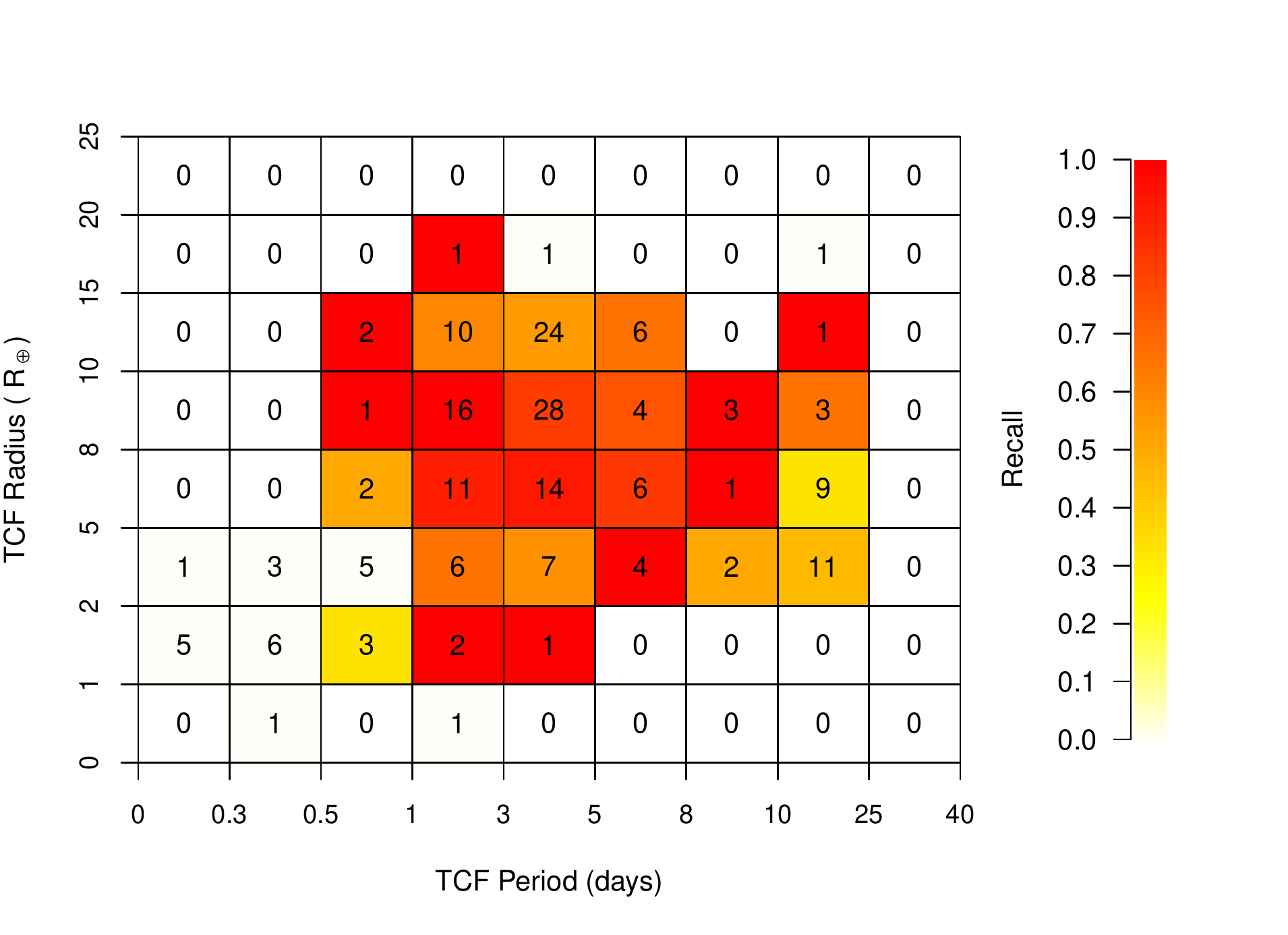}\\
   \includegraphics[width=0.75\textwidth]{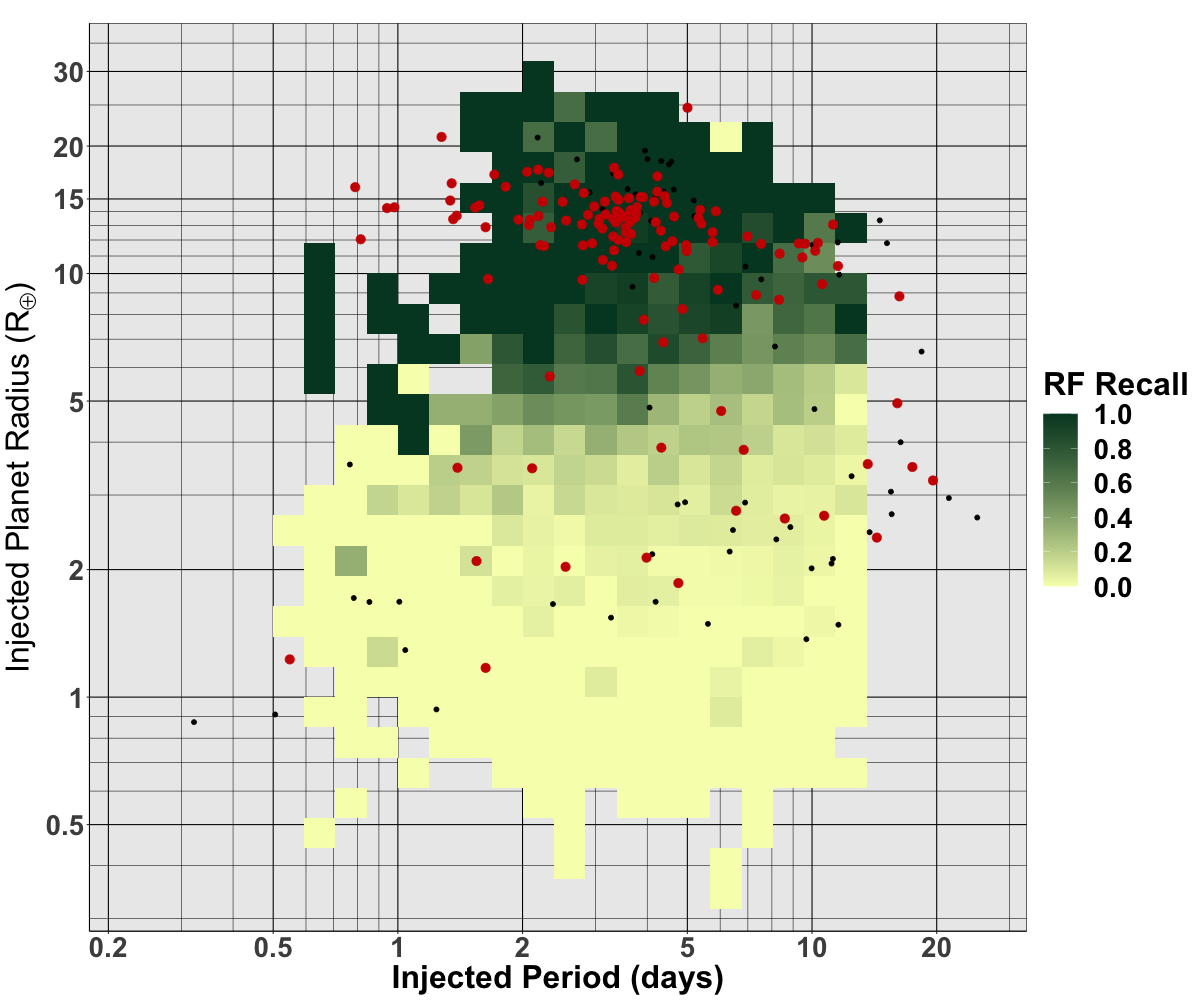}
   \caption{Recall rates by the Random Forest classifier for 202 Confirmed Planets in the NASA Exoplanet Archive \citep{NEA-CP}. (Top) Heat map as a function of planet radius and orbital period found with TCF. (Bottom) Confirmed planets from the NASA Exoplanet Archive superposed on the DTARPS-S heat map for injected planets from Figure \ref{fig:inj_recall_inj}. Red circles are recovered objects and black points are not recovered.}
   \label{fig:recall_KP}
\end{figure}

Of 3,616 `Confirmed Planet' or `Known Planet' systems in the NASA Exoplanet Archive \citep{NEA-CP} or TOI$^1$ lists (accessed March 15, 2022), 202 were in the DIAmante data set classified by our RF classifier. TCF correctly matched the period for 166 (82\%) Confirmed Planets. Of the 36 Confirmed Planets that failed the DTARPS-S selection criteria, 9 have periods $>$13.5 days. But 7 other Confirmed Planets with periods $13.5-30$~days were recovered, and the 1/3 harmonic of one Confirmed Planet with period $>$30 days was found. Even though the injected planetary signals have periods restricted to $< 13.5$ days, DTARPS-S is still capable of matching the periods of long period planets. The $P_{RF} = 0.300$ threshold of the DTARPS-S RF classifier identified 130 of the 202 Confirmed Planets giving a 64\% recall for known exoplanets. All have correctly matched periods. 

Figure \ref{fig:recall_KP} shows DTARPS-S has a strong recall rate ($>50$\%) for periods between 1 and 10 days, up to radii of 10 $R_{\oplus}$, similar to the DIAmante planet candidates. DTARPS-S has poor recall rates for Confirmed Planets in the lower-left bins with periods $<$1 day and radii $<$5 $R_{\oplus}$ likely due to the overpopulation of short TCF periods (\S \ref{sec:recovered_prop}). Of the 22 Confirmed Planets in these bins, only 3 have TCF periods that match the reported period. Recall exceeds 60\% for periods $0.5-1.0$~day and radii $5-15$~R$_{\oplus}$, but often the recovered periods are incorrect.  DTARPS-S thus has poor recovery overall of Confirmed Planets with periods less than 1 day.

\subsection{TESS Objects of Interest}

The recall coverage of the \TESS TOI confirmed planets and planet candidates are included in \S \ref{sec:recall_KP} and \S \ref{sec:recall_PC}, so they are not presented here again. The DTARPS-S sample contains 846 \TESS TOIs of which 140 have reported periods $>13.5$ days. DTARPS-S matches the periods for 566 of 706 of the remaining TOI planets and candidates (80\%).  The recall rate of the TOI confirmed planets and planet candidates is 51\% with 433 of 846 TOIs.

\subsection{Candidate Planets in Other Surveys \label{sec:recall_PC}}

\begin{figure}[tb!]
  \centering
  \includegraphics[width=0.75\textwidth]{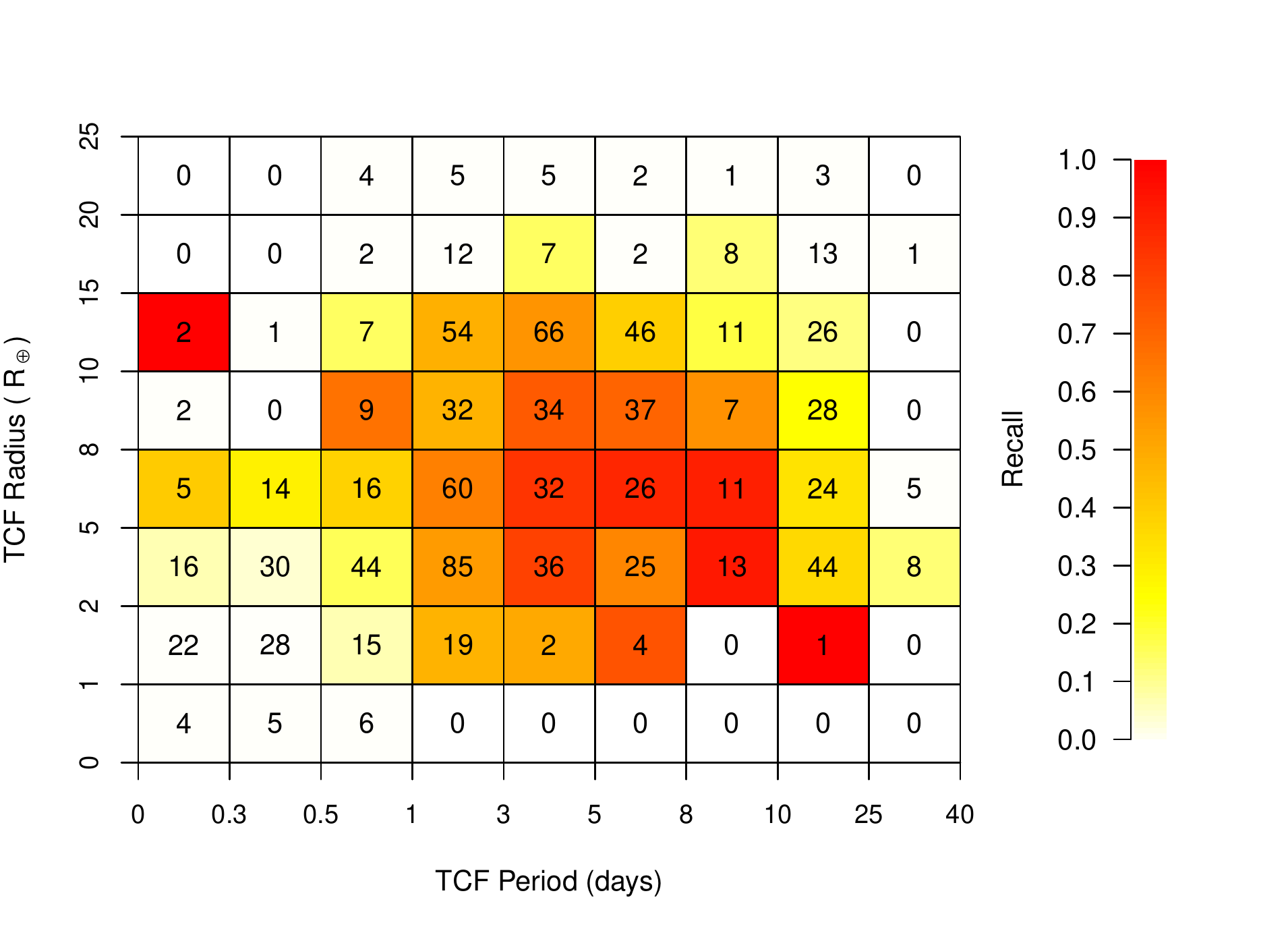}\\
  \includegraphics[width=0.75\textwidth]{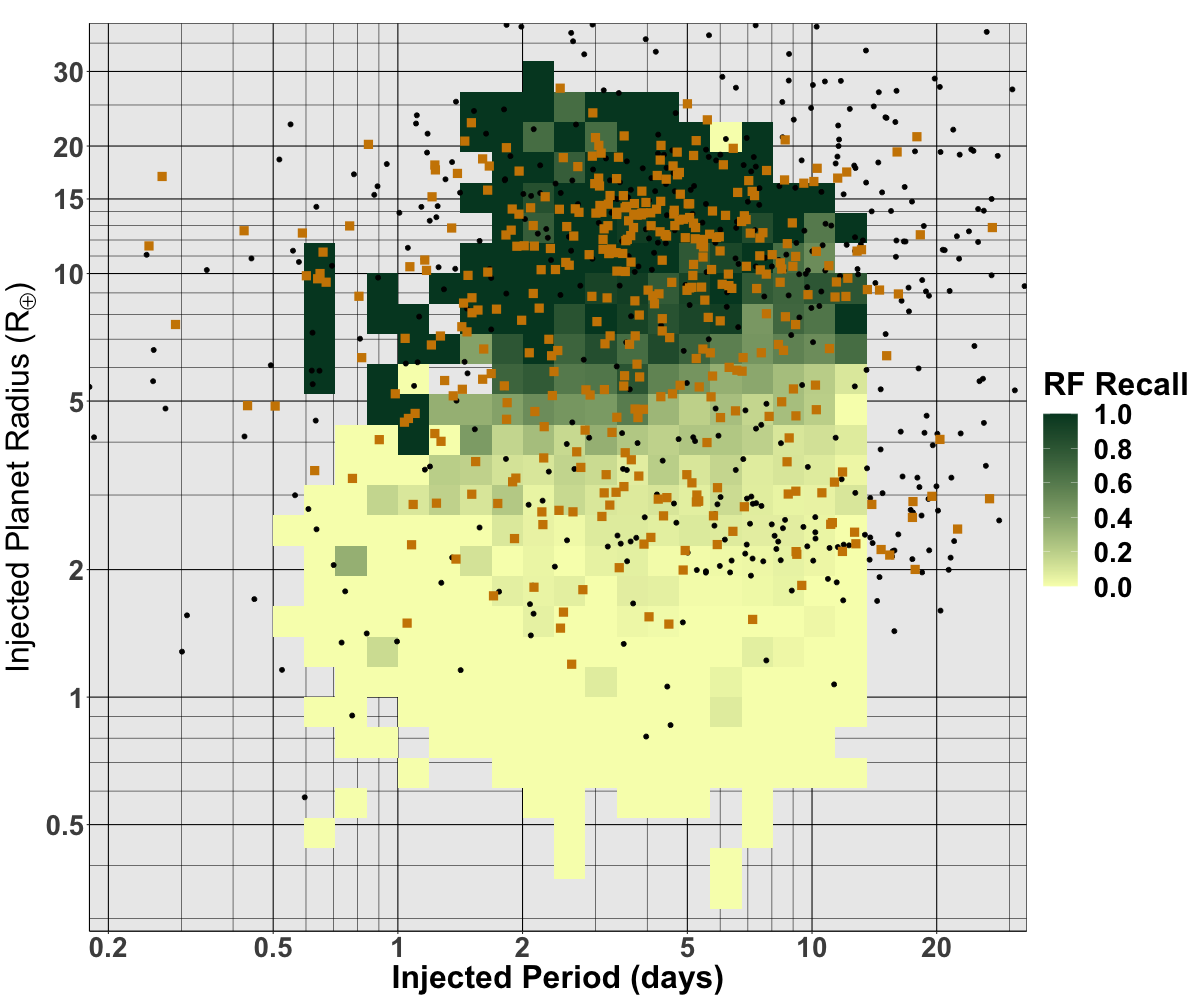}
   \caption{Recall rates by the Random Forest classifier for 1,042 previously identified planet candidates from various surveys. (Top) Heat map as a function of planet radius and orbital period found with TCF. (Bottom) Candidate planets from various surveys superposed on the DTARPS-S heat map for injected planets from Figure \ref{fig:inj_recall_inj}. Gold squares are recovered objects and black points are not recovered.}
   \label{fig:recall_PC}
\end{figure}

Figure \ref{fig:recall_PC} shows recall dependencies for the planet candidates in the surveys/lists listed in \S\ref{sec:other} above. The DIAmante data set contains light curves corresponding to 1,042 stars in these candidate planet samples and has an overall recall rate of 41\% for previously identified planet candidates.  The results are similar to the Confirmed Planets and DIAmante sample. DTARPS-S has strong recall for TCF periods $1-10$ days and for TCF radii $1-10$ $R_{\oplus}$. Recall rate drops to 13\% for TCF periods $<$1 day. At large planet radii, the recall rate drops to 46\% ($10-15$ $R_{\oplus}$) and 4\% ($>$15 $R_{\oplus}$). Note that a handful of planets are recovered with $15-30$~day periods despite the absence of injected planet training for these long periods. 

As these reported candidates are not yet confirmed by spectroscopic observations, these recovery rates do not reflect true planetary populations. Rather, the value of DTARPS-S is to add confidence to the reality of candidates it confirms, and to cast some doubt on those it does not confirm. Appendix A gives more detail on the overlaps between DTARPS-S and these external surveys/lists.

\subsection{False Positives \label{sec:recall_FP}}

A list of False Positives was created by combining False Positives from the TOI list$^1$ available from the NASA Exoplanet Archive \citep{NEA-CP} (accessed March 15, 2022), the cTOI list$^2$ available from the TESS Follow-Up Program website (accessed March 15, 2022), and independent studies conducted by \citet{Affer12}, \citet{Collins18}, \citet{Dressing19}, \citet{Eisner21}, \citet{Feinstein19}, \citet{Kostov19}, \citet{Kruse19}, \citet{Mayo18}, \citet{Olmschenk21}, \citet{Schanche19}, \citet{Tu20}, \citet{vonBoetticher19}, and \citet{Yu19}. Objects labeled as false alarms, flare stars, flare stars with a nearby M companion, False Positives, giant stars, probable eclipsing binaries, probable False Positives, pulsing stars, rotating stars, and planetary candidates with no discernible corresponding radial velocity signal were all considered as False Positives for this analysis. If an object was labeled as a planetary candidate by one study and a false positive by another, than it was considered to be a false positive object.

\begin{figure}[t]
  \centering
   \includegraphics[width=0.75\textwidth]{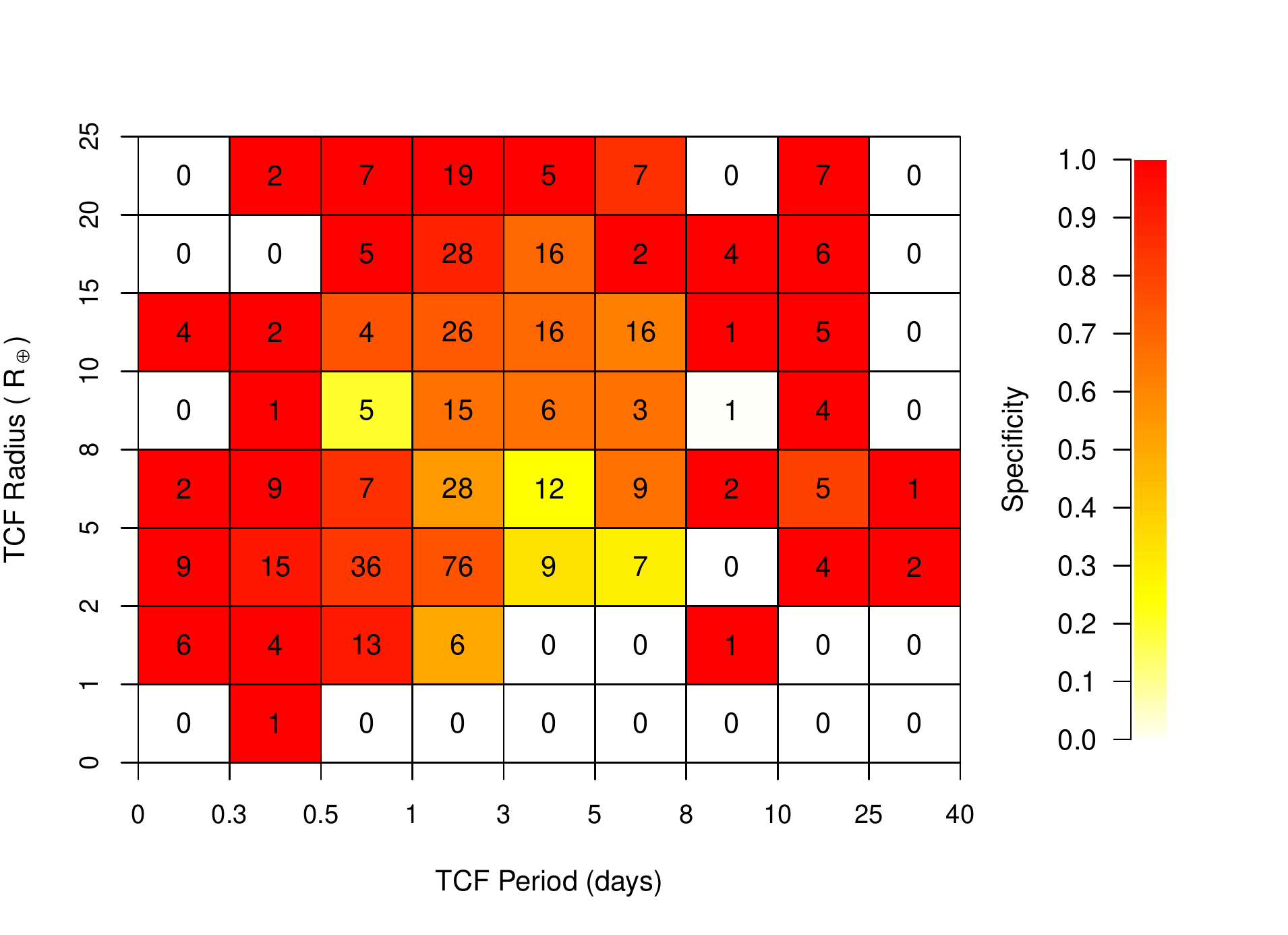}
   \caption{Specificity of DTARPS-S RF classifier for 513 previously identified False Positives. A red colored bin represents a high percentage of False Positives with correct classification as a non-planetary signal. The number of False Positives in each bin are labeled.}
   \label{fig:recall_FP}
 \end{figure}

There are 513 previously identified False Positives in the full DIAmante data set of which 390 (76\%) have TCF best period matching the reported period. The DTARPS-S RF classifier correctly identified 400 of the 513 (78\%) false positive signals as non-exoplanet candidates. In the parlance of statistical classification, this 78\% is the specificity or the True Negative Rate, the probability of obtaining a negative test result given that the object is truly negative. 

Figure \ref{fig:recall_FP} shows the specificity of the RF classifier for different bins in transit signal radius-period space. DTARPS-S has excellent specificity for most periods and radii, even at the extreme values: 90\% for periods $<$1 day, 97\% for periods $>$10 days; 93\% for strong signals with associated radii $>$15 $R_{\oplus}$. {\it DTARPS-S thus effectively removes known False Positives over a wide range of parameters.} 

The specificity in Figure \ref{fig:recall_FP} is similar to a hypothetical inverted Figure \ref{fig:recall_PC} heat map.  DTARPS-S has high specificity where Figure \ref{fig:recall_PC} has low recall rates and vise versa.  The poorer specificity for periods between 1 and 8 days and radii between 2 and 15 $R_{\oplus}$ is likely due to the abundance of injected planetary signals in this region (Figure \ref{fig:inj_orbital_params}c).  There are a number of Confirmed Planets and previously identified planet candidates in this region as well, confirming the concentration of planetary signals.  Although the RF classifier utilizes many features during classification, both the radius and orbital period are important features,  They affect the specificity in this region and the recall rate of Confirmed Planets and planetary candidates at the extreme values.

\subsection{Random Forest Recall Dependence on ARIMA and TCF}

The overall recall rates reported in \S \ref{sec:recall_DIA}-\S \ref{sec:recall_PC} and the completeness heat map in Figure \ref{fig:inj_recall_inj} are dependent on two aspects of the DTARPS-S process: the ability of the TCF periodogram to recover the previously identified signal from ARIMA residuals, and the ability of the RF classifier to recognize the planetary transit signal. The RF classifier was trained only using injected planet signals whose TCF peak period matched the injected period. So the RF classifier is dependent (as with all planetary classification systems) on the TCF peak period.

As shown in \S \ref{sec:TCF_acc}, the accuracy of the best TCF periodogram peak orbital parameters depends heavily on the best TCF peak period matching the injected (or in these cases the existing) period.  The ability of the RF classifier to identify Confirmed Planets and previously identified planet candidates can, therefore, be more accurately reported by calculating the recall rate only for Confirmed Planets and planet candidates whose TCF peak periods match their reported periods. Table \ref{tab:recall_comp} compares the recall rates reported in \S \ref{sec:recall_DIA} $-$ \S \ref{sec:recall_PC} with recall rates calculated using only objects in each category with TCF periods that matched their reported period.  The biggest improvement in the recall rate was the recall rate for the Confirmed Planets and the planet candidates in the TOI list. The recall rate for the full set represents what DTARPS-S can recover from the DIAmante data set and the recall rates for the matched periods represent what DTARPS-S can recover from the processed DIAmante data whose best TCF peak corresponds to real periodicity in the light curve. 

\begin{deluxetable}{rcccc}[tb]
  \centering
    \caption{Recall Rates for the DTARPS-S (Full Set) and the RF classifier alone (Matching Period)} \label{tab:recall_comp}
\tablehead{
          & \colhead{M20} & \colhead{Confirmed} & \colhead{\TESS Objects} & \colhead{Combined} \\
          &  \colhead{Candidates} &  \colhead{Planets} & \colhead{of Interest} & \colhead{Candidates\tablenotemark{a}}
}
\startdata
        Recall for Full Set (\%) & 58.5 & 64.4 & 51.2 & 41.2\\
        Recall for Matched Periods Only (\%) & 64.0 & 78.8 & 68.2 & 54.7\\
\enddata
\tablenotetext{a}{Planet Candidates from the TOI list$^1$, cTOI list$^2$, \citet{Mayo18}, \citet{Dressing19}, \citet{Feinstein19}, \citet{Kostov19}, \citet{Kruse19}, \citet{Yu19}, M20, \citet{Dong21}, \citet{Eisner21}, and \citet{Olmschenk21}
}
\end{deluxetable}

\section{Discussion} \label{sec:summary1} 

\subsection{Overview of the study \label{sec:summary}}

This study is based on the premise that existing searches for transiting exoplanets $-$ even those conducted by the \TESS Science Office producing the TOI lists$-$ have not identified the full detectable population of planetary systems in \TESS FFF light curves (\S\ref{sec:challenges}). The sensitivity and reliability of transit search depends critically on the development and refinement of statistical methodology focused on the complexities of this scientific problem.  The issues are challenging: a wide variety of contaminating stellar and instrumental signatures in the light curves; a highly imbalanced classification problem with imperfect training sets; and limited telescope time to validate the resulting planetary candidates.  

We adopt and refine the AutoRegressive Planet Search (ARPS) procedure developed by \citet{Caceres19b} in an effort called {\it DIAmante TESS AutoRegressive Planet Search} for the southern ecliptic hemisphere, or DTARPS-S. It combines  the time series extraction and preprocessing from the DIAmante project \citep[][M20]{Montalto20}, Box-Jenkins analysis (ARIMA modeling) for light curve detrending, our Transit Comb Filter (TCF) periodogram for transit discovery, and machine learning with Random Forest for optimizing True Positive and minimizing False Positive classifications.  We apply the procedure to $\sim 0.9$ million \TESS Year 1 Full Field Image (FFI) light curves for brighter stars in the southern ecliptic hemisphere (\S\ref{sec:DIAmante}). Best fit ARIMA models, fitted by maximum likelihood estimation with optimized model complexity, are subtracted from the DIAmante light curves (\S\ref{sec:ARIMAi}). The TCF periodogram is then calculated to identify and characterize periodic transit-like behavior in the light curves (\S\ref{sec:TCF}). Statistical virtues of TCF have been eludicated by \citet{Gondhalekar23}. The most likely transit signal period was chosen as the TCF periodogram peak with the greatest robust signal-to-noise ratio after detrending the periodogram.  

Considerable effort was expended to tune a Random Forest (RF) machine learning classifier that identifies exoplanet transit candidates while reducing various sources of contamination (\S\ref{sec:rf_training}). The positive training set was constructed from synthetic planetary injections augmented from confirmed planets in the 4-year \Kepler survey, and the negative training set of random light curves was supplemented with synthetic eclipsing binary (EB) and short period variable injections from M20. Several dozen features drawn from every stage of the analysis were combined with stellar metadata to construct a RF classifier; the final classifier has 37 features with different weights (\S\ref{sec:rf_final}).  

After choice of a threshold of RF prediction value, we produce a list of 7,377 objects called the DTARPS-S Analysis List (\S\ref{sec:DAL}). While it has high recall of True Positives, it is particularly optimized to have a small False Positive Rate. The classifier performance is summarized in Figures \ref{fig:rf_ROC_prec_recall}, \ref{fig:rf_conf_matrix}, and \ref{fig:rf_pred}.  The list has a True Positive Rate of 92.5\% with respect to injections of simulated planets, and False Positive Rate of 0.43\% with respect to simulated astrophysical False Positives (mostly EBs) and random light curves (\S\ref{sec:recall_KP}).  We compare the DTARPS-S Analysis List to other southern hemisphere samples: NASA Exoplanet Archive Confirmed Planet \citep{NEA-CP}, \TESS Objects of Interest, and other transit surveys (\S\ref{sec:other}, Appendix \ref{app:surveys}).

Our classifier has imperfections.  Smaller injected planets are not recovered by the TCF fitting algorithm and the TCF transit depth (scaling to planet radius) is somewhat underestimated. This effect is from overfitting by the ARIMA modeling and underfitting by the TCF matched filter procedure (\S\ref{sec:TCF_acc}). In addition, large-radii stellar companions are fitted with planetay-radii signals, thereby biasing the RF classifier against longer period ($> 8$ days) Jovian planets and short period ($< 1$ day) planets. 

The DTARPS-S completeness heat map of the injected planetary signals for the RF classifier (Figure~\ref{fig:inj_recall_inj}) shows that the DTARPS-S method has poor recall for radii $<$2 $R_{\oplus}$ or periods $<$1 day, low completeness for planets with radii between 2 and 4 $R_{\oplus}$, and high completeness for planes with radii between 8 and 30 $R_{\oplus}$ and periods between 0.6 and 13 days. The distribution of the recall for the confirmed planets, M20 candidates, and previously identified candidates generally follow the results of the injected planet completeness map (\S\ref{sec:complete}).

The principal product of this paper, the DTARPS-S Analysis List (Table~\ref{tab:DTARPS_DAL} available as a Machine Readable Table),  optimizes recall (completeness) at the expense of precision (acceptance of False Positives).  It serves three purposes:
\begin{description}
\item[Potential transiting planets for spectroscopic followup]  This would proceed with the understanding that more than half are likely to be False Alarms (no real periodicity) and False Positives (non-planetary periodicity).  It may be particularly useful for subsets such as very bright host stars.
\item[Intermediate list ready for vetting]  Vetting will increase precision, greatly reducing False Alarms and False Positives, but with reduced recall  (completeness.  This how we proceed in Paper II; see \S\ref{sec:vetting} for discussion.
\item[Support for other surveys]  If a star in the DTARPS-S Analysis List was independently found to be an unconfirmed planetary candidates in the TOI list or another transit search procedure, then confidence in its planetary nature is increased. 
\end{description}

\subsection{Improvements to DTARPS-S methodology}  \label{sec:improvements}

The statistical issues arising in reliable planetary transit identification are complex and differ with each survey.  The of our DTARPS-S effort is based on the ARPS procedures developed in \citet{Caceres19b} and applied to the \Kepler dataset by \citet{Caceres19a}.  We institute a variety of improvements to their methods in our application to \TESS Year 1 light curves: injection-based training sets for both planetary transits and eclipsing binaries with sophisticated data augmentation procedures (\S\ref{sec:kep_inj}); optimized Random Forest algorithm for imbalanced training set (\S\ref{sec:rf_optimize}); extensive engineering of feature selection including \emph{Gaia} information (\S\ref{sec:rf_optimize}); multiple metrics for classification performance (\S\ref{sec:rf_class_metric}); classification training and validation using both injections and Confirmed Planet samples (\S\ref{sec:rf_training}, \S\ref{sec:complete}); and completeness heat maps (\S\ref{sec:complete}).  Based on the results presented here for application to \TESS, a number of further improvement can be envisioned:
\begin{description}

\item[Stellar variation removal]  The linear ARIMA model seems effective in removing  autocorrelated trend for $\gtrsim 90$\% of the DIAmante preprocessed \TESS light curves considered here (Figure~\ref{fig:lb_test_comp}). However, ARIMA was less effective over the 4-year Kepler data, where only 47\% of the ARIMA residuals were consistent with white noise \citep{Caceres19a}.  More elaborate nonlinear autoregressive models might be better for \TESS lightcurves near the ecliptic poles or multi-year lightcurves.  However, detailed examination is needed to reduce overfitting of ARIMA and ARIMAX models for blended eclipsing binaries.  Overfitting may disguise photometric behaviors arising from tidally distorted and mutually illuminating close binaries and erroneously lead to classification as transiting planets. 

\item[Transit depth estimation] The ARIMAX modeling often gave biased estimates based on the ARIMA residuals because some of the planetary signal is incorporated into the ARIMA model (\S\ref{sec:ARIMAXi}). This may not have affected the Random Forest classifier greatly as we use the ARIMAX SNR, rather than the ARIMAX depth, as a feature. But it does affect the astronomical interpretation, and corrections to the estimated planet radius are therefore instituted in Paper II. Also, the trapezoidal-shaped model used as the exogenous variable is probably too simple. A more accurate exoplanet transit with curved ingress and egress is likely to be more effective, as in the Transit Least Squares procedure for identifying exoplanet transits \citep{Hippke19}, at the expense of adding astrophysical parameters to the statistical model.  

\item[TCF sensitivity] For \Kepler 4-year light curves, the Transit Comb Filter periodogram appears to be more sensitive to smaller planets than other periodicity search methods such as the Box-Least Squares periodogram \citep[Figures 9-10 in Caceres et al. 2019a; Figure 10 in Caceres et al. 2019b;][]{Gondhalekar23}.  However, in our \TESS FFI application, the TCF periodogram had a low recall rate for injected planets with radii $\lesssim 4$~R$_\oplus$. The reason for this difference needs to be elucidated. Is it a product of the number of points in the light curve available or is there another factor? Further investigation of TCF, BLS and other periodograms is needed. This can lead to discovery of smaller planets in a given data set which is a driving goal of the \TESS and upcoming PLATO missions \citep{Rauer14}.

\item [Multiple planet systems] Currently, the ARPS procedure only treats the TCF periodogram peak with the strongest SNR and does not search for nor consider multiple transiting planets. Multi-planet systems could be searched for by iterative `pre-whitening' procedure: the strongest planetary signal can be subtracted from the lightcurve, and ARIMA and TCF can be reapplied.  The procedure would be repeated until the TCF peak  effective signal-to-noise falls below the threshold indicated in Figure~\ref{fig:eff_SNR_comp}.  

\item[Classifier features] New features can be added to the Random Forest classifier to better identify true exoplanets or astrophysical False Positives. A feature that quantifies the difference between the TCF periodogram peak and the spurious spike of non-exoplanet transits with a period between 13.5 and 15 days due to the TESS satellite orbit (Figure~\ref{fig:rf_pred}) might mitigate the bias against longer period exoplanets seen in the current DTARPS-S classifier. Features that characterize an EB secondary eclipse or tidal distortions may help reduce EB contamination and reduce the human vetting effort.

\item[Classifier training set] The training set planet properties were derived from injections based on confirmed planets in the \Kepler sample.  But with 4 year lightcurves, rather than 1 month typical of \TESS FFIs, most of the injected planets are too small.  The large number of undetectable planets in the positive training set may have distorted the classifier.  A better match to \TESS sensitivity might improve classifier performance.  In addition,  a larger number of planet injections may be helpful in regions of the period-radius diagram where the recall rate is transitioning between low and high ($2-5$~R$_\oplus$), where true hot Jupiters compete with False Positive EBs, and near the edges of the heat map (Figure~\ref{fig:inj_recall_inj}). Finally, the distribution of injected EB signals might be adjusted to approximate the expected dilution in blended systems.  

\item[Specialized classifiers] Random Forest or other classifiers might be trained for particular sub-populations of \TESS FFI stars such as Sun-like stars, lower mass K and M stars, subgiants, or stars in the continuous observing zone near the ecliptic poles. This would require new training sets of injections on light curves of just these stellar host sub-populations. 

\item[Classifier type] The Random Forest classifier developed in \S\ref{sec:rf_optimize} is highly effective, but the 0.43\% False Positive rate is high given the huge class imbalance. This necessitates a  complex vetting process (Paper II). Improved performance might be achieved with different machine learning classifiers such as XGBoost, LightGBM, GOSDT, or Explainable Boosting Machines. XGBoost \citep{Chen16} is very similar to Random Forest but, rather than building decision trees independently, new trees are built iteratively to minimize classification error. LightGBM \citep{Ke17} is also similar to Random Forest and XGBoost but grows trees leaf-wise rather than level-wise. Explainable Boosting Machines \citep{Lou12} use small forests of decision trees for each feature in a linear regression ensemble. GOSDT \citep{Lin20} is a modern approach to fast decision tree optimization. 

\end{description}

\subsection{The Need for Vetting} \label{sec:vetting}

A False Positive Rate of 0.43\% multiplied by a test sample of 823,099 DIAmante light curves predicts that $\sim 3500$ False Positives, about half of 7,377, will be present in the DTARPS-S Analysis List of stars passing the classifier threshold. This includes possible False Alarms (i.e. cases where no significant periodicity is present despite the high RF classification probability) or False Positive signals from unmodeled stellar or instrumental variability. Despite our careful statistical efforts, the DTARPS-S Analysis List  is dominated by non-planetary signals. The full list thus cannot be accepted for reliable calculation of exoplanet populations, such as converting the completeness heat maps into planetary occurrence rates, and it is an inefficient list for follow-up observations and spectroscopy with valuable telescope resources.

\begin{figure}[t]
  \centering
  \includegraphics[width=0.48\textwidth]{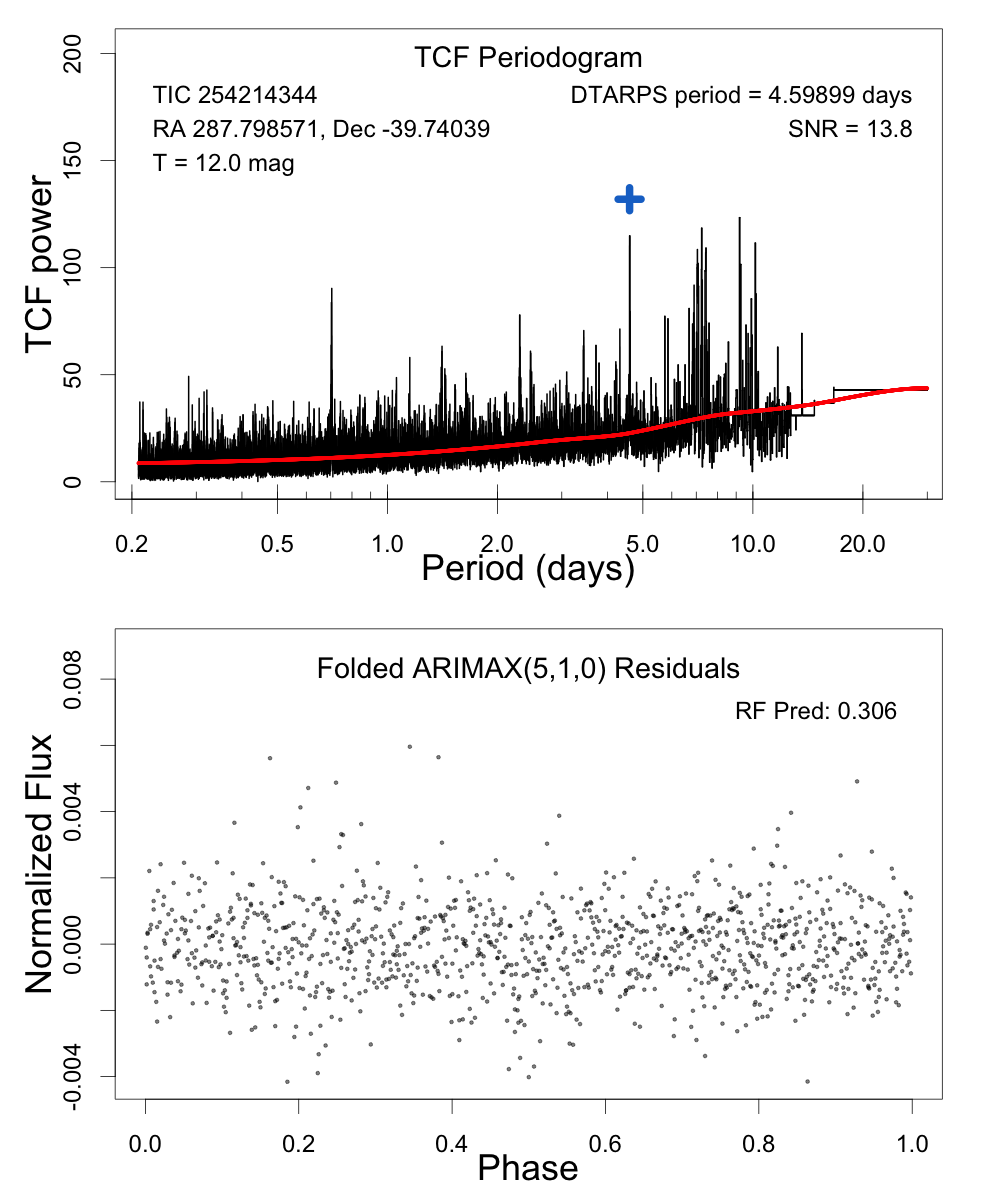}
  \includegraphics[width=0.48\textwidth]{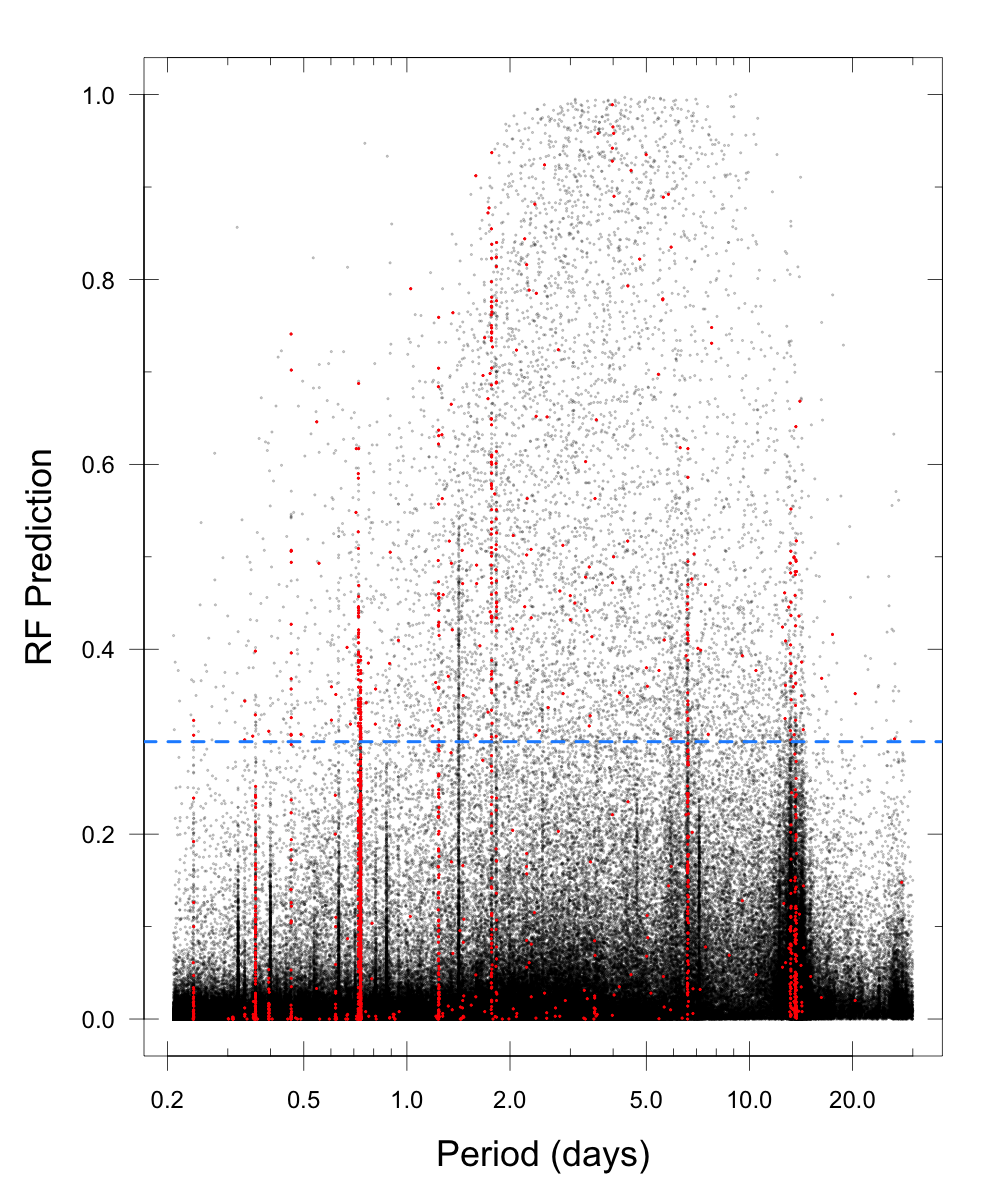}
  \caption{Two sources of contamination that affect the DTARPS-S Analysis List of potential planetary transits. (Left) TCF Periodogram and folded ARIMAX residual light curve of TIC 254214344 that shows an unconvincing peak in the TCF periodogram with no clear transit in the folded light curve. (Right) Random Forest prediction score for the full set of DIAmante extracted light curves showing clusters of stars subject to ephemeris matching and periodic satellite operations as red points.}
  \label{fig:rf_pred_full_sample}
\end{figure}

Fortunately, a suite of `vetting' operations can be conducted to cull many of the False Positives and False Alarms. Two examples of the effects to be removed are shown in Figure~\ref{fig:rf_pred_full_sample}: A TCF periodogram and folded light curve without convincing signal that nonetheless passed the RF threshold; and groups of stars with similar periods due to satellite operations associated with the 13.7 day orbit or light contamination from a bright star \citep[this is known as the `ephemeris matching' problem;][]{Coughlin14}. Other effects to be removed include centroid wobbling in the FFI image, possible photometric binaries in Gaia photometry, and deep secondary eclipses inconsistent with planetary radii. These and other vetting operations are applied in Paper II to give a more reliable, though less complete, catalog of DTARPS-S Candidate Planets for the Year 1 \TESS FFI DIAmante light curves.

\begin{acknowledgements}

E.J.M. and E.D.F. benefit from the vibrant community of Penn State's Center for Exoplanets and Habitable Worlds that is supported by the Pennsylvania State University and the Eberly College of Science.  This study is also a product of the Center for Astrostatistics supported by the Eberly College of Science.  We benefited from comments on the manuscript by members of these Centers: Ian Czekela, Rebekah Dawson, Hyungsuk Tak, Jason Wright, as well as Joel Hartman (Princeton). We also appreciate comments by a thoughtful anonymous referee. 

This paper includes data collected by the TESS mission. Funding for the TESS mission is provided by NASA’s Science Mission directorate. We acknowledge the use of public TOI Release data from pipelines at the TESS Science Office and at the TESS Science Processing Operations Center.  This research also uses the NASA Exoplanet Archive \citep{NEA-CP} operated by the California Institute of Technology, under contract with the National Aeronautics and Space Administration under the Exoplanet Exploration Program. Data from the European Space Agency mission {\it Gaia}, processed by the {\it Gaia} Data Processing and Analysis Consortium (DPAC). Funding for the DPAC has been provided by national institutions, in particular the institutions participating in the {\it Gaia} Multilateral Agreement.  

We acknowledge use of {\it Lightkurve} software, a Python package for Kepler and TESS data analysis \citep{lightkurve18}, as well as Astropy, a community-developed core Python package for Astronomy \citep{astropy13, astropy18}.

\end{acknowledgements}

\software{R core \citep{Rcore}: \emph{forecast} \citep{Hyndman21}, \emph{goftest} \citep{goftest19}, \emph{moments} \citep{Lukasz15}, \emph{nortest} \citep{nortest15}, \emph{randomForestSRC} \citep{Ishwaran22}, \emph{ROCR} \citep{Sing05}, \emph{smotefamily} \citep{smotefamily19}, \emph{tseries} \citep{tseries19}, \emph{KernSmooth} \citep{kernsmooth20}; 
Python: \texttt{Jupyter} \citep{Jupyter16}, \texttt{LightKurve} \citep{lightkurve18}, \texttt{Astropy} \citep{astropy13, astropy18}, \texttt{Astroquery} \citep{astroquery19}, \texttt{tesscut} \citep{tesscut19}, \texttt{NumPy} \citep{numpy11, numpy20}, \texttt{Matplotlib} \citep{matplotlib07, matplotlib16}
}

\facility{\TESS, NASA Exoplanet Archive, \emph{Gaia}}

\bibliography{Paper_i_revised.bib}{}
\bibliographystyle{aasjournal}

\appendix

\section{Other Planet Searches \label{app:surveys}}

External lists and surveys that are compared to DIAmante samples and DTARPS-S Analysis List are briefly described here.  The list of confirmed planets were taken from the list of confirmed planets on the NASA Exoplanet Archive (NEA, accessed March 15, 2022), some of which are \TESS Objects of Interest$^1$ (TOIs).  Previously identified exoplanet candidates were found in the list of community TOIs$^2$ (cTOIs) at the TESS EXOFOP webste (accessed March 15, 2022) with contributions from \citet{Mayo18}, \citet{Dressing19}, \citet{Feinstein19}, \citet{Kostov19}, \citet{Kruse19}, \citet{Yu19}, M20, \citet{Dong21}, \citet{Eisner21} and \citet{Olmschenk21}. Previously identified astrophysical False Positives were found in \citet{Affer12}, \citet{Collins18}, \citet{Schanche19}, \citet{vonBoetticher19}, and \citet{Tu20}.

\subsection{Confirmed Planets from NEA}

The Confirmed Planets used here includes planets with published refereed planet confirmation papers. The NEA Confirmed Planets in the DIAmante data set were identified with the TIC ID number. In the case of multi-planet systems, we used the planet whose reported period best matched the TCF peak period. For planets with multiple entries, an average of reported orbital parameters were used. In total, 184 Confirmed Planet hosts on the NASA Exoplanet Archive were matched with objects in the DIAmante data set.

\subsection{TOI List}

The TOI list$^1$ reports dispositions of known planet (KP), confirmed planet (CP), planetary candidate (PC), ambiguous planetary candidate (APC), or false alarm or false positive (FP). We combined CPs and KPs to be confirmed planets.  Objects with recent confirmation in unpublished papers on arXiv are considered to be Confirmed Planets. We combined the APCs and PCs when considering planet candidates, and combined FAs and FPs in when considering False Positives.

Of the 1,036 objects in the TOI catalog that overlap with the DIAmante data set, 185 were labeled confirmed planets, 670 were labeled planetary candidates, and 181 were labeled False Positives.

\subsection{cTOI List}

The cTOI list$^2$ has a wide range in the quality of planet candidates: followup EXOfop examination shows some are False Positives while others are promoted to planetary candidates on the TOI list. The DIAmante sample has 566 cTOIs; 364 are planetary candidates from M20 discussed below.

\subsection{Affer et al. 2012}

\citet{Affer12} measured rotation and binarity of field stars from the COnvection ROtation and planetary Transits (CoRoT) satellite for stars in the solar neighborhood. Forty objects in Table 2 of \citet{Affer12} were matched with objects in the DIAmante data set, one of which is in our  DTARPS-S Analysis List. Affer et al. report a rotation period of 72 days for TIC 234091431 and we report a TCF period of 2.76592 days.  Of the other 39 objects, only three have rotational or pulsational periods or pulsation periods that matched the TCF peak period.  We abel these as False Positives in the DIAmante data set. 

\subsection{Collins et al 2018} 

\citet{Collins18} identified and classified False Positives in Kilodegree Extremely Little Telescope (KELT) light curves. They classified over one thousand transit like signals in KELT as False Positives through photometric and spectroscopic  observations in several classes: single-line spectroscopic binaries, multi-line spectroscopic binaries, spectroscopic giant stars, eclipsing binaries, blended eclipsing binaries, variable stars, nearby eclipsing binaries (blended in the KELT aperture), and stars with no significant radial velocity detected. The DIAmante samples has 156 objects matched in \citet{Collins18}, 19 of which are in the DTARPS-S Analysis List. We consider these to be previously identified False Positives.

\subsection{Mayo et al. 2018}

\citet{Mayo18} identified 275 planet candidates in the NASA's K2 mission, Campaigns 0-10, and estimated False Positive probability with the \emph{vespa} package.  The DIAmante samle has 21 objects examined by \citet{Mayo18}, one of which, TIC 21184505, is in the DTARPS-S Analysis List. Another object, TIC 68694240, was a probable eclipsing binary that we label as a False Positives.

\subsection{Dressing et al. 2019}

\citet{Dressing19} performed spectroscopic and photometric characterization for 172 K2 target stars identified as candidate hosts of transiting planets. They identified giants, likely eclipsing binaries, and cool dwarf stars. The DIAmante sample matches 8 of these stars with one, TIC 438338723, in the DTARPS-S Analysis List. It is a probable eclipsing binary that we lael as a False Positive.

\subsection{Feinstein et al. 2019}

\citet{Feinstein19} developed \texttt{elenor}, an open-source tool for extracting light curves from TESS FFIs. They applied the method to \TESS Sector 1 Year 1 data and vetted by visual examination. The DIAmante sample matches 16 of their objects, three of which are in the DTARPS-S Analysis List: TIC 159835004 and TIC 299780329 previously identified as planetary candidates,  and TIC 38813184 is identified as an eclipsing binary.  The reported EB period for TIC 38813184 matches the TCF peak period. We include TIC 38813184 in the previously identified False Positive list.

\subsection{Kostov et al. 2019}

\citet{Kostov19} created an open source automatic vetting pipeline for K2 data called Discovery and Vetting of Exoplanets (DAVE). They applied DAVE to 772 planet candidates from K2 and vetted the candidates either as planet candidates or False Positives. Of the 30 objects that match the DIAmante stars, TIC 21184505, TIC 294301883 and TIC 366443576 in the DTARPS-S Analysis List were labeled  as planetary candidates by DAVE.  All three objects had a TCF peak period that matched the reported DAVE period.

\subsection{Kruse et al. 2019}

\citet{Kruse19} identified 818 planetary candidates and 1060 eclipsing binary systems in Campaigns 0-8 of the K2 mission usong the EVEREST pipeline.  The DIAmante samples matches 44 objects, two of which are in the DTARPS-S Analysis List: planetary candidate TIC 294301883 and eclipsing binary TIC 438338723.  The reported periods from Kruse et al. match the TCF peak period.   

\subsection{Schanche et al. 2019}

\citet{Schanche19} presented a catalog of 1,041 False Positives from the SuperWASP survey of the northern hemisphere that had been identified as potential planetary candidates previously and rejected after follow-up observations. The False Positives were classified as eclipsing binaries, blended eclipsing binaries, and low mass eclipsing binaries. The DIAmante sample matches 47 objects, 12 of which lie in the DTARPS-S Analysis List with the following classifications: TIC 16490297 as an eclipsing binary system; TIC 61069470, TIC 117549305, TIC 13675776, TIC 271269442 and TIC 271374913 as blended eclipsing binaries; TIC 9433212, TIC 12529950, TIC 264537668, TIC 277712294, TIC 443618156, and TIC 449050248 were labeled as low mass eclipsing binary systems. Most, but not all, have TCF periods matching the SuperWASP periods.   

\subsection{von Boetticher et al. 2019}

\citet{vonBoetticher19} characterized 10 low mass stars part of low mass eclipsing binary systems as part of the EBLM project. The DIAmante sample has six of these systems with TCF period matching the reported period. Four lie in the DTARPS-S Analysis List: TIC 101395259, TIC 277712294, TIC 350480660 and TIC 734505581.

\subsection{Yu et al. 2019}

\citet{Yu19} modified an neural network classifier, developed by \citet{Shallue18} for identifying \Kepler planet candidates, for \TESS data. Applying the classifier to Year 1 Sector 6 \TESS data and, accompanied by visual vetting, 288 new planetary candidates were identified.  The DIAmante sample matches 140 objects of which 65 are in the DTARPS-S Analysis List. In all cases, the TCF peak period agreed with the period reported in Yu et al. We lavel these as previously identified planetary candidates.

\subsection{Montalto et al. 2020}

Of the 394 candidates identified by M20 in the DIAmante study, 364 were in the set of light curves classified by the DTARPS-S Random Forest. These are identified by a flag in the DTARPS-S Analysis List; see \S\ref{sec:recall_DIA} for details.  These include 221 in the NEA Confirmed Planet list, the TOI list on the NEA, or in other external surveys. Altogether 82 are  identified as Confirmed Planets. The M20 objects were placed on the cTOI list: 82 are labeled as planet candidates, 18 as ambiguous planet candidates, and 26 as False Positive. These include 13 DIAmante candidates independently listed as planetary candidates by \citet{Yu19}, and 2 identified as low mass EBs by \citet{vonBoetticher19}.

\subsection{Tu et al. 2020}

\citet{Tu20} studied superflares and other properties of 400 solar-type stars in  \TESS Year 1 data.  Of the 277 stars in the DIAmante data set that had flares identified by Tu et al., only 57 had TCF peak periods that matched their stellar rotational periods. Six flare stars are in the DTARPS-S Analysis List. In the two cases where stellar rotational period matched the TCF peak period,  TIC 121048789 and TIC 373844472, DTARPS-S is likely identifying the rotational period rather than a transiting planetary period. 
 
\subsection{Dong et al. 2021}
 
\citet{Dong21} identified and characterized 55 Warm Jupiters in \TESS Year 1 FFIs. The DIAmante sample has 40 of these systems of which 21 lie in the DTARPS-S Analysis List.  Of these, 20 had TCF periods matching those reported by Dong et al.; the exception is TIC 73038411.
 
\subsection{Eisner et al. 2021}
 
\citet{Eisner21} presented results from the Planet Hunters \TESS citizen science project for the first two years of the \TESS survey. They identified 90 new planetary candidates of which 18 lie in the DIAmante sample.  However, none of the overlapped objects have a TCF peak period that matches the reported period from Eisner et al.. This is partly due to their single transit events where the period of the planet was estimated from the transit duration. Two of their objects are in the DTARPS-S Analysis List. TIC 142087638 and TIC 404518509 were identified as single transit events by \citet{Eisner21}; the TCF periodogram gives accurate periods within the error bars of their single transit event estimate.  
 
\subsection{Olmschenk et al. 2021}
 
\citet{Olmschenk21} applied a convolutional neural network to \TESS FFI light curves to identify planetary candidates followed by visual vetting. Of their 185 planet candidates, 25 overlap with the DIAmante sample, all of which have TCF peak periods that match their reported periods. The DTARPS-S Analysis List recovers 15 of their planet candidates.

\end{document}